\documentclass[aps,amssymb,superscriptaddress,nofootinbib]{revtex4}
\usepackage{setspace}
\usepackage{graphicx}
\usepackage{psfrag}

\newcommand{\kstar}{{k^*}}
\newcommand{\Screll}{{\mathcal L}}
\newcommand{\Scrim}{{\mathcal M}}

\def\be{\begin{equation}}
\def\ee{\end{equation}  }
\def\bea{\begin{eqnarray}}
\def\eea{\end{eqnarray}  }
\def\rg{\sqrt{-g}}

\newcommand{\msun}{{M}_\ensuremath{\odot}}

\date{August 12 2007}

\begin{document}
\title{Binary Black Hole Coalescence}

\author{Frans Pretorius}
\affiliation{Department of Physics, Princeton University, Princeton, NJ 08544}
\begin{abstract}
The two-body problem in general relativity is reviewed, focusing
on the final stages of the coalescence of the black holes as uncovered
by recent successes in numerical solution of the field equations.
\end{abstract}
\maketitle
\vspace{-.15in}
\tableofcontents
\section{Introduction}
A black hole is one of the most fascinating and enigmatic
predictions of Einstein's theory of general relativity.
Its interior can have rich structure and is intrinsically
dynamical, where space and time itself are inexorably 
led to a singular state. The exterior of an isolated black hole
is, on the other hand, remarkably simple, described uniquely by the stationary
Kerr solution. The dynamics of black holes are governed
by laws analogous to the laws of thermodynamics, and indeed
when quantum processes are included, emit Hawking radiation
with a characteristic thermal spectrum. Most remarkable however,
is that black holes, ``discovered'' purely through thought and 
the mathematical exploration of a theory far removed from every 
day experience, appear to be ubiquitous objects in our universe.

The evidence that black holes exist, though circumstantial,
is quite strong~\cite{Narayan:2005ie}. The high luminosity of quasars
and other active galactic nuclei (AGN) can be explain
by gravitational binding energy released through gas accretion onto 
supermassive ($10^6-10^9 \msun$) black holes at the centers
of the galaxies~\cite{Rees:1984si,Ferrarese:2004qr}, several dozen X-ray binary
systems discovered to date have compact members too massive
to be neutron stars and exhibit phenomena consistent with
matter interactions originating in the strong gravity
regime of an inner accretion disk~\cite{McClintock:2003gx}, and the
dynamical motion of stars and gas about the centers of nearby galaxies
and our Milky Way Galaxy infer the presence of very massive, compact
objects there, the most plausible explanation being supermassive
black holes~\cite{Gebhardt:2000fk,Schodel:2002py,Ghez:2003qj}.

To conclusively prove that black holes exist one needs to ``see'' them,
or conversely see the compact objects masquerading as black holes.
The only direct way of observing black holes is via the gravitational
waves they emit when interacting with other matter/energy (an isolated
black hole does not radiate). The quadrupole formula says that
the typical magnitude $h$ of the gravitational waves emitted
by a binary with reduced mass $\mu$ on a circular orbit 
measured a distance $r$ from the source is (for a review
of gravitational wave theory see~\cite{Flanagan:2005yc})
\be
h=\frac{16 \mu v^2}{r},
\ee
where $v$ is the average tangential speed of the two members
in the binary (and geometric units are used---Newton's
constant $G=1$ and the speed of light $c=1$). This formula
suggests that the strongest sources of gravitational waves 
are simply the most massive objects that move the fastest.
To reach large velocities in orbit, the binary separation
has to be small; black holes, being the most compact objects
allowed in the theory, can reach
the closest possible separations and hence largest orbital
velocities. Therefore, modulo questions about source populations
in the universe, a binary black hole interaction offers 
one of most promising venues of observing black holes through
gravitational wave emission.

Joseph Weber pioneered the science of gravitational
wave detection with the construction of resonant bar detectors.
Weber claimed to have detected gravitational waves~\cite{Weber:1969bz}, though
no similar detectors constructed following his claims
were able to observe the putative (or any other) source, and the general 
consensus is that given the sensitivity of Weber's
detector and expected strengths of sources it is very
unlikely that it was a true detection~\cite{Thorne:1980rt}.
Note that the {\em existence} of gravitational waves is
not in doubt---the observed spin down rate of the
Hulse-Taylor binary pulsar~\cite{Hulse:1974eb}
 and several others discovered
since, is in complete accord with the general relativistic
prediction of spin down via gravitational wave emission.
Today a new generation of gravitational wave detectors
are operational, including laser 
interferometers (LIGO~\cite{LIGO}, VIRGO~\cite{VIRGO},
GEO600~\cite{GEO}, TAMA~\cite{TAMA}) and resonant bar detectors
(NAUTILUS~\cite{NAUTILUS}, EXPLORER~\cite{EXPLORER}, 
AURIGA~\cite{AURIGA}, ALLEGRO~\cite{ALLEGRO}, NIOBE~\cite{NIOBE}).
A future space-based observatory is planned (LISA~\cite{LISA}),
and pulsar timing and cosmic microwave background polarization
measurements also offer the promise of acting as
gravitational wave ``detectors'' (for reviews 
see~\cite{Maggiore:1999vm,Cutler:2002me}).
The ultimate success of gravitational wave detectors, in particular
with regards to using them as more that simply detectors,
but tools to observe and understand the universe, 
relies on source modeling to predict the structure of the waves
emitted during some event. Even if an event is detected with a
high signal-to-noise ratio (SNR), there simply is not enough information
contained in such a one dimensional time series to ``invert'' it to
reconstruct the event; rather template banks of theoretical waveforms from
plausible sources need to be built and used to decode the signal. 
In rare cases an electromagnetic counterpart may be 
detected, for example during a binary neutron star merger if
this is a source of short gamma ray bursts, which could identify
the event without the need for a template. Though even in such
a case, to extract information about the event, its environment, etc.
requires source modeling.

Gravitational wave detectors have therefore provided much of the 
impetus for trying to understand the nature of binary
black hole collisions, and the gravitational waves emitted 
during the process. However, from a theoretical perspective
black hole collisions are fantastic probes of the dynamical,
strong-field regime of general relativity. What is already
know about this regime---the inevitability of spacetime
singularities in gravitational collapse via the singularity
theorems of Penrose and Hawking~\cite{Hawking:1969sw,hawking_ellis}; the spacelike,
chaotic ``mixmaster'' nature of these
singularities conjectured by Belinsky, Khalatnikov and 
Lifshitz (BKL)~\cite{Berger:1998us};
the null, mass-inflation singularity discovered by Poisson
and Israel~\cite{Poisson:1990eh} that, together with regions of BKL
singularities could generically describe the interiors
of black hole; the rather surprising discovery of
critical phenomena in gravitational collapse by Choptuik~\cite{Choptuik:1992jv,Gundlach:1999cu};
etc---together with the sparsity
of solutions (exact, numerical or perturbative), suggests
there is potentially a vast landscape of undiscovered phenomena.
Of particular interest, and potential application to 
high energy particle collision experiments, are
ultra-relativistic black hole collisions. It is
beyond the scope of this article to delve 
much into these aspects of black hole coalescence,
though a brief overview of this will be given
in Sec.~\ref{sec_he}.

The two body problem in general relativity, introduced
in more detail in Sec.~\ref{sec_2bdy}, is a very rich
and complicated problem, with no known closed-form 
solution.
Perturbative analytic techniques have been developed
to deal with certain stages of the problem, in particular
the inspiral prior to merger and ringdown after merger.
Numerical solution of the full field equations are
required during the merger, and this aspect
of the problem is the main focus of this
article. Much effort has been expended by the community
over the past 15-20 years to numerically
solve for merger spacetimes, and within the last
two years an understanding of
this phase of the two body problem is finally being attained. 
Sec.~\ref{sec_num}
summaries the difficulties in discretizing the field
equations, and describes the methods known at
present that work for black hole collisions, namely
{\em generalized harmonic coordinates} and {\em BSSN} 
(Baumgarte-Shapiro-Shibata-Nakamura) with moving
punctures. Preliminary results are discussed in Sec.~\ref{sec_res},
though given the rapid pace at which the field
is developing much of this will probably be dated in
short order. Sec.~\ref{sec_imp} concludes with a 
discussion of some astrophysical
and other implications of the results.

\section{The two-body problem in general relativity}\label{sec_2bdy}

Consider the classical two body problem of finding the
motion of two masses interacting only
via the Newtonian gravitational force, and given the initial
positions and velocities of the objects. The solution is
well known---in the center of mass frame each body travels
within the same plane along a conic section with a focus at the center of mass,
and the type of conic (ellipse, hyperbola, parabola) depends on the
net energy of the system (bound, unbound, marginally unbound).
In Newtonian gravity this setup is an idealization to the dynamics
of two ``real'' objects in that one treats them as point sources
without any internal structure. If one were to extend the problem
to include the structure of the bodies there would be an infinite
class of two body problems depending on the nature of the material
composition of the objects. 

In general relativity (GR) the two body
problem is on one hand a significantly more challenging problem
than its Newtonian counterpart due to the complexity of solving
the Einstein field equations, yet if attention is restricted to the
vacuum case is much simpler in that one can formulate the {\em exact
problem without idealization}: given an initial spacelike slice of a 
vacuum spacetime containing two black holes, what is the subsequent evolution
of the spacetime exterior to the event horizon?\footnote{This is not 
a technically precise definition, as the global
structure of the spacetime is being ignored, and to capture the spirit
of the two body problem in a technically precise manner applicable
to situations in our universe would probably require defining
it using the concepts of isolated horizons~\cite{Ashtekar:2004cn},
and furthermore restrict the solution to
the future domain of dependence of an initial finite volume of the 
spacetime.}. If Penrose's cosmic censorship conjecture holds the solution will be unique
and entirely independent of the interior structure of the black holes due
to the causal structure of the spacetime. A wrinkle in this clean picture
of the two body problem in GR is that now, rather than a simple set
of mass, position and velocity parameters, there are infinitely many
degrees of freedom required to describe the initial conditions. These
include the initial properties of each black hole and the gravitational
wave content of the spacetime. To constrain the possibilities one could
restrict the class of initial conditions to black holes that were,
at some time in the past, sufficiently separated to each be well described
by a Kerr metric with given mass and spin vector, require the initial
spacetime slice to have ``minimal'' gravitational wave content and possess an 
asymptotic structure such that the black hole positions and relative velocities 
can unambiguously be defined. This class of initial conditions will 
cover the vast majority of conceivable astrophysical black hole 
binary configurations, and the black hole scattering problem setup
discussed later on.

So what makes the two body problem so interesting in GR? First of all,
it almost goes without saying that as gravity is one of the fundamental forces
influencing our existence and shaping the structure of the cosmos,
and since GR is the best theory of gravity at our disposal,
we want to understand all the details of
of the more basic interactions in GR. A less prosaic reason to study
this problem is the rich and fascinating phenomenology of solutions: what
in Newtonian gravity is entirely describable by the mathematics
of conic sections is now a problem that is unlikely to a have
a closed form solution in any but the most trivial scenarios,
featuring regimes with complicated orbital dynamics,
and is accompanied by the emission of gravitational 
radiation. It is this latter feature which has the most profound 
implication: the two body problem in GR for any bound system
is {\em unstable}, and will eventually result in the decay
of the orbit and collision of the two black holes. If cosmic censorship
holds the collision will always result in a single black hole,
and then, from the ``no hair'' theorems of
Israel and Carter~\cite{Israel:1967wq,Carter:1971zc,Carter:1997im},
one knows that the exterior
structure of the remnant black hole will eventually settle down 
to the Kerr solution. For an idea of how unstable
orbits are to gravitational radiation, the time to merger
$t_m$ in units of the Hubble time $t_H$ for an equal
mass binary with each black hole having a mass $M$, initially separated
by $R_0$ times the Schwarzschild radius $R_s=2GM/c^2$, is
roughly
\be\label{thub}
\left(\frac{t_m}{t_H}\right) \approx \left(\frac{M}{\msun}\right) 
\left(\frac{R_0}{10^6 R_s}\right)^4
\ee
For example, two solar mass black holes initially a million
Schwarzschild radii, or $\approx 3\times 10^6 km$ apart, will
merge within a Hubble time; two $10^9 \msun$ supermassive
black holes need to be within $\approx 6\times 10^3$ of their 
Schwarzschild radii, or roughly 1 parsec, to merge within the
age of the universe.

In the following section the qualitative features
of the two body interaction in general relativity are described.

\subsection{Stages of a merger}

This article is primarily concerned with numerical solution of
the field equations as a tool to study the two body problem.
However, it is only in the final stages of coalescence where
full numerical solution is required to obtain an accurate
depiction of the spacetime. This stage of a merger occurs
on a very short time scale compared to other phases of the
two body interaction, which is fortunate, for due to the
computational complexity of solving the field equations
it is not feasible to evolve the spacetime for times much longer
than this. In the following two sections a more detailed discussion
of the various stages of a merger is given, in particular
to set the scope for the remainder of the article and to highlight
how much interesting phenomenology in the two body problem
is {\em not} addressed by full numerical simulation. We break the
discussion up into two classes of merger, ``astrophysical'', and
the black hole scattering problem. A merger scenario is considered
astrophysical if, to some approximation and non-negligible likelyhood
the initial conditions could be realized by a binary system 
in our universe.
The latter classification deals with the gedanken experiment
of colliding two black holes with ultra-relativistic initial
velocities and with an impact parameter of order
the total energy of the system or less.

One reason for this classification is that we {\em might} expect
very different qualitative behavior of the spacetime in 
these two cases. Consider two black holes of mass $m_1$ and
$m_2$ with net ADM~\cite{ADM} energy $E$ in the
center of mass frame\footnote{Here we consider $m_1$ and $m_2$
to be the total BH mass including spin energy, so not
the irreducible mass.}. All astrophysical mergers are expected
to take place in the rest-mass dominated regime where 
$(m_1 + m_2)/E \approx 1$, while
in the scattering problem the kinetic energy of the black 
holes will dominate so that $(m_1+m_2)/E\approx 0$. 
In the latter regime the geometry of each black hole
also gets length contracted into a pancake-like region,
with the actual black holes occupying an ever smaller region
of the non-trivial geometry as the boost factor increases.
In fact, eventually it does not matter that it was 
a black hole that was boosted to large energies---any compact source will
produce the same geometry in the limit. The ultra-relativistic
limit might also be an interesting place to
look for violations of cosmic censorship---the collision
of plane-fronted gravitational waves generically leads
to the formation of naked singularities~\cite{Khan:1971vh},
and though not exactly analogous to high energy black hole
collisions, there are enough similarities that it would
be worth while to explore this regime of the two body problem in some detail.
Note that the particular value of $E$ is not
relevant; there
is no intrinsic length scale in the field equations of general
relativity, and any solution with energy $E_0$
can trivially be re-scaled to a ``new'' solution with arbitrary
energy $E$. 

\subsubsection{Astrophysical binaries}

Here the astrophysical merger scenario is broken down into four 
stages: {\em Newtonian, inspiral, plunge/merger} and {\em ringdown}.

{\bf Newtonian:} In this stage the two black holes are sufficiently
far apart that gravitational wave emission will be too weak
to cause the binary to merge within a Hubble time $t_H$ (\ref{thub}).
Thus, to have any hope of observing mergers of binaries formed in this
stage, other ``Newtonian'', non-two-body processes need to
operate. For example, in the stellar mass range, it
is unlikely that a close black hole binary could be formed 
as the end point of the evolution of a massive binary star system.
The reason is that at the requisite separations for a subsequent
gravitational wave driven inspiral within $t_H$, the stars
will most likely evolve through a common envelope phase, and
recent results have suggested this will cause a merger
of the stellar cores before a binary black hole system
could be formed~\cite{Belczynski:2006zi}. A likely mechanism
then to produce hard binaries is through n-body interactions
that occur in dense cluster 
environments~\cite{Portegies Zwart:1999nm,O'Leary:2005tb}. 
For supermassive
black hole binaries, which are thought to form during galaxy
mergers in the hierarchical structure formation 
scenario~\cite{White:1977jf,Springel:2005nw}, gas interactions,
dynamical friction and other n-body processes are thought
cable of driving most black holes close enough so 
that gravitational wave emission can take over and
cause a merger~\cite{mmmm,Colpi:2007bh}. If such processes did not 
operate efficiently it would be in apparent contradiction with the observations
that most galaxies harbor supermassive black holes at 
their centers~\cite{Rees:1984si,Ferrarese:2004qr}. 

{\bf Inspiral:} In the inspiral regime gravitational wave
emission becomes the dominant process driving the black holes
to closer separation, though the orbital time scale is still
much shorter than the time scale over which orbital parameters change.
The majority of non-extreme mass ratio binaries are expected to ``form''
with sufficiently large semi-major axis that the orbit will 
circularize via gravitational wave emission long 
before the binaries merge~\cite{Peters:1963ux,Peters2}. Some exceptional cases might
be stellar and intermediate mass binaries in dense star clusters,
where numerous interactions with neighbors could frequently
perturb the orbit, or triple systems where the Kozai mechanism 
operates~\cite{Miller:2002pg,Wen:2002km,Portegies Zwart:2004vm,
Gultekin:2005fd,Mandel:2007hi,Aarseth:2007wv}.
On the other hand, the majority of extreme
mass ratio systems that will merge within the Hubble time are expected
to have sizable eccentricities at merger\cite{Gair:2004iv,AmaroSeoane:2007aw}.
Note however that much of the theory behind the formation mechanisms and 
environments of binary black holes are not well known (indeed, no candidate
binary black hole system has yet been observed), which offers gravitational
wave detection a fantastic opportunity to help decipher some of these
interesting questions.

The inspiral phase is well modeled by post-Newtonian (PN) 
methods~\cite{Blanchet:2002av}. For initially non-spinning, zero
eccentricity binaries the higher order PN approximations 
and {\em effective one body} (EOB)
resummations~\cite{Buonanno:1998gg}
give waveforms that are surprisingly close to full numerical 
results even until very close to merger, well beyond
when naive arguments suggest they should 
fail~\cite{Buonanno:2006ui,Baker:2006ha,Pan:2007nw,Hannam:2007ik,Buonanno:2007pf};
comparisons for more generic scenarios have yet to be made.
Extreme mass ratio inspirals (EMRIs) can also be well described
by geodesic motion in a black hole background together
with prescriptions for computing the gravitational wave
emission and effects of radiation 
reaction~\cite{Teukolsky:1973ha,ruffini_sasaki,Quinn:1996am,Mino:1996nk,Hughes:2000pf,Glampedakis:2002cb,Glampedakis:2005cf,gkl}.
Generic (non-equatorial) orbits about a Kerr black hole will not lie
in a plane due to precession and frame-dragging effects, and thus
during the lengthy course of an EMRI, which could be in LISA-band
for thousands of cycles, the small black hole will ``sample'' much of the geometry
of the background spacetime. The structure of the corresponding gravitational waves
emitted will therefore contain a map of this geometry, and so EMRIs offer
a remarkable opportunity to probe the geometry of a black hole,
and will be able to confirm whether it is indeed described
by the Kerr metric~\cite{Ryan:1995wh,AmaroSeoane:2007aw}.

{\bf Plunge/merger:} Here, for non extreme-mass-ratio systems,
gravitational wave emission becomes
strong enough that the evolution of the orbit is no longer adiabatic,
and the black holes plunge together to form a single black hole. 
Understanding this phase requires full numerical simulations,
and it is only within the last couple of years that such simulations
have become available. The interesting picture that is now emerging
is that this phase is very short, lasting on
the order of one to two gravitational wave cycles. To get an idea
for the time scale of this regime, the Keplarian orbital angular 
frequency $\omega$ for an equal mass quasi-circular inspiral is
\be
\frac{\omega}{2\pi} = \frac{1}{2\pi}\sqrt{\frac{M}{R^3}} \ \ \
\approx 11 {\rm kHz} \frac{\msun}{M} \left(\frac{R_s}{R}\right)^{3/2},\label{omega_orbit}
\ee
where $M$ is the total mass of the binary with corresponding
Schwarzschild radius $R_S$, $R$ is the separation,
and the plunge/merger
happens as $R_s/R\rightarrow 1$. Note that the frequency of the
dominant quadrupolar ($\ell=2,m=2$)
component of the gravitational wave that is emitted is twice the 
orbital frequency.
The structure of the waveform is quite simple, however, this is the time of
strongest gravitational wave emission, with the luminosity 
approaching on the order of one-hundredth of the Planck 
luminosity of $10^{59}ergs/s$,
making black hole mergers by far the most energetic events in the
post-big-bang era of the universe. Furthermore, the 
frequency of the emitted wave rapidly grows to that of the dominant
quasinormal mode frequency of the final black hole, causing
the spectrum of the plunge/merger phase to occupy a large
region of the frequency domain. For equal mass systems upwards
of $3\%$ of the rest mass energy of the system is radiate away here.
A more detailed discussion of this phase is give in Sec.~\ref{sec_res}.

If cosmic censorship holds (and there are no signs
that it is violated in any merger simulations to date),
then Hawking's ``no-bifurcation'' theorem~\cite{Hawking:1971tu}
states that a single black hole must result as the
consequence of a merger. The uniqueness, or ``no hair''
theorems~\cite{Israel:1967wq,Carter:1971zc,Carter:1997im}
further imply that the newly formed black hole must
eventually settly down to the Kerr solution in
the so-called ringdown phase.

{\bf Ringdown:} 
The ringdown is the phase
where the remnant black hole can be described as a
perturbed Kerr spacetime. A more precise definition
might be the time afterwhich the gravitational waves
emitted from the merger can, to good precision, be written
entirely as a superposition 
of {\em quasi-normal modes} (QNMs)~\cite{CVV,Davis,Press,QNR,Leaver:1986gd,Kokkotas:1999bd}
of the final black hole\footnote{The QNM spectrum is not 
complete,
and so it is conceivable that they might not be able to {\em exactly}
describe the wave structure.}. Given appropriate initial conditions,
the ringdown phase could be calculated using perturbative
techniques, in particular using the so-called {\em close limit}
approximation~\cite{Price:1994pm}.
As discussed more in Sec.~\ref{sec_res},
the early simulation results suggests 
this description is already adequate
very shortly after formation of the common apparent horizon, which 
roughly coincides with the time of peak luminosity. Very shortly
after ringdown begins, the waveform ($|h|\propto e^{-t/\tau_{22}}\sin(\omega_{22} t)$) is dominated by 
the least damped (fundamental harmonic) quadrupolar QNM,
with angular frequency $\omega_{22}$ and decay time $\tau_{22}$,
given approximately by the following fitting formulas ~\cite{E89,Berti:2005ys}
\bea
\frac{\omega_{22}}{2\pi} &\approx& \frac{1}{2\pi M}\left[1-0.63(1-j)^{0.3}\right] \ \ \ 
\approx \ \ 32 {\rm kHz} \frac{\msun}{M} \left[1-0.63(1-j)^{0.3}\right] \\
\tau_{22} &\approx& \frac{4M (1-j)^{-0.45}}{1-0.63(1-j)^{0.3}} \ \ \ 
\approx \ \ 20 \mu {\rm s} \frac{M}{\msun}\frac{(1-j)^{-0.45}}{1-0.63(1-j)^{0.3}},
\eea
where $M$ and $J=jM^2$ are the total mass and angular momentum 
of the final black hole (with $|j|\leq 1$). The dominant ringdown
frequency is several times higher than the orbital frequency
in the last few inspiral cycles, and the decay time is quite short,
so the majority of the energy lost during ringdown
($1\%-2\%$ of the rest mass) is emitted quite rapidly. 
Waves propagating in a curved spacetime like Kerr are
back-scattered off the curvature, producing
so-called {\em power law tails}~\cite{Price:1971fb}.
They decay by 
integer powers of time, and so even though initially of much smaller
amplitude than the ringdown waves, they will eventually dominate
the late-time structure of the gravitational wave. Given their small
amplitude it is unlikely that the tails could be detected 
by ground-based detectors.

The relative simplicity of the plunge/merger waveform, together
with how short this phase is, suggests it may be possible
to build effective analytic template banks of merger waveforms
by stitching together PN inspiral waveforms with ringdown
waveforms. Numerical simulations of the plunger/merger phase
can provide instruction on exactly how this stitching 
should be performed, i.e., how long the transition region is, 
which set of quasi-normal modes are excited, how does the
waveform interpolate between inspiral and ringdown modes, etc.
In fact, this kind of prescription for constructing templates for
mergers was already proposed several years ago by proponents of
the effective-one-body (EOB) approach to binary 
dynamics~\cite{Buonanno:1998gg,Buonanno:2000ef},
and was recently demonstrated to work well for the extreme
mass ratio problem~\cite{Damour:2007xr}
and a range of non-spinning comparable mass mergers~\cite{Buonanno:2007pf}. 
Why might such a ``simple''
approach work so well for the merger phase, which was anticipated 
to be a showcase of the complexity and non-linearity of the field 
equations? First of all, recall that the PN approaches (including the EOB)
are hardly simple, having required the dedicated effort of many
researchers over a couple of decades to push to the orders 
presently known~\cite{Blanchet:2002av}---$(v/c)^7$ beyond Newtonian
order for non-spinning
binaries, and $(v/c)^5$ if spin is included.
At such high orders in $v/c$ it is not too surprising that much
of the essential physics is already being captured, and the only question
becomes how far the approach can be trusted. As the velocity of the
black holes increase toward the merger one would expect the expansions
to become increasingly inaccurate\footnote{For the qausi-circular equal
mass inspirals the coordinate velocities of the apparent horizons
only approach around $v=0.3$ prior to 
formation of the common horizon.}. Though, at the same time the black
holes are falling deeper into what is becoming the effective potential
of the final black hole spacetime, and eventually details of the local
dynamical geometry that may be poorly described by PN expansions will
have little effect on the radiated gravitational wave structure.
Also, a black hole by itself is not a simple, ``linear'' object, and
thus perturbations thereof could also be expected to capture
much of the late time physics of a merger. 

\subsubsection{The black hole scattering problem}\label{sec_bhs}

Consider the collision of two black holes in the center of mass frame 
with masses $m_1$ and $m_2$, Kerr spin parameters $a_1$ and $a_2$,
and initially moving toward each other with 
impact parameter $b$ and (large) Lorentz $\gamma$-factors 
$\gamma_1$ and $\gamma_2$. At present
very little is know about all the possible outcomes as a function
of $(b,\gamma_{1,2},m_{1,2},a_{1,2})$, though one can speculate
about several distinct stages, that will be classified here as
{\em Lorentz}, {\em collision/ringdown}, {\em scatter} 
and {\em threshold}. Note that in contrast to the
rest-mass dominated case, there is not necessarily such
a straight-forward progression through the phases.
In particular, there
could be a range of impact parameters where the black holes do 
not merge during the initial encounter, but have lost sufficient
energy that they now form a bound system. Then subsequent evolution
of the system will follow the stages of the astrophysical
binaries outlined in the preceding section.

{\bf Lorentz}: With sufficiently large $\gamma$ factors
the initial non-trivial geometry of each black hole
is Lorentz contracted into a thin ``pan-cake'' (or plane-wave) 
transverse to the direction of propagation, and close to Minkowski
spacetime on either side\footnote{In the
the limit $\gamma\rightarrow\infty$ and $m\rightarrow 0$ 
with $m\gamma=E$ kept constant, one obtains 
the Aichelburg-Sexl solution\cite{Aichelburg:1970dh}, and the spacetime
becomes exactly Minkowski on either side of a propagating
$C^0$ kink in the geometry.}.

{\bf collision/ringdown}: As suggested by studies of colliding black
holes in the infinite $\gamma$ limit\cite{penrose,D'Eath:1976ri,D'Eath:1992hb,D'Eath:1992hd,D'Eath:1992qu,Eardley:2002re,yoshino_nambu,berti_et_al,yoshino_rychkov,cardoso_et_al}, 
if the impact parameter is close to zero there will not be
any phase analogous to inspiral; rather an encompassing 
apparent horizon forms at the moment of collision, and this will
presumably settle down to a Kerr black hole. Estimates based on the
size of the initial apparent horizon place
an upper limit of $30\%$ on the net
energy of the spacetime that could be radiated in a head-on collision,
though these limits weaken as the impact parameter increases.
For the head-on collision case,
perturbative studies suggest the energy radiated is close 
to $16\%$\cite{D'Eath:1992qu}. 

{\bf scatter}: For larger values of the impact parameter there will
be a deflection of the two black hole trajectories, accompanied by a 
burst of radiation, afterwhich they will move apart and the spacetime near
each black hole will settle down to the {\em Lorentz} phase again.
It has been suggested that there may even be a regime
where a {\em third} or more black holes are formed during the interaction
of the two black holes before they scatter, essentially due to
the strong focusing of gravitational waves by the shock-fronts representing
the boosted black holes~\cite{choptuik_3bh}. This would
be an astonishing addition to the phenomenology of the two body
problem in general relativity if the scenario can be realized.

{\bf threshold}: At intermediate values of the impact parameter
there could be threshold-type behavior as seen when fine-tuning
eccentric orbits in the rest-mass dominated 
regime~\cite{cqg_review,Pretorius:2007jn}.
Namely, approaching a critical value $b=b^*$ of the impact parameter,
the two black holes settle into the analogue of an unstable 
circular geodesic orbit, whirl
around for an amount of time proportional to $-\ln|b-b^*|$, then
either fly apart or plunge together (this is described in 
more detail in Sec.\ref{sec_ecc}). During this phase copious
amounts of energy could be radiated in gravitational waves; in fact,
at threshold
it is conceivable that essentially {\em all} of the kinetic energy
of the system is radiated as gravitational waves in $O(10)$ orbits.
If the black holes merge after the whirl phase, there will be a 
plunge/merger and ringdown phase similar to astrophysical binaries.
If they separate and have lost enough kinetic energy to form a bound
system they will enter the {\em inspiral} phase of an astrophysical
binary, otherwise they will fly apart as in the {\em scatter} phase above.
It is tempting to speculate that exactly at threshold, $b=b^*$,
the spacetime may approach a self-similar solution (see the discussion
in Sec.~\ref{sec_scatt}).

\section{Contemporary successful numerical solution methods}\label{sec_num}
This section describes the two methods of formulating
the field equations presently known that are amenable to stable numerical
integration of binary black hole spacetimes\footnote{At least for the regions
of parameter space studied to date, which are all in the rest-mass
dominated regime.}, namely {\em generalized harmonic coordinates with
constraint damping} (GHC) and 
the {\em Baumgarte-Shapiro-Shibata-Nakamura} (BSSN) formalism with
moving punctures~\cite{paper2,utb,nasa}.
It is beyond the scope of this article to discuss
either method in much detail, or all the variations and details
of particular codes; rather the equations will be presented
and briefly discussed to provide the reader with some appreciation
for the similarities and differences between them. Note also
that if a code produces an apparently stable, convergent solution,
it is much more likely that the method actually {\em is} stable,
compared to the opposite situation where a simulation ``crashes''
and from which one would like to conclude that the {\em method} is
unstable. This is simply because bugs are easy to make, more difficult
to find, and almost never ``help'' in any interpretation
of the word. The point of this discussion is that there have been
numerous good ideas and formulations of the field equations proposed over
the past several years (see for example ~\cite{Reula:1998ty,Lehner:2001wq}), 
and only in a few cases were they
studied with sufficient detail to conclude that they were unstable; thus
that we now know of two methods that are stable does not imply
that all earlier methods are not.
A case-and-point might be the 
$Z4$ formalism~\cite{Bona:2002ft,Bona:2003fj} proposed several
years ago, which is quite similar in some respects to generalized harmonic
coordinates, and to which the same constraint damping mechanism 
can be applied~\cite{gundlach_et_al}. On the other hand, the fact that ``zero's''
need to be added to the equations in just the right way to make
things stable also tells us that the Einstein equations are even 
more subtle and intricate than previously thought.

\subsection{Historical background}

The first attempt at a numerical solution to the field equations
for a binary black hole spacetime was carried out by Hahn and Lindquist~\cite{hahn_lindquit}
in 1964. This was even before the word ``black hole'' had
been coined by Wheeler, and they evolved what was then called
the ``worm hole'' initial data of Misner~\cite{misner}. 
They considered a time-symmetric scenario in axisymmetry,
and reported a run performed on an IBM 7090 computer, 
using a 51x151 mesh. It took about
4 hours to complete 50 time steps, after which they
concluded that errors had grown too large to warrant
further evolution. This corresponded to a time of $~m/2$, with
$m=\sqrt{A/16\pi}$, $A$ being the area of the throat. Given the 
short run time, not much physics could be extracted from the
simulation, yet even so there was no motivation to explore gravitational
wave emission.
In 1975 Smarr~\cite{smarr_thesis}, and shortly afterwards Eppley~\cite{eppley_thesis}, 
again simulated the head-on collision of two black holes now, 
with one of the primary goals being to compute the gravitational
waves emitted in the process.
Despite still being an axisymmetric simulation and almost a
decade after Hahn and Lindquist, it was still beyond the
capabilities of computers of the time to integrate the field equations
with sufficient resolution to obtain very accurate results. Nevertheless,
they were able to extract gravitational waveforms from the solutions,
calculating (with uncertainties of a factor of 2) that upwards of $0.1\%$ of the
rest mass energy is released in gravitational waves in a time-symmetric
case where the initial proper separation between the
two throats is $9.6M$~\cite{smarr_sources}. Primarily because
of the stringent computational requirements for numerical solution,
no further work on the problem was carried out until
the early 1990's, when prospects for the construction
of LIGO became solid. LIGO was the impetus for returning
to the two body problem as it was realized early on~\cite{Thorne:1980rt} that
given a practical design for the instrument, together
with the estimated density of sources in our universe, matched-filtering
would be an essential data analysis tool to allow a
decent detection rate within a several year time-frame. 
Matched-filtering looks for known signals
in a noisy data stream by convolving theoretical templates of the signals with the 
data. To be successfull it is therefore imperative to understand the
gravitational wave emission properties of the source with sufficient
detail to construct the template libraries. 

The early expectations following a revisit of the head-on collision 
case~\cite{anninos_col1}
was that although certain issues about the generic merger problem
had yet to be fully addressed and could be complicated, such as
having well behaved coordinates and providing astrophysically relevant initial conditions, 
a fair consensus was that the most significant hurdle to the problem
was lack of computer power~\cite{Anninos:1994yw,Anninos:1995am}. 
Certainly a portion of the
difficulties encountered may be traceable to attempting to find solutions with
insufficient resolution, though it turns out that a host of additional issues had to be 
``discovered'', understood and overcome to reach the state where the field
is today. 

The review of the history of the numerical two body problem
now continues, though switching to a non-traditional format: instead of
trying to follow events in chronological order the
ingredients needed for a successful simulation will be summarized, noting 
contributions that offered insights or solutions to the various 
problems\footnote{And my apologies in advance to authors that I have missed here.}

\subsection{Historical background continued---ingredients 
to assemble a successful numerical 2-body code}
The list of ingredients described below certainly ``makes sense'',
and so one might wonder, why not satisfy all of them to begin
with? First of all, many of the issues, such as choosing
well behaved coordinates, are quite complicated, and one does
not expect general solutions applicable to all 
spacetimes of interest. Second, it was perhaps not fully appreciated
how vast the landscape of free-evolution schemes are, i.e. systems
of equations that give solutions to the Einstein equations for
only a restricted subset of initial and boundary
conditions, and how important the behavior of these equations are 
for initial and boundary conditions that do {\em not} exactly satisfy those requirements.
This is particularly so because it is numerical truncation error that
sources ``constraint violations'', which is not {\em a priori}
a problem as truncation error is a well understood, part-and-parcel
component of any numerical solution. The surprising thing then
is that, in dynamical systems language, it appears as if for the vast majority 
of free evolution formulations of the Einstein equations, trajectories
through phase space denoting solutions to Einsteins equations
form an {\em unstable} manifold in the space of all solution trajectories.
A third reason is the ADM formulation, in the form
popularized by York~\cite{york_78}, certainly also ``makes sense'',
and seems to be a very reasonable and intuitive approach to an
initial boundary value formulation of the field equations. Furthermore, given
the success of the ADM equations in early evolutions of many
symmetry reduced spacetimes, there was not much reason to suspect
problems with it.

\subsubsection{Fix the character of the differential equations}\label{sec_fix}
The Einstein field equations\footnote{Again, using {\em geometric units}
where the speed of light $c=1$ and Newton's constant $G=1$.}
\be\label{efe}
G_{ab}=8\pi T_{ab},
\ee
when expanded verbatim in terms of the metric $g_{ab}$
\be\label{gdef}
ds^2 = g_{ab} dx^a dx^b,
\ee
results in a coupled system of $10$, quasilinear, second order partial 
differential equations for the $10$ independent components of the metric, depending
on the four spacetime coordinates $x^a$. However, these equations have
{\em no} definite mathematical character---hyperbolic, parabolic or elliptic---and 
moreover, do not admit a well posed initial value problem. This in large part is
due to the gauge invariance of the field equations: for a given, {\em unique}
physical spacetime there are infinitely many {\em different} metric tensors
describing it and all satisfying the same equations(\ref{efe}). The first
step towards obtaining a well posed system of equations is to specify
enough of the gauge to fix the character of each of the
four spacetime coordinates $x^a$. There are several possibilities, the most
common being to choose one coordinate ($t$) to be timelike, and
the remaining three ($x,y,z$) to be spacelike. After a bit more
work, outlined in the following item, one comes up with a system
of elliptic/hyperbolic equations. This ``space-plus-time'' (or 3+1) approach
will be the exclusive focus of the remainder of the article, after briefly
mentioned one alternative, {\em characteristic} or {\em null} 
coordinates (for more
details, see for example \cite{Lehner:2001wq,Winicour:2001kx}). Here, 
one (single null) or two (double null) coordinates are chosen
to be lightlike, and the rest of the coordinates spacelike. 
Single null evolution schemes
have been very successful in evolving single black hole spacetimes\cite{Gomez:1998uj}.
Part of the reason for pursuing null evolution is that it is easy to extend
the domain to future null infinity, which is ideally where one would
want to measure gravitational waves. The difficulty with this
approach is preventing or treating caustics that can form along
the null coordinate in non-trivial, dynamical spacetimes, and
no viable mechanism has yet been proposed that might be applied
to a binary black hole spacetime. Hybrid null---$3+1$ schemes
(often called {\em Cauchy-characteristic matching}) 
have been proposed(see~\cite{Winicour:2001kx} and the references therein), whereby
a 3+1 scheme is used to evolve the spacetime in the
vicinity of the binary, a characteristic scheme far 
from the binary, and the solutions mapped to one another in an intermediate zone 
where the coordinate systems overlap. The
matching process is non-trivial, and to date the method has only
been applied to single black hole spacetimes~\cite{Bishop:1997xi,Gomez:1997nq,Barreto:2004fn}. 

\subsubsection{Find a formulation admitting a numerically well-posed initial
boundary value problem}\label{sec_form2}

To obtain a well-posed $3+1$ formulation of the Einstein equations more
work needs to be done than merely choosing one timelike and three spacelike
coordinates. The choice of which set of fields to treat as the dependent
variables of the system of PDEs, the gauge conditions that will be used, 
as well as modification of the equations by the addition of constraint 
terms (i.e. terms that are identically zero for any solution of 
the Einstein equations) all play an important role in determining 
the ultimate stability of the system.
The ``traditional ADM'' approach, as outlined by York~\cite{york_78},
is based on a Hamiltonian formulation of the field equations due to 
Arnowitt, Deser and Misner~\cite{ADM}. The end result is that the
equations are rewritten in terms of quantities either {\em intrinsic}
or {\em extrinsic} to $t={\rm const.}$
slices of the geometry. First, the four dimensional metric is
decomposed as 
\be\label{adm}
ds^2 = -\alpha^2 dt^2 + h_{ij} \left(dx^i + \beta^idt\right)\left(dx^j + \beta^jdt\right),
\ee
where $h_{ij}$ is the spatial metric of the $t={\rm const.}$ hypersurface,
$\alpha$ is the {\em lapse function} measuring the rate at which proper time
flows relative to $t$ for a hypersurface-normal observer, and $\beta^i$
is the spatial {\em shift} vector describing how the spatial coordinate label
for such an observer changes with time $t$. In other words, the time flow
vector $(\partial/\partial t)^a$ is related to the unit 
hypersurface normal vector $n^a$ by
\be\label{nv}
\left(\frac{\partial}{\partial t}\right)^a = \alpha n^a + \beta^a
\ee
In this way of describing
the four-geometry the lapse and shift naturally represent the coordinate
degrees of freedom in the theory. Second, the manner in which $h_{ij}$
is embedded into the four dimensional space is described
by the {\em extrinsic curvature} tensor $K_{ij}$ $\ $\footnote{In the Hamiltonian
picture the momentum $\pi_{ij}$ canonically conjugate to $h_{ij}$
is $\pi_{ij}=\sqrt{h}(K_{ij} - K h_{ij})$, where $h$ is the determinant
of $h_{ij}$ and $K$ is the trace of $K_{ij}$}
\bea\label{ex_curv}
K_{ij} &\equiv& - h_i{}^a h_j{}^b \nabla_b n_a \\
       &=& -\frac{1}{2\alpha} \left( \frac{\partial h_{ij}}{\partial t}
          - \Screll_\beta h_{ij}\right).
\eea
In terms of the variables ($h_{ij},K_{ij},\alpha,\beta^i$) the field
equations can be written as a set of 12 independent hyperbolic evolution equations
for ($h_{ij},K_{ij}$), 4 {\em constraint} equations that do not contain
any time derivatives of $K_{ij}$, and need to be augmented with
evolution equations for the gauge quantities ($\alpha,\beta^i$) (see
~\cite{Lehner:2001wq} for an overview of oft-used choices).
A common way to proceed to solve these equations is 
by {\em free evolution} (see~\cite{TP-80} for a discussion of
the general alternatives):
the constraints are only solved at the initial time, 
and the remaining equations are then used to evolve the variables
with time. In a consistent discretization scheme the constraint equation
will remain zero to within numerical truncation error, and hence,
as mentioned before, that the constraints are not strictly enforced 
is not necessarily a problem.
However, in general scenarios, i.e. when there are no symmetries
that can be used to simplify the equations, it turns out that the standard
ADM form of the equations just outlined is only {\em weakly hyperbolic},
and this implies that one cannot in general find a fully consistent, hence stable
discretization scheme for the system~\cite{GKO}.
This problem began to be appreciate by the numerical relativity
community in the mid-90's, which sprouted a cottage industry 
of finding symmetric-hyperbolic reductions or various more ``ad-hoc'' 
modifications
of the field equations~\cite{bona_masso,Bona:1994dr,frittelli_reula,
Abrahams:1995nj,Friedrich:1996hq,
Frittelli:1996wr,Abrahams:1996hh,ChoquetBruhat:1996ak,Bona:1997hp,
Iriondo:1997yr,Yoneda:1998qr,Anderson:1999qx,Yoneda:1999uy,Arbona:1999ym,Alcubierre:1999wj,
Frittelli:1999sj,Kidder:2001tz,Shinkai:2001nd,ChoquetBruhat:2001pm,
Friedrich:2002xz,Alvi:2002hu,Bona:2002fq,Bona:2002ft,
Buchman:2003sq,Nagy:2004td,Reula:2004xd,Chee:2004dt,
Paschalidis:2005as,Paschalidis:2007ey}.
Unfortunately, even though some of these methods were successfully applied
to single black hole spacetimes, they offered only
marginal improvements at best compared to ADM codes for the binary 
black hole problem~\cite{Brugmann:1996kz,Cook:1997na,Scheel:1997kb,
Bona:1998dp,Scheel:1998uv,Alcubierre:1999rt,Brugmann:1999we,
Brugmann:1997uc,Camarda:1997qv,Allen:1998wy,Camarda:1998wf,Alcubierre:2000xu,Alcubierre:2000ke,Kelly:2001kj,
Yo:2002bm,Scheel:2002yj,Tiglio:2003xm,
Sperhake:2003fc,Shoemaker:2003td,Anderson:2003dz,Alcubierre:2004hr,
Buchman:2005ub,Diener:2005mg}, with the arguable 
exception of the BSSN formulation,
which showed success in binary neutron star evolutions(see~\cite{Baumgarte:2002jm}
and the works cited therein), and 
set the record for the longest binary black hole evolution~\cite{bruegmann_et_al}
prior to the breakthroughs of 2005~\cite{paper2,utb,nasa}. 
One reason why some of these methods, even though provably stable,
can still ``fail'' in practice, is if the truncation error grows too rapidly
with time. The truncation error $f_{te}(t,x^i)$ for a variable $f$ will,
to leading order in the mesh spacing $h$ in an $n^{th}$ order discretization
scheme, have the form $f_{te}(t,x^i) = e(t,x^i) h^n$. Formal stability
only requires that the $h$-independent error term $e(t,x^i)$ does not 
grow {\em faster} than exponential, though with sufficiently rapid 
exponential growth
it might be impractical to give high enough resolution (small $h$) to keep
the error term small for the desired run-time. Another (and somewhat related)
reason why stable codes could fail, and which actually appears to 
be {\em the} problem
in most free evolution schemes, are ``constraint violating modes'' 
discussed next.

\subsubsection{Curb truncation-error-induced growth of constraints}
Constraint violating modes (CVMs) are {\em continuum solutions} to the
{\em subset} of Einstein equations that are evolved during a free 
evolution, but do not satisfy the constraint equations. These constraints
could either be the Hamiltonian and momentum constraints inherent 
to the Einstein equations, or constraints arising from first order reductions
or similar redefinitions of the underlying fields. Note that truncation 
error is not a CVM by this definition, however since in general
truncation error will not satisfy the constraints it will be a source
of CVMs in any free evolution scheme. For CVMs to be benign their growth
rate must be sufficiently small to remain of comparable magnitude to truncation
error during the evolution. At present only two formulations of the field 
equations appear to have this desired property ``off the constraint manifold'' for
binary black hole evolutions---{\em generalized harmonic coordinates} with 
{\em constraint damping}~\cite{paper2}, and variants of {\em BSSN} with appropriate gauge
choices and methods for dealing with the black hole singularities~\cite{utb,nasa}. These two approaches will be described in more detail in 
Secs.~\ref{sec_harm} and ~\ref{sec_bssn}. Constraint damping adds a particular 
function of the constraint equations to the Einstein equations to try
to curb the growth of CVMs for solutions close to the desired one. This
is not a very new idea~\cite{ChoquetBruhat:1983vs,bona_masso,frittelli_reula,bs,
Abrahams:1995nj,Frittelli:1996wr,Anderson:1999qx,
lambda_ref99,Kidder:2001tz,Kelly:2001kj,
Sarbach:2002gr,Friedrich:1996hq,Gentle:2003hs,Bona:2003fj}
though the particular method that works
for harmonic evolution was only recently proposed by Gundlach et al.~\cite{gundlach_et_al}.
They were able to prove that their damping terms could curb all finite-wavelength 
constraint violating perturbations of Minkowski space.
There is no mathematical proof that this 
should work for the binary black hole
problem, and the evidence that the CVMs are adequately under control is 
entirely empirical. However, experience suggests there may never
be a ``black box'' solver for the Einstein equations applicable to solving
for arbitrary spacetimes; rather, schemes need to tailored to the particular
scenario, and constraint damping is probably no exception.
The nature of the evolution of the constraints in BSSN is even less well 
understood. 

Given all the problems with the constraints, an obvious alternative
would be {\em constrained evolution}, whereby the constraints
are solved at each time step in lieu of a subset of evolution 
equations. Such methods have worked very well in symmetry
reduced situations, though with the exception of~\cite{bonazzola_et_al}
have not yet been attempted in 3D. Part of the reason is
that solving the constraints involves solving elliptic equations,
which many people in the community have been reluctant to attempt.
Also, it is not clear in a general 3D setting which degrees
of freedom to constrain, and which to freely evolve. In~\cite{bonazzola_et_al},
a spherical polar coordinate system is used, which does
allow for a ``natural'' decomposition into free vs. constrained
variables; such a coordinate system is not well adaptive
to studying binary black hole spacetimes. Several years ago Andersson and 
Moncrief~\cite{Andersson:2001kw} discussed an elliptic-hyperbolic formulation
of the field equations that appears to be ideally suited for
3D constrained evolution, though to date no implementations of this system
have been carried out. A related idea is {\em constraint projection}
(similar to ``divergence cleaning'' in the solution of Maxwell's equations),
whereby a free evolution system is used, then periodically 
the constraints are re-solved, modifying a subset of the variables
accordingly. This technique was shown to have promise
in a single black hole spacetime~\cite{Anderson:2003dz}, though
in that code excision boundary problems (apparently) prevented 
long-time stable evolution. In~\cite{Holst:2004wt} a Langrange multiplier
method was proposed to optimally project out the constraints;
it was successfully implemented for scalar field evolution, though has
not yet been applied to the full Einstein equations. One might
think that another option for dealing with the constraints
is at the numerical level via something akin to the
{\em constrained transport}~\cite{Evans:1988qd} scheme
used in some magnetohydrodynamic codes, however Meier~\cite{Meier:2003bn} showed
that similar finite-difference based techniques will not work for
the Einstein equations due to the non-linearity of the equations.

\subsubsection{Provide well behaved dynamical coordinates conditions}
It almost goes without saying that covering the spacetime 
manifold with a well behaved, non-singular coordinate system
is a necessary condition for stable evolution. 
The difficulty is that in a Cauchy evolution the dynamics
of the fields describing the geometry are intimately linked
to the coordinates, and thus when solving for a new spacetime
where the future geometric structure is unknown, the future
behavior of the coordinates is just as uncertain. A large number of 
analytic solutions discovered throughout the history of relativity have,
in their original form, been riddled with coordinate 
pathologies (the most famous example of course being the event 
horizon of the Schwarzschild solution); dealing with them
involved first understanding the nature of the pathology, then
constructing a coordinate transformation to remove it. In {\em principle}
this is an approach that could be applied in a numerical evolution:
evolve to the point of a coordinate singularity, understand it,
apply a coordinate transformation to remove it, and continue the
evolution. However, given the nature of a numerical solution,
i.e. discrete meshes of numbers representing either field
values or coefficients of basis functions, this would be a
very challenging endeavour in all but the simplest spacetimes.
Thus, the universal approach in numerical relativity to try to avoid
coordinate problems has been to devise coordinate {\em conditions}
that typically either make the coordinates satisfy certain 
properties (e.g. constant mean curvature, or CMC slicing 
where $t={\rm const.}$ is a space of constant
mean curvature, or harmonic coordinates described in Sec.\ref{sec_harm}),
or conditions that force the variables to satisfy certain
constraints (e.g. the unit-determinant condition
on the conformal metric in the BSSN approach discussed
in Sec.~\ref{sec_bssn}). Coordinate conditions usually take
the form of algebraic or differential operators acting on the
``gauge'' fields of the formalism, which are most commonly
the lapse and shift. It is beyond the scope of this
article to describe the numerous coordinate conditions
proposed over the years related to the binary
hole problem (see ~\cite{smarr_sources,york_78,Lehner:2001wq} for
more details),
though in Secs.~\ref{sec_harm} and~\ref{sec_bssn} we will described 
those that have been instrumental in the recent successful 
binary black hole simulations.

\subsubsection{Specify good outer boundary conditions}
``Good'' outer boundary conditions for evolved fields
in a simulation must have three properties:
(1) be mathematically well posed, (2) be consistent with
the constraints, and (3) be consistent with the physics
being modeled, which here is asymptotic 
flatness\footnote{For the purposes of modeling the
local geometry of a merger and extracting the resultant gravitational
waves in the far-zone there is little practical
distinction between an asymptotically flat versus
Friedman-Robertson-Walker universe. The effects of a wave propagating
across cosmological distances in an expanding universe
can readily be accounted for 
analytically, as the wave amplitude decays
as $1/D_L$ and its wavelength increases by a
factor $1+z$, where $z$ is the redshift and $D_L$ the luminosity distance
to the source---see for example ~\cite{peebles}.}
and no incoming gravitational 
radiation. The class of boundary conditions (3)
will form a subset of (2), which in turn is a subset
of conditions (1). A common approach is 
to apply either exact or some approximation
to {\em maximally dissipative} boundary conditions,
where the incoming characteristics of all fields
normal to the boundary are set to zero.
Though mathematically
well-posed this in general is neither consistent with
the constraints nor prevents outgoing waves
of the solution from being reflected back. Much effort
has been spent over the past several years devising
constraint preserving boundary conditions (CPBCs) for various
formulations~\cite{Calabrese:2001kj,szilagyi_et_al,Sarbach:2002gr,Calabrese:2002xy,
Frittelli:2003ym,Frittelli:2003yc,Gundlach:2004jp,Bona:2004ky,
Sarbach:2004rv,Kidder:2004rw,Arnold:2006vv,Babiuc:2006ik,Buchman:2006xf,
Rinne:2006vv,Kreiss:2007cc,Buchman:2007pj,Babiuc:2007rr,Rinne:2007ui}. 
By themselves CPBCs do not alleviate
the problem of spurious incoming radiation, and more recently
research in CPBCs has begun to focus on subsets
of CPBCs that do address this issue.

An alternative approach to outer boundary conditions
is to extend the computational domain to infinity, where
the exact Minkowski spacetime boundary conditions can
be placed. As mentioned in Sec.~\ref{sec_fix} Cauchy-characteristic
matching effectively extends the domain to future null 
infinity. The matching procedure is non-trivial however,
and this technique has yet to be applied to a binary
black hole merger scenario. Another option is to compactify
the coordinates to spatial infinity, which is the
approach used in the generalized harmonic evolutions
in~\cite{paper2}. This rather straight-forwardly
solves all issues (1)-(3), though introduces potential
numerical complications in that outgoing waves suffer 
an ever increasing blue shift as they travel toward the
outer boundary~\footnote{The blue shift is infinite
in the limit, though it would take an infinite amount of
time for the waves to reach the boundary.}. Either 
increasing resolution and hence computational resources
must be used to resolve the waves, or once they have passed
the desired wave extraction radius be allowed to 
blue-shift to coarse resolution. With the latter option
the numerical technique must therefore be robust to 
the introduction of high-frequency solution components; in the generalized
harmonic evolution code this is achieved using Kreiss-Oliger
style dissipation~\cite{KO}. Note that this kind of
dissipation is {\em not} akin to artificial viscosity sometimes
used in hydrodynamical simulations, as Kreiss-Oliger
dissipation modifies the difference equations at the
level of the truncation error terms, and thus converges
away in the continuum limit. 

A couple of alternative methods of compactification include
conformal compactification~\cite{Friedrich:2002xz,Misner:2005yz,van Meter:2006mv}, and using 
asymptotically hyperboloidal or null slices~\cite{Misner:2004wx,Calabrese:2005rs}.

\subsubsection{Deal with black hole singularities}
By the singularity theorems of Hawking and Penrose~\cite{Hawking:1969sw,hawking_ellis}
we know that all black holes contain true (geometric)
singularities, which in a simulation will manifest as various
field quantities diverging as the spacetime slice
approaches the singularity. Infinities
cannot be dealt with in a numerical code, and must
be ``regularized'' in some fashion. The two contemporary
successful approaches to deal with the singularities 
are {\em excision} and {\em punctures}.
Both techniques rely on the causal property of 
a black hole spacetime that no information
can flow out of the event horizon, and that
cosmic censorship is valid,
namely that all the geometric singularities that
might exists in the spacetime are always inside
the event horizon. If cosmic censorship were violated
in a particular evolution, the codes would ``crash'';
thus a stable evolution is confirmation that in that instance
cosmic censorship was not violated.

With excision, a 2-sphere {\em inside} the black hole and enclosing the
singularity is designated as a boundary of the computational
domain. By the assumed causal properties of the black
hole there will always exist a class of such boundaries
where all characteristics of the
fields are directed {\em toward} the boundary, i.e. out of the
computational domain. Mathematical theory (and common sense)
says one is only allowed to place boundary conditions on 
the incoming components of fields satisfying hyperbolic equations.
Thus {\em no} boundary conditions are specified on the excision
surface; rather, the difference equations are simply solved
there\footnote{In a finite difference code this implies
replacing centered derivative operators with ``sideways''
operators as appropriate to avoid referencing regions
of the domain inside the excision boundary.}. The formal definition
of a black hole event horizon is the boundary of the causal past
of future null infinity, which is not a local property of
spacetime and cannot be found during evolution.
Instead, the {\em apparent horizon}---a marginally outer-trapped
surface, which is surface from which ``outward'' traveling 
photons have zero expansion---is used to determine where to 
excise. Excision surfaces on or inside the apparent horizon
can also satisfy the requirement that all field-characteristics
are directed outside of the domain, since, if cosmic censorship 
holds, the apparent horizon will always be
{\em inside} the event horizon (see for example~\cite{Wald}).

Originally, a {\em puncture} was the singular point
inside a maximally extended vacuum black
hole spacetime representing the spatial infinity reached
by a conformally mapped slice passing through
an Einstein-Rosen bridge into a second asymptotically
flat universe~\cite{Brandt:1997tf}. Therefore a puncture is a coordinate rather
than geometric singularity. The manner in which the metric diverges
approaching the puncture is known analytically, 
and can be factored out. Punctures were originally
used to construct initial data for the binary black hole
problem~\cite{Brandt:1997tf}, though soon afterwards it was shown that punctures
can be used in evolution~\cite{Brugmann:1997uc}. The metric
at the puncture was regularized by diving out a time-independent
conformal factor, however the extrinsic curvature was not regularized.
Thus, to avoid problems this was anticipated to cause, the
punctures were placed ``between'' grid points, and coordinate
conditions were chosen to make the shift vector zero at the
punctures so that derivatives of the extrinsic curvature across
the punctures would not be needed. The vanishing of the shift
vector implies the puncture locations are fixed in the grid.
Maximal slicing was used for the lapse.
The breakthrough in puncture
evolutions discovered recently is to relax the condition on the
shift vector, allowing the punctures to move through
the domain. At the same time the slicing condition
is altered to force the lapse to zero at the puncture,
essentially ``freezing'' evolution there (see Sec.~\ref{sec_punct_gauge}
for more on these coordinate conditions). This, remarkably, 
causes the codes to remain stable despite the irregular
nature of the solution about the punctures.
There have been several studies
since attempting to understand geometrically what a moving
puncture represents~\cite{Hannam:2006vv,Hannam:2006xw,Thornburg:2007hu,
Baumgarte:2007ht,Brown:2007tb}; a couple of competing viewpoints
at present are that 1) the puncture remains attached
to spatial infinity of the alternate 
universe~\cite{Brown:2007tb}, and 2)
the spacetime slice quickly evolves so that the alternate
universe is ``pinched-off', and the puncture effectively
becomes a single excised point inside the black 
hole~\cite{Hannam:2006vv,Hannam:2006xw,Baumgarte:2007ht}. Though
 from the perspective of seeking solutions to the field
equations exterior to the event horizons of the black holes,
the question of what a puncture represents is academic.

\subsubsection{Provide consistent and relevant initial data}\label{sec_id}
The initial data problem for binary black holes
is not trivial. First, the initial conditions
must satisfy the constraints, which typically evolves
solving systems of coupled, non-linear elliptic 
equations. Second, providing
astrophysically relevant initial data is quite challenging, as
for practical considerations the evolution must begin
within several or tens of orbits before merger. This implies
that there should already be a non-negligible amount
of gravitational radiation from the prior inspiral of the
black holes present in the initial data. Also, the closer
the black holes are the more difficult it becomes
to unambiguously associate relevant orbital parameters
to the spacetime, for example the orbital eccentricity,
binary separation, orbital frequency, etc. It is beyond
the scope of this article to describe these 
issues---see~\cite{Cook:2000vr,Pfeiffer:2004nc,Gourgoulhon:2007tn}
for review articles on contemporary initial data construction
methods, and~\cite{Nissanke:2005kp,Yunes:2005nn,Yunes:2006iw,Kelly:2007uc} for suggestions to incorporate realistic
initial conditions motivated by post-Newtonian expansions.

\subsection{Generalized harmonic evolution}\label{sec_harm}

{\em Generalized harmonic evolution}, as its name implies,
is an evolution scheme based on a generalization of {\em harmonic
coordinates}. Harmonic coordinates are a set of gauge conditions
that require each spacetime coordinate $x^a$ to independently satisfy the 
covariant scalar wave equation:
\be\label{harm_def}
\Box x^a = \frac{1}{\rg}\partial_b\left(\rg g^{bc} \partial_c x^a\right) \equiv 0,
\ee
where $g$ is the determinant of the spacetime metric (\ref{gdef}).
The use of these coordinate conditions have a long and celebrated
history in relativity, including DeDonder's analysis of the characteristic
structure of general relativity~\cite{dedonder}, Fock's study of 
gravitational waves~\cite{fock} and proofs of uniqueness and existence
of solutions to the field equations by Choquet-Bruhat~\cite{choquet}
and Fischer and Marsden~\cite{fischer_marsden}. In fact, as early
as 1912 Einstein used harmonic coordinates, then known as isothermal
coordinates, in his search for a relativistic theory of 
gravitation~\cite{renn_sauer}. One of the key properties of harmonic
coordinates that make them so useful in these studies
is that when (\ref{harm_def}) is substituted into the field
equations, the principal part of the resultant equation for each 
metric element becomes a scalar wave equation for that particular
metric element, with all non-linearities and couplings between
the equations relegated to lower order terms. This has obvious
benefits for formal analysis of the field equations, and is also
a natural system to study the radiative degrees of freedom in 
the theory. Also, given that there are simple and effective numerical 
solution techniques available to solve wave equations, it would
seem that harmonic coordinates would be a natural starting point
for a numerical code.

However, only recently in numerical relativity have harmonic coordinates
been used as the basis for numerical 
evolution schemes\cite{garfinkle,szilagyi_winicour,paper1,new_lindblom_et_al,babiuc_et_al,szilagyi_et_al_b}, meaning 
discretizing the field equations {\em after} the harmonic
conditions have been used to re-express the system
as a set of wave-like equations. Prior to this harmonic
coordinates had been advocated
and used within the more traditional ADM space-plus-time 
formulation of the field 
equations\cite{Bona:1988zd,Bona:1994dn,sn,Anninos:1995am,
Abrahams:1995nj,Alcubierre:1996su}, where harmonic gauge (or variants of it) are
imposed via conditions on the lapse function
and shift vector. In such a decomposition the wave-like character
of the field equations in not manifest, and the
primary reason quoted for using harmonic gauge (in particular harmonic time 
slicing) was for its geometric ``singularity avoiding'' properties.
However even within ADM evolutions harmonic coordinates were
seldom used due to the notion that they would generically
lead to the formation 
of ``coordinate shocks''~\cite{Alcubierre:1996su,Alcubierre:1997ee,Hern:2000rp,Alcubierre:2005gw,Garfinkle:2007yt}. An 
in-principle solution to this problem noted by Garfinke~\cite{garfinkle}
(and see an earlier discussion of this by Hern~\cite{Hern:2000rp})
was to use {\em generalized harmonic coordinates} (GHC), first introduced
by Friedrich~\cite{friedrich}. Here, a set of arbitrary 
{\em source functions} are added to (\ref{harm_def}):
\be\label{ghc_def}
\Box x^a \equiv H^a.
\ee
To see that this can avoid coordinate pathologies, note that
(\ref{ghc_def}) can be regarded as a {\em definition} of the
source functions; in other words, take {\em any} metric in {\em any}
(well behaved) coordinate system, and (\ref{ghc_def}) tells
one what the corresponding source functions for the metric
in GHC are. When imposing GHC in a Cauchy evolution,
the $H^a$ must be treated as independent functions to 
allow (\ref{ghc_def}) to reduce the principal parts of the
field equations to the desired wave-like equations. Thus additional
evolution equations must be supplied for $H^a$ to close the
system, and so the issue of finding well-behaved coordinates
for a particular dynamical spacetime becomes one of finding
the appropriate evolution equations for $H^a$.

For concreteness, below an explicit form of
the Einstein equations in GH form with constraint damping terms
will be given, using the covariant
metric elements and covariant source functions ($H_a = g_{ab} H^b$)
as the fundamental variables. This is certainly not the
only way to proceed---for a symmetric hyperbolic
first order reduction see~\cite{new_lindblom_et_al} (and see~\cite{Owen:2007bh} for how the
constraints introduced via auxiliary variables are kept under
control), and versions using
the densitized contravariant metric elements see~\cite{szilagyi_winicour,babiuc_et_al,Szilagyi:2006qy}.
Consider the Einstein equations in trace-reversed form
\be\label{efe_r}
R_{ab}=4\pi\left(2 T_{ab}-g_{ab} T\right),
\ee
where $R_{ab}$ is the Ricci tensor
\be\label{ricci}
R_{ab}=\Gamma^{d}_{ab,d} -
       \Gamma^{d}_{db,a}
      +\Gamma^{e}_{ab}\Gamma^{d}_{ed}
      -\Gamma^{e}_{db}\Gamma^{d}_{ea},
\ee
$\Gamma^{g}_{ab}$ are the Christoffel symbols of the second kind
\be
\Gamma^{g}_{ab}=\frac{1}{2}
g^{ge}\left[g_{ae,b}+g_{be,a}-g_{ab,e},
\right]
\ee
$T_{ab}$ is the stress energy tensor with trace $T$, and 
a comma is used to denote partial differentiation.
Using the definition of GHC (\ref{ghc_def}) and its first derivative (\ref{efe_r})
can be written out explicitly as
\bea
\frac{1}{2} g^{cd}g_{ab,cd} &+& \label{efe_princ} \\
g^{cd}{}_{(,a} g_{b)d,c} 
+ H_{(a,b)} 
&-& H_d \Gamma^d_{ab} 
+ \Gamma^c_{bd}\Gamma^d_{ac} \label{efe_rest} \\
+ \kappa[n_{(a} C_{b)} &-& \frac{1}{2}g_{ab} n^d C_d] \label{efe_cd} \\
&=& - 8\pi\left(T_{ab}-\frac{1}{2}g_{ab} T\right)\label{efe_set}.
\eea
Line (\ref{efe_princ}) shows the principal, hyperbolic part
of the equations, line (\ref{efe_rest}) are the rest
of the terms coming from (\ref{efe_r}) where all the
couplings and non-linearities reside, line (\ref{efe_cd}) are
the constraint damping terms with adjustable parameter
$\kappa$ and unit timelike vector $n^a$ normal to
$t=\rm{const.}$ hypersurfaces\footnote{In \cite{gundlach_et_al} it was
suggested that an arbitrary timelike vector can be 
used in the constraint damping terms, though in all situations studied to date
$n^a$ has been chosen as the hypersurface normal unit timelike vector.},
and line (\ref{efe_set}) contains the coupling to
matter. Here, the constraints are simply the definition
of GHC
\be
C_a \equiv g_{ab} \left(H^a - \Box x^a\right),
\ee
and are thus zero for any solution of the field equations.
The relationship between the GH constraints and the more familiar 
form of the constraints of the Einstein equations,
written as a one-form $\Scrim_a$
\be
\Scrim_a \equiv (R_{ab} - \frac{1}{2}g_{ab} R -8\pi T_{ab}) n^b,
\ee
where the time-component $\Scrim_a n^a$ is the Hamiltonian constraint
and the momentum constraints are the components of the 
spatial projection $\Scrim_a(\delta^a{}_b+n^a n_b)$, is~\cite{new_lindblom_et_al}
\be
\Scrim_a = \nabla_{(a}C_{b)} n^b -\frac{1}{2} n_a \nabla_b C^b.
\ee
Furthermore, one can show that if the metric is
evolved using (\ref{efe_princ}-\ref{efe_set}), the constraints 
will satisfy the following
evolution equation
\be
\Box C^a = - R^a{}_b C^b + 2\kappa\nabla_b\left[n^{(b}C^{a)}\right].\label{new_cp}
\ee
From this it easy to see that, at the continuum level, a solution
that initially satisfies the constraints will always do so
if constraint-preserving boundary conditions are used during evolution.
Part of the constraint damping modification in (\ref{new_cp})---namely
the term proportional to $n^b \nabla_b C^a$---is a wave-equation damping term,
so one might reasonably then expect that (\ref{new_cp}) will {\em not}
admit exponentially growing solutions given small (truncation-error-sourced)
initial conditions. This ``expectation'' has been proven for small,
finite-wavelength constrain-violating perturbations of
Minkowski spacetime~\cite{gundlach_et_al}, though not yet for general spacetimes.

\subsubsection{Source function evolution}
To close the system of equations~(\ref{efe_princ}-\ref{efe_set}), 
an additional set of evolution equations must be specified
for the source functions, written schematically as
\be
\Screll_a H_a = 0 \ \ \ ({\rm no \ summation}).
\ee
$\Screll_a$ is a differential operator that in general can dependent
upon the spacetime coordinates, the metric and its derivatives, and the
source functions and their derivatives. The source functions
directly encode the coordinate degrees of freedom of general
relativity, as can be seen by writing the definition of GHC (\ref{ghc_def})
in terms of ADM variables (\ref{adm}):
\bea
H_a \ n^a &=& -K - \partial_{\nu} (\ln\alpha) n^\nu \label{hdotn} \\
H_b \ h^{ab} &=& - \bar{\Gamma}^a_{jk} h^{jk}
  + \partial_j (\ln\alpha) h^{a j} + \frac{1}{\alpha}\partial_b \beta^a n^b \label{hdoth},
\eea
where $\bar{\Gamma}^i_{jk}$ is the connection associated 
with spatial metric $h_{ij}\equiv g_{ij}+n_i n_j$. Thus,
the temporal source function $H_a n^a$ is related to the
time derivative of the lapse $\alpha$, whereas a spatial
source function $H_b h^{ab}$ is related to the time derivative
of the corresponding component of the shift vector $\beta^a$.
Not much research has been done on finding source function
evolution equations to achieve a particular slicing or
satisfy some coordinate conditions directly within the
GH framework, though the above relationship
between $H^a$ and the lapse and shift allows many of the ideas
developed over the years for ADM evolutions to be adopted in 
a GH evolution~\cite{paper1,lee_grg18}.
We end this section by showing one example of a
set of source evolution equations, used 
in~\cite{paper2}:
\be
\Box H_t = - \xi_1 \frac{\alpha-1}{\alpha^\eta} + \xi_2 H_{t,\nu} n^\nu\label{t_gauge}, \ \ \ H_i=0.
\ee
This equation
for $H_t$ is a damped wave equation with a forcing function
designed to prevent the lapse $\alpha$
from deviating too far from its Minkowski value of
$1$, which helps alleviate an apparent instability
in the code of~\cite{paper1} that sets in when the lapse 
drops close to zero inside a black hole\footnote{Note 
that $\alpha$ is not an independent
variable in the formalism---in the code it is replaced
by its definition in terms of the metric $g_{ab}$.}.
In (\ref{t_gauge})
the parameter $\xi_2$ controls the damping term,
and $\xi_1,\eta$ regulate the forcing term. Ranges of
useful parameter values are discussed in \cite{cqg_review}.

\subsection{BSSN with `moving punctures'}\label{sec_bssn}
The BSSN formulation of the field equations~\cite{nok,sn,bs}
begins with the ADM (\ref{adm}) decomposition of spacetime,
then continues by performing a York-Lichnerowicz-like 
conformal decomposition of the spatial metric and
extrinsic curvature~\cite{YL}. 
The conformal metric is defined via
\be\label{bssn_cm}
\tilde{h}_{ij} \equiv e^{-4\phi} h_{ij}
\ee
and is {\em chosen} to have unit determinant, so that 
\be\label{bssn_cf}
e^{4\phi} = h^{1/3},
\ee
where $h$ is the determinant of $h_{ij}$. Continuing,
the trace $K$, and conformal, trace-free part of the extrinsic
curvature (\ref{ex_curv})
\be
\tilde{A}_{ij}\equiv e^{-4\phi} (K_{ij} - \frac{1}{3} h_{ij} K),
\ee
are treated as fundamental variables.
The final ingredient
in the BSSN formalism is to also evolve the conformal
connection coefficients
\be\label{bssn_g}
\tilde\Gamma^i\equiv \tilde{h}^{jk} \tilde\Gamma^i_{jk} = - \tilde{h}^{ij}{}_{,j}
\ee
independently, where $\tilde\Gamma^i_{jk}$ is the Christoffel
symbol of the conformal spatial metric. In summary then,
$\phi$, $\tilde{h}_{ij}$, $K$, $\tilde{A}_{ij}$,
$\Gamma^i$, $\alpha$ and $\beta^i$ are the fundamental
variables of the BSSN formalism. The evolution
equations for $\phi$, $\tilde{h}_{ij}$ and
$\Gamma^i$ derive from their definitions
\bea
\frac{d}{dt} \phi &=& - \frac{1}{6} \alpha K, \\
\frac{d}{dt} \tilde h_{ij} &=& - 2 \alpha \tilde A_{ij},\\
\frac{\partial}{\partial t} \tilde \Gamma^i
 = &-& 2 \tilde A^{ij} \alpha_{,j} + 2 \alpha \Big(
   \tilde \Gamma^i_{jk} \tilde A^{kj} -  \nonumber
   \frac{2}{3} \tilde h^{ij} K_{,j}
   - \tilde h^{ij} S_j + 6 \tilde A^{ij} \phi_{,j} \Big)
   \nonumber \\
&+&   \frac{\partial}{\partial x^j} \Big(
   \beta^l \tilde h^{ij}_{~~,l}
   - 2 \tilde h^{m(j} \beta^{i)}_{~,m}
   + \frac{2}{3} \tilde h^{ij} \beta^l_{~,l} \Big).
\eea
and the evolution equations for $K$ and $\tilde{A}_{ij}$
come from the Einstein equations
\bea
\frac{d}{dt} K &=& - h^{ij} D_j D_i \alpha +
   \alpha(\tilde A_{ij} \tilde A^{ij}
   + \frac{1}{3} K^2) + \frac{1}{2} \alpha (\rho + S),\\
\frac{d}{dt} \tilde A_{ij} & = & e^{- 4 \phi} \left(
   - ( D_i D_j \alpha )^{TF}  +
   \alpha ( R_{ij}^{TF} - S_{ij}^{TF} ) \right)
   + \alpha (K \tilde A_{ij} - 2 \tilde A_{il} \tilde A^l_{~j}),
\eea
with 
\bea
R_{ij} &=& \tilde R_{ij} + R_{ij}^{\phi},\\
R^{\phi}_{ij} & = & - 2 \tilde D_i \tilde D_j \phi -
   2 \tilde h_{ij} \tilde D^l \tilde D_l \phi
    + 4 (\tilde D_i \phi)(\tilde D_j \phi)
   - 4 \tilde h_{ij} (\tilde D^l \phi) (\tilde D_l \phi), \\
\tilde R_{ij} & = & - \frac{1}{2} \tilde h^{lm} \label{rij}
   \tilde h_{ij,lm}
   + \tilde h_{k(i} \partial_{j)} \tilde \Gamma^k
   + \tilde \Gamma^k \tilde \Gamma_{(ij)k}  + 
   \tilde h^{lm} \left( 2 \tilde \Gamma^k_{l(i}
   \tilde \Gamma_{j)km} + \tilde \Gamma^k_{im} \tilde \Gamma_{klj}
   \right)
\eea
and matter projections
\bea
\rho & = & n_{a} n_{b} T^{ab}, \\
S_i  & = & - h_{i a} n_{b} T^{ab}, \\
S_{ij} & = & h_{i a} h_{j b} T^{ab}.
\eea
The gauge variables $\alpha$ and $\beta^i$ are freely specifiable.
In the above the operator $d/dt$ is defined to be
\be
\frac{d}{dt} \equiv \frac{\partial}{\partial t} - \Screll_\beta,
\ee
where $\Screll_\beta$ is the Lie derivative with respect to
the shift vector $\beta^i$ (and note that $\tilde h_{ij}$ and
$\tilde A_{ij}$ are tensor densities of weight $-2/3$), 
$D_i$($\tilde{D}_i$) is the covariant
derivative operator with respect to $h_{ij}$($\tilde{h}_{ij}$), and
${TF}$ denotes the trace-free part of the expression.
The BSSN equations listed above were taken from \cite{bs};
some of the actual implementations use
slightly different variables (for example $\chi\equiv e^{-4\phi}$ 
is used instead of $\phi$ in \cite{utb}), 
differ in whether and/or how certain algebraic
constraints in the formalism are enforced (such as the trace-free nature
of $\tilde A_{ij}$ or that $\tilde{h}_{ij}$ has unit determinant), 
replace undifferentiated
occurrences of $\tilde\Gamma^i$ with its definition (\ref{bssn_g}),
or adds multiples of the constraints inferred by (\ref{bssn_g}) to
the evolution equation 
for $\tilde\Gamma^i$\cite{Alcubierre:2000yz,Yo:2002bm,Laguna:2002zc,Baker:2006yw}.

There are several reasons often quoted as
motivation behind the BSSN formalism. First, the conformal
decomposition in part separates the extrinsic curvature
into ``radiative'' versus ``non-radiative'' degrees
of freedom (though within the York-Lichnerowicz formalism
it is the {\em transverse} trace-free
part of the extrinsic curvature that represents the
radiative degrees of freedom). Second, the constraint
equations are used to eliminate certain terms
from the ``bare'' evolution equations (in
particular the Hamiltonian constraint is used
to eliminate a Ricci scalar term from the
evolution equation for $K$,
and the momentum constraints to eliminate a 
divergence of $\tilde{A}_{ij}$ term from the
evolution equation of $\tilde\Gamma^i$), and so in a sense this
is a partially constrained evolution system~\cite{Gentle:2007ix}.
Third, with appropriate gauge conditions the BSSN
system of equations is 
hyperbolic~\cite{Sarbach:2002bt,Reula:2004xd,Beyer:2004sv,Gundlach:2005ta,Gundlach:2006tw}.
An important step in achieving hyperbolicity is treating
the connection functions $\tilde\Gamma^i$ as independent quantities,
which makes the principle part of the differential operator
acting on the conformal metric in (\ref{rij}) elliptic.
Incidentally, this is {\em exactly} what would be done
if one were to express the spatial conformal metric in generalized
harmonic form, with $\tilde\Gamma^i$ 
being the source functions. 

\subsubsection{Moving punctures}\label{sec_punct_gauge}

An important element in achieving stable evolution of binary
black hole spacetimes with the BSSN formulation is using
coordinates that allow the punctures hiding the black hole
singularities to move through the grid, yet do not allow
any evolution {\em at} the puncture point itself (i.e., the
lapse is forced to go zero at the puncture, though not the shift vector,
hence the ``frozen'' puncture can be advected through the domain).
The conditions that have so far proven
successful are modifications to the so-called {\em 1+log slicing}
and {\em Gamma-driver shift} conditions~\cite{Bona:1994dr,Alcubierre:2002kk}:
\bea
\frac{d}{dt} &=& -2\alpha K\\
\partial_t\beta^i &\equiv& \xi B_i, \ \ \ 
\partial_t B_i = \chi \partial_t \tilde\Gamma^i - \eta B^i
-\zeta \beta^j\partial_j\tilde\Gamma^i.
\eea
In the above, $\xi,\chi,\eta$ and $\zeta$ are parameters (that are
required to be within certain ranges for stable evolution, though do
not require fine-tuning); 
a couple of examples for typical choices: 
$(\xi=3\alpha/4,\chi=1,\eta=4,\zeta=1$)~\cite{Herrmann:2006ks}
and $(\xi=1,\chi=1,\eta=1,\zeta=0$)~\cite{Sperhake:2006cy}.
Common initial conditions are $\beta^i=B^i=0$, and
$\alpha=1/\psi_{BL}^2$, where 
$\psi_{BL} = 1 +\sum_i m_i/2|\vec{r}-\vec{r}_i|$ 
is the Brill-Linquist
conformal factor for the initial data containing
black holes with mass parameter $m_i$ at coordinate
location $\vec{r}_i$.

It is uncertain exactly why these coordinate conditions work as
well as they do (similar to why the rather ad-hoc equations used in the
generalized harmonic scheme shown in (\ref{t_gauge}) 
improve the evolution);
an alternative way of phrasing this is that it
is not known why {\em fixed} puncture evolutions are prone
to instabilities. 
Recently in~\cite{Garfinkle:2007yt} it
was suggested that 1+log slicing could generically lead to the
formation of gauge shocks near the punctures as
anticipated in~\cite{Alcubierre:1996su}, and that
these have not yet been observed in current 3D simulations
due to poor resolution about the punctures (though again,
as long as stability can be maintained this in
theory is not problematic for studying the geometry
exterior to the horizon). 

\subsection{Comparison of the two techniques}
After discussion of the two evolution formalism, generalized
harmonic and BSSN, a ``required'' section deals with a comparison
of the methods. That section is here, though there really
is not much to say on the matter. Personal preferences and
aesthetics aside, {\em both} methods are capable of finding
discrete solutions describing similar physical processes 
within the context of the {\em same} theory---general 
relativity---and thus are equivalent from a scientific perspective.
In terms of technical issues, one could argue that
moving punctures are much easier to get working than excision.
However, this is more an issue of dealing with black hole
singularities, and in principle either method could
be implemented within either formalism. A technical
issue of some relevance to numerical implementation
is that there is (presently) no known fully first order, symmetric
hyperbolic reduction of the BSSN equations, which would
be a requirement for a spectral implementation using the
methods of the Caltech/Cornell group.

\subsection{Numerical algorithms}

It is beyond the scope of this article to delve into computational
issues involved in solving the field equations---here a few references
to related material in the literature is given. With the exception
of the Caltech/Cornel pseudo-spectral code~\cite{Kidder:2000yq,Boyle:2006ne},
all contemporary binary black hole evolution codes use 
finite-difference methods (for a broader
view of the use of spectral methods in relativity see~\cite{Grandclement:2007sb}).
The complexity of the field equations and the physical set-up of the
binary black hole problem requires solution in a parallel computing
environment, and adaptive mesh refinement (AMR) to adequately
resolve all the relevant length scales (the only code at present not
employing AMR is the {\em LazEv} code~\cite{Zlochower:2005bj}, however
there a non-linear ``fisheye'' coordinate transformation is used to
resolve the length scales in the vicinity of the binary). Some of the
parallel/AMR software presently used is the Cactus Computational
Toolkit~\cite{cactus} with the {\em Carpet} thorn for AMR~\cite{Schnetter:2003rb},
{\em paramesh}~\cite{paramesh}, {\em PAMR/AMRD}~\cite{pamr_amrd}, 
{\em HAD}~\cite{Neilsen:2007ua} and {\em BAM}~\cite{Marronetti:2007ya}.
Descriptions of some of the more computational aspects of the merger codes can be 
found in~\cite{Imbiriba:2004tp,paper1,Zlochower:2005bj,Herrmann:2006ks,szilagyi_et_al_b,
new_lindblom_et_al,Marronetti:2007ya,Husa:2007hp}.

\section{Results from black hole merger simulations}\label{sec_res}
In this section some results from recent merger simulations
are discussed. This is a rapidly evolving field, and much of what
is said could be dated in short-order. Also, in many respects
this is still a very young field, and though there has been 
a flurry of early results, systematic, in-depth studies
are sparse. We break the discussion up into the following
classes of binary: a) equal mass, minimal eccentricity and spin,
b) unequal mass, minimal eccentricity and spin, 
c) equal mass, non-negligible spin, minimal eccentricity,
d) equal mass,
large eccentricity, minimal spin, and e) generic. 

\subsection{Equal mass, minimal eccentricity and spin}
The equal mass, minimal eccentricity and spin case
is one of the simplest configurations in that there is
no precession of the orbital plane, and no recoil imparted
to the final black hole.
Thus, the key parameters that characterize the
merger are essentially only the final mass and spin of
the remnant black hole. 
Recent 
simulations~\cite{utb,nasa,Campanelli:2006gf,Baker:2006yw,Baker:2007fb,Pfeiffer:2007yz,Hannam:2007ik,Husa:2007rh} of
this scenario have used either
Cook-Pfeiffer~\cite{Cook:2004kt} or Bowen-York~\cite{Bowen:1980yu} initial
data. These results indicate that the energy emitted during the last 
orbit, plunge/merger and ringdown is $3.5\%$($\pm0.2\%$) of 
the total total energy of the
system, resulting in a Kerr black hole with 
$a=0.69$ ($\pm 0.02$).
Based on the binding energy of the initial
data configurations, or PN/EOB estimates of the energy radiated during
the inspiral (see for example~\cite{Cook:2004kt,Buonanno:1998gg}), 
an additional $1.5\%$ ($\pm0.2\%$) of the available energy is radiated
prior to this, implying that an equal mass, non-spinning inspiral 
beginning at infinite radial
separation looses $5.0\%$($\pm0.4\%$) of its total rest-mass energy 
to gravitational
waves during the entire merger event\footnote{The uncertainties 
reflect the authors
best conservative ``guess'' based on the various results published in the
literature to date; the uncertainty in the PN inspiral value is that the
results usually quoted are for the integrated energy up to the
ISCO (innermost stable circular orbit), which only
approximately corresponds to the ``last'' orbit of the numerical
results.}.
As mentioned in the discussion in Sec.\ref{sec_id},
these families of initial data do not exactly capture
the conditions of the equivalent astrophysical scenario,
and though it is difficult at present to estimate precisely
what the effects of this are, systematic studies
suggest the artifacts are small. In particular, the
initial data lacks the correct initial gravitational
radiation content, though within roughly an orbital
light-crossing time this ``junk'' radiation leaves the
vicinity of the orbit and is quickly replaced by radiation
emitted by the binary motion. The energy content of the
junk radiation also appears to be negligible to within
the quoted uncertainties. Other noticeable ``artifacts''
in some of the cited simulation results is a small amount of orbital
eccentricity (due to the choice of initial data having zero initial
radial momentum), and the black holes are initially co-rotating
for the Cook-Pfeiffer data presently in use; again, the effect
on the waveforms appear to be small, and can also be removed
without much effort~\cite{Pfeiffer:2007yz,Husa:2007rh}. There have
been several suggestions for how the correct radiation
content can be inserted into initial fields~\cite{Nissanke:2005kp,Yunes:2005nn,Yunes:2006iw,Kelly:2007uc}, though none
of these methods have yet been implemented.

\begin{figure}
\includegraphics[width=4.25in,clip]{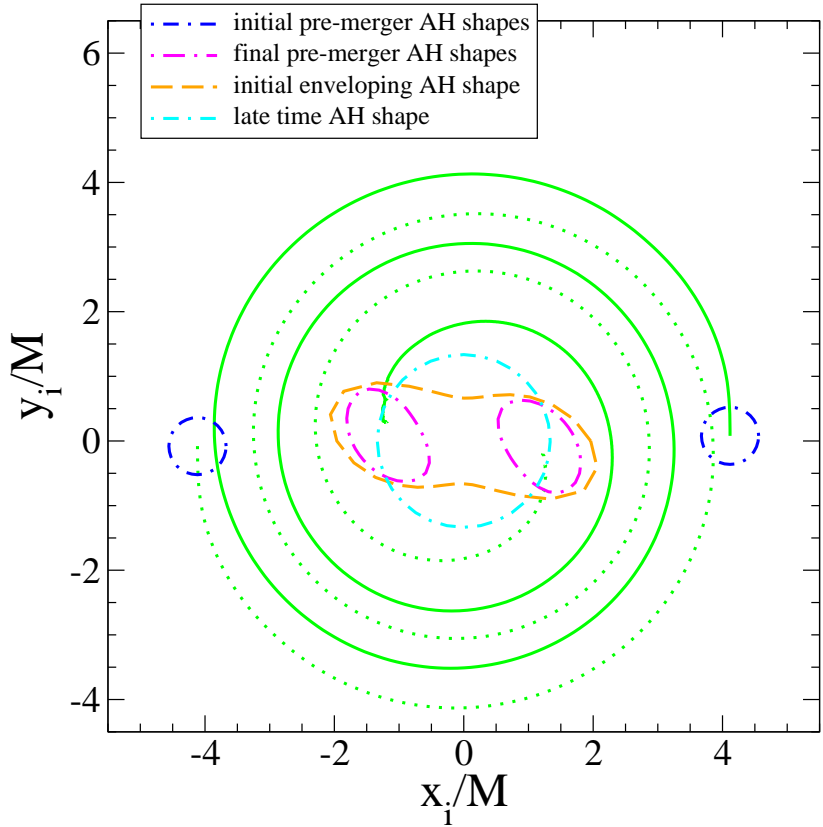}
\caption{A depiction of the trajectories of the black holes from
a merger simulation (the ``d=16'' Cook-Pfeiffer case, from~\cite{Buonanno:2006ui}).
The green lines are the centers of the apparent horizons of each 
black hole. The trajectories end once a common horizon is found.
Also shown are the coordinate shapes of the apparent horizons at several key moments.
\label{orbits}}
\end{figure}

\begin{figure}
\includegraphics[width=1.27in,clip]{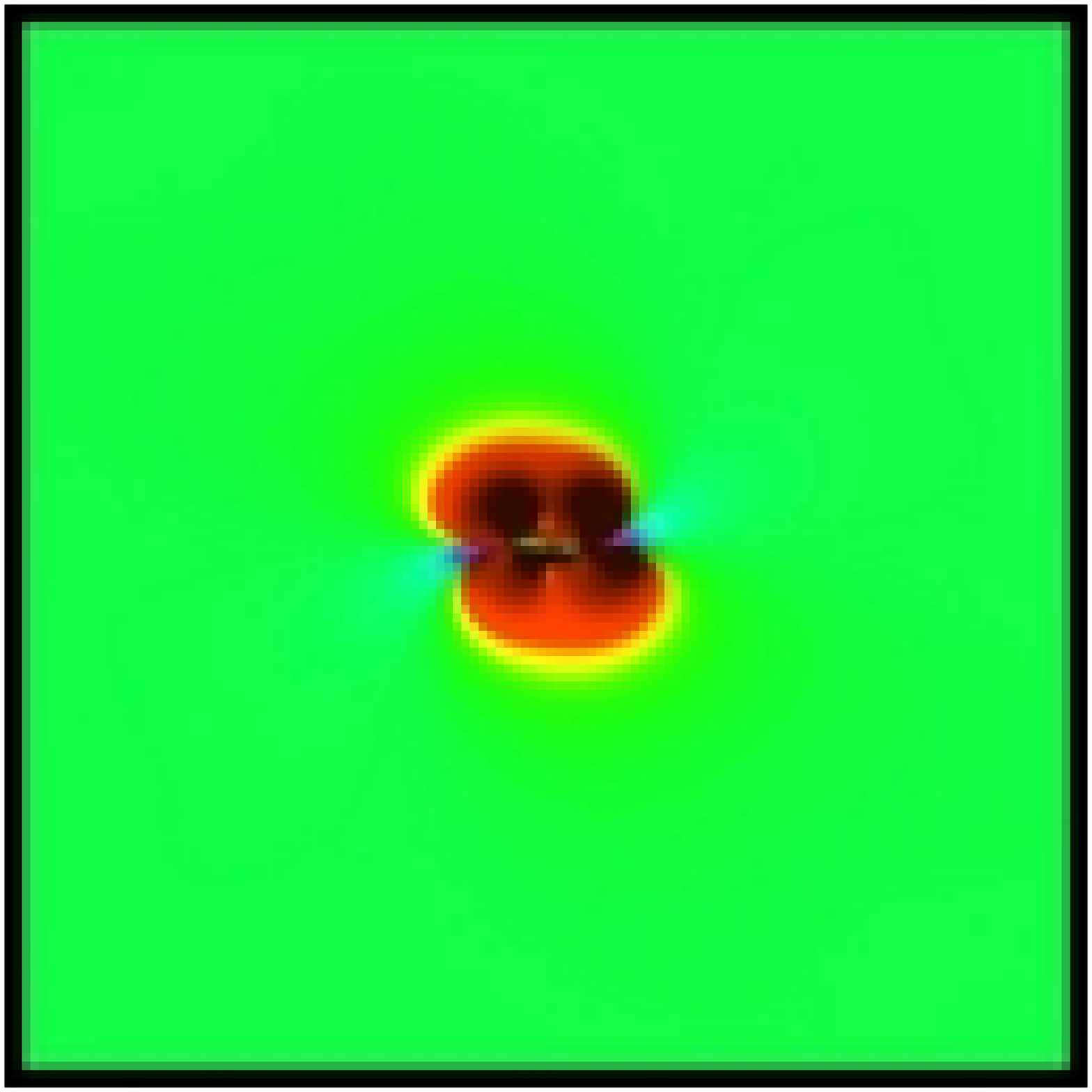} \hspace{-0.4cm}
\includegraphics[width=1.27in,clip]{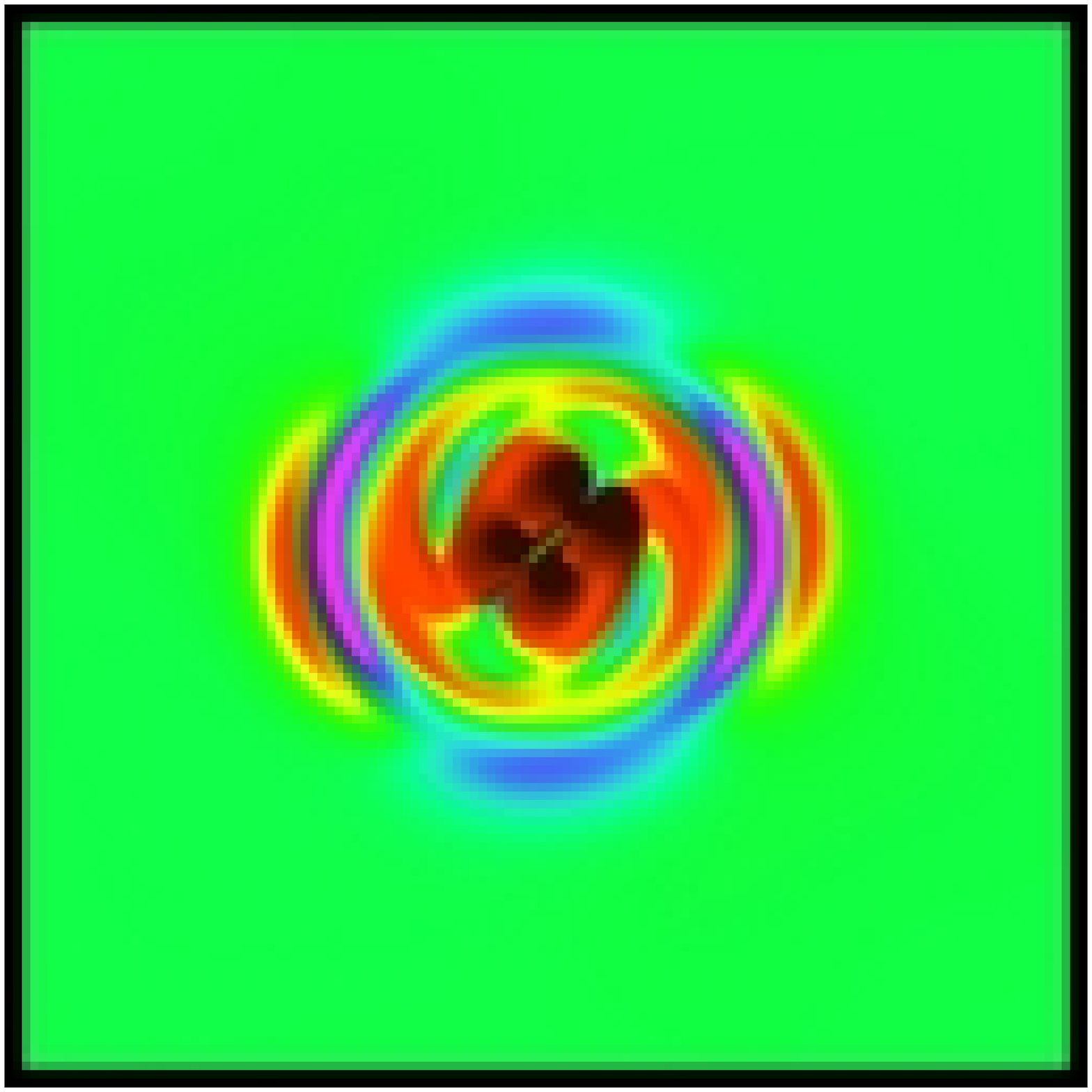} \hspace{-0.4cm}
\includegraphics[width=1.27in,clip]{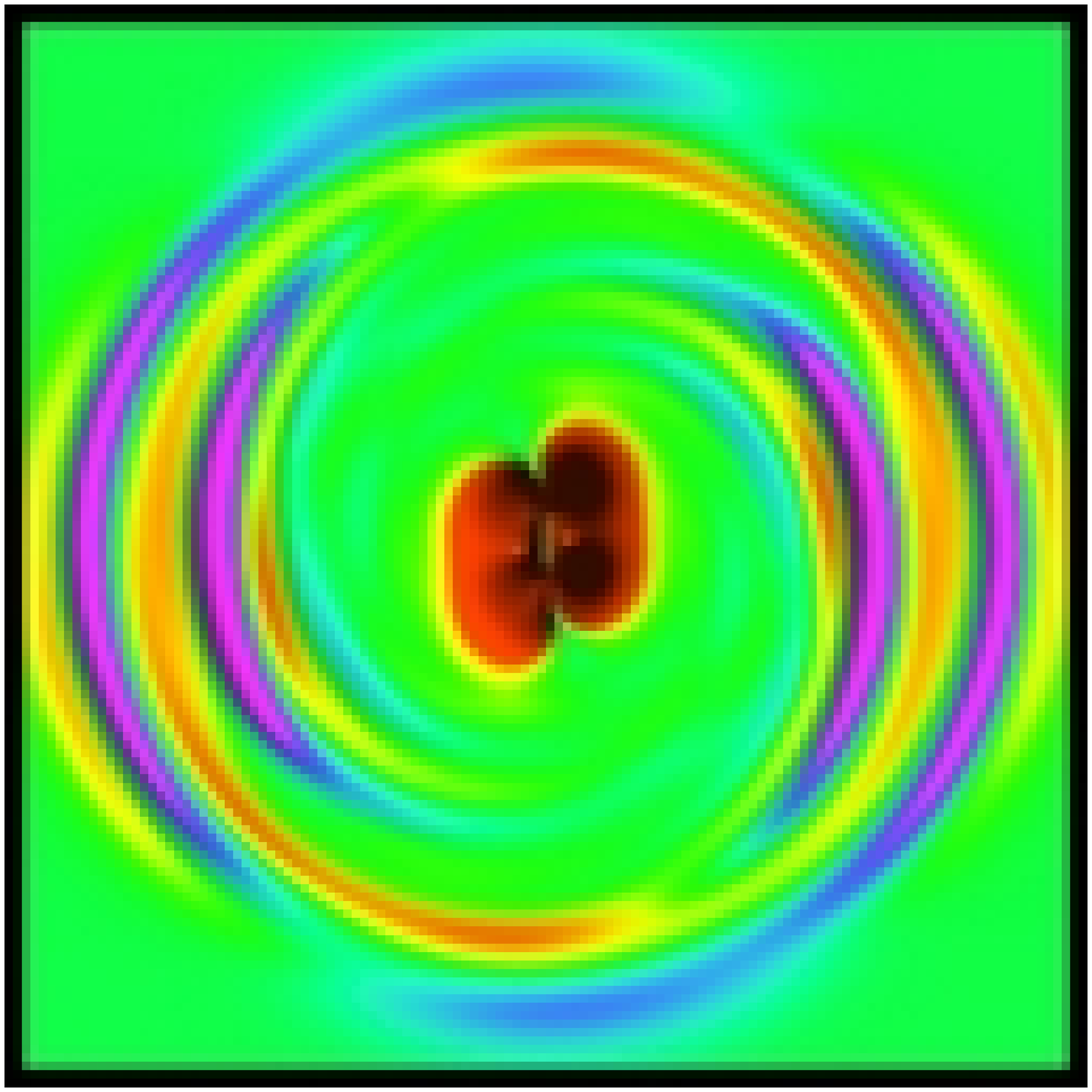} \hspace{-0.4cm}
\includegraphics[width=1.27in,clip]{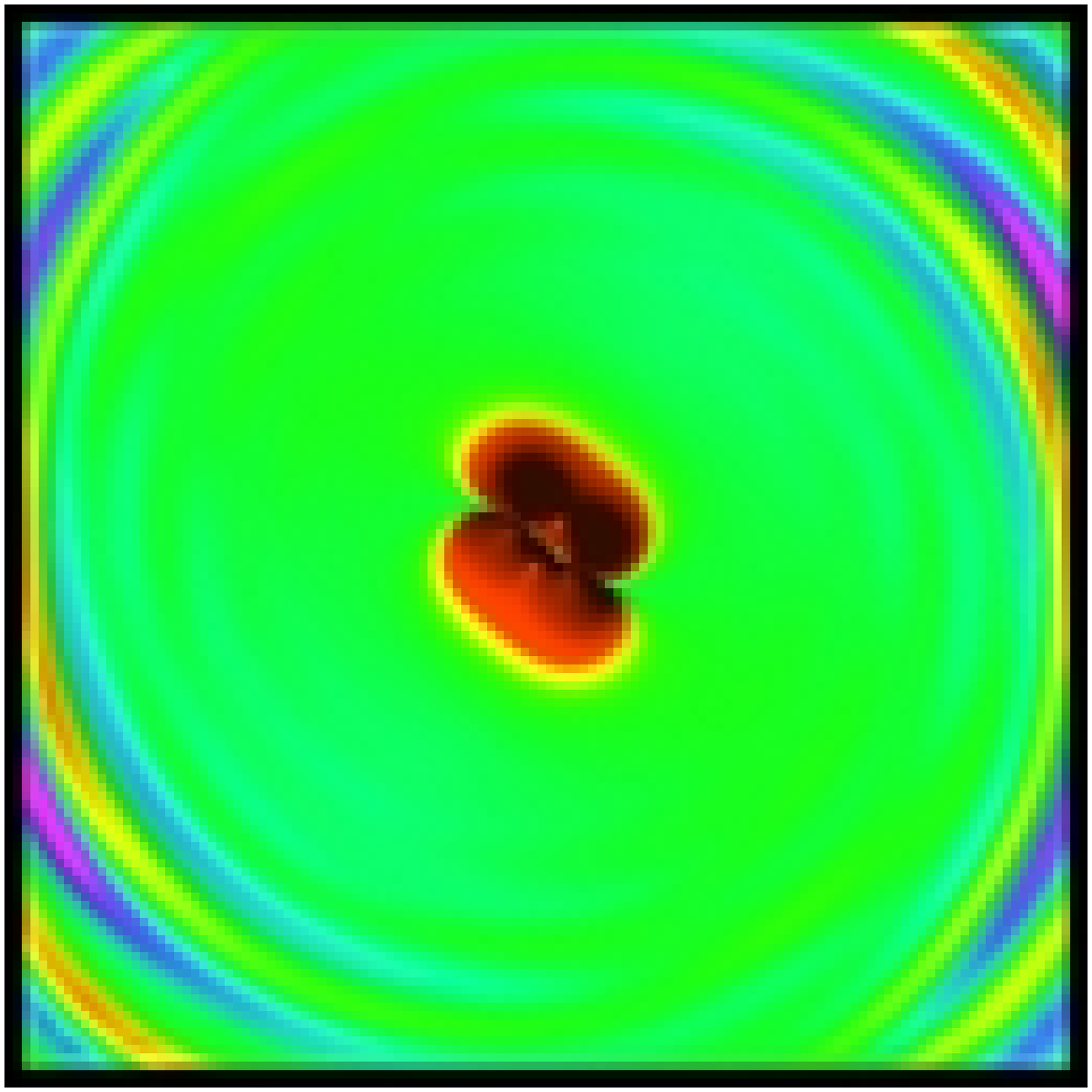} \hspace{-0.4cm}
\includegraphics[width=1.27in,clip]{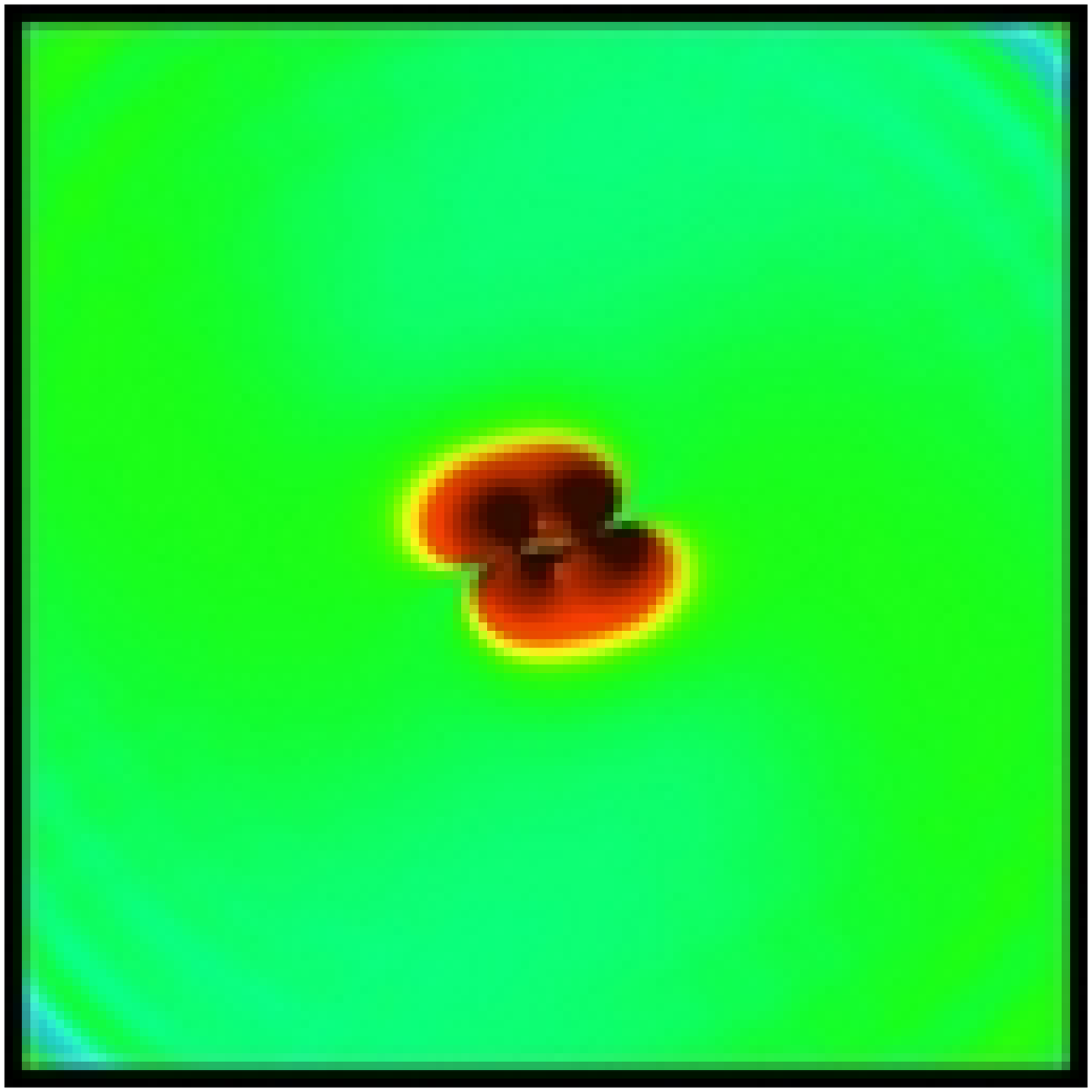} \\\vspace{-0.1cm}
\includegraphics[width=1.27in,clip]{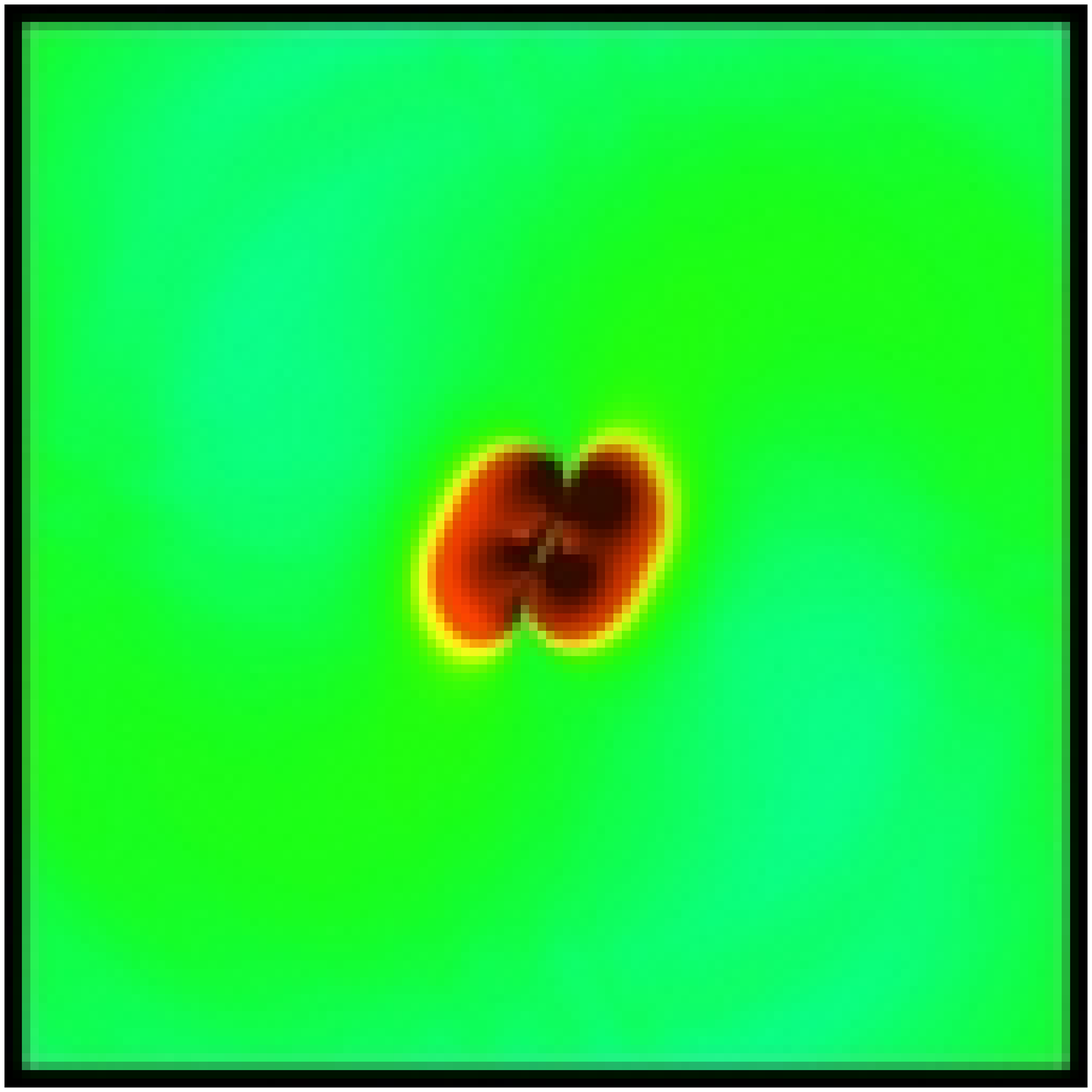} \hspace{-0.4cm}
\includegraphics[width=1.27in,clip]{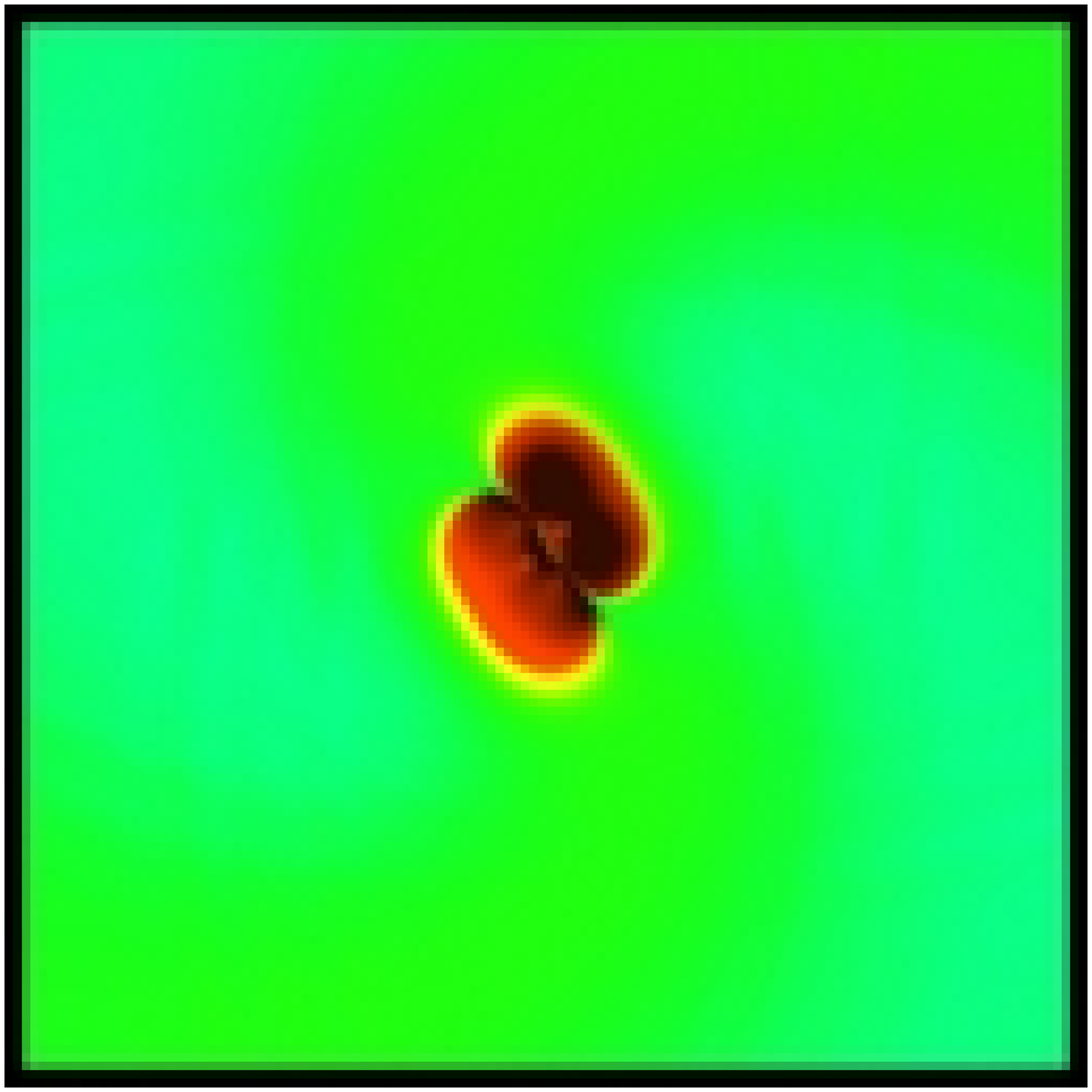} \hspace{-0.4cm}
\includegraphics[width=1.27in,clip]{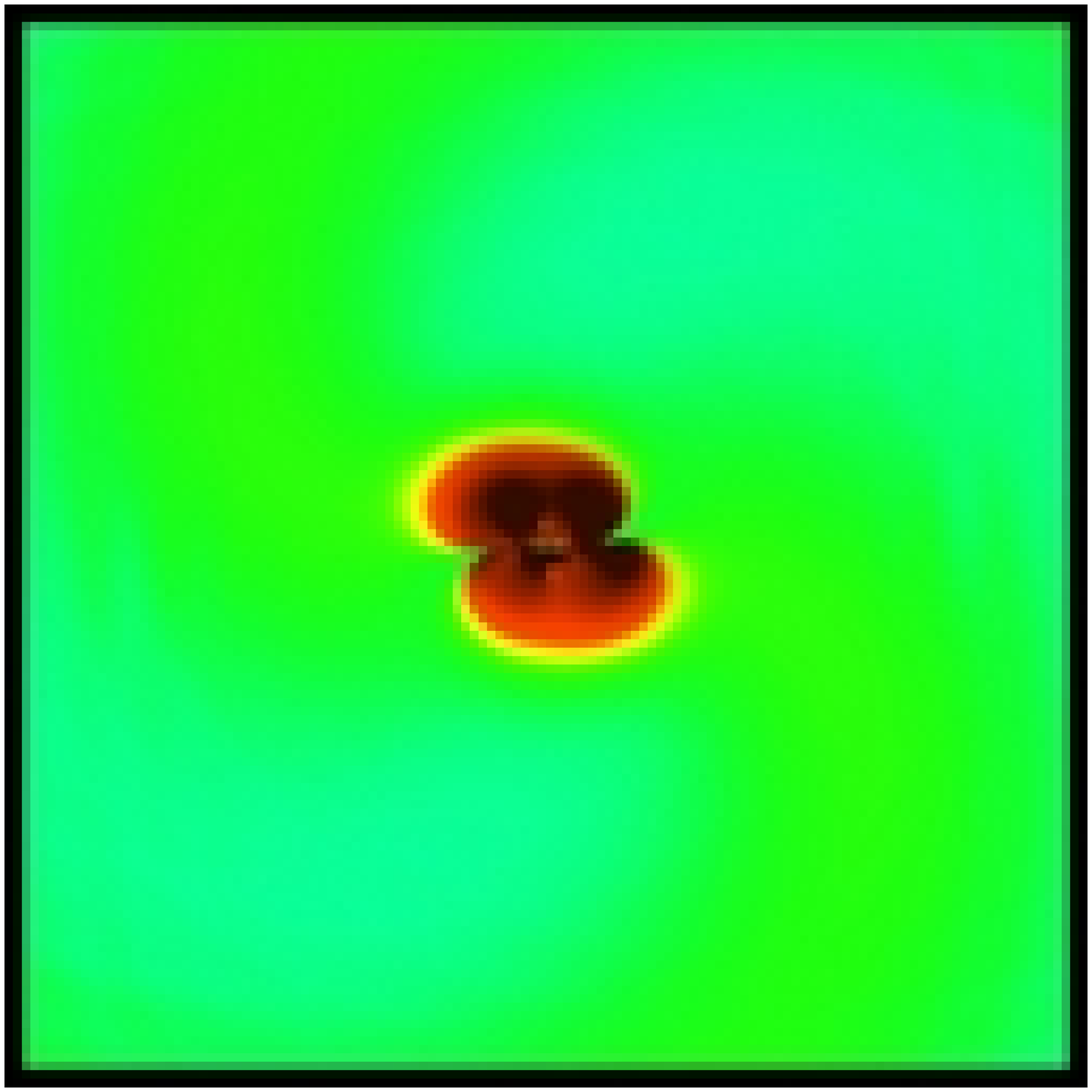} \hspace{-0.4cm}
\includegraphics[width=1.27in,clip]{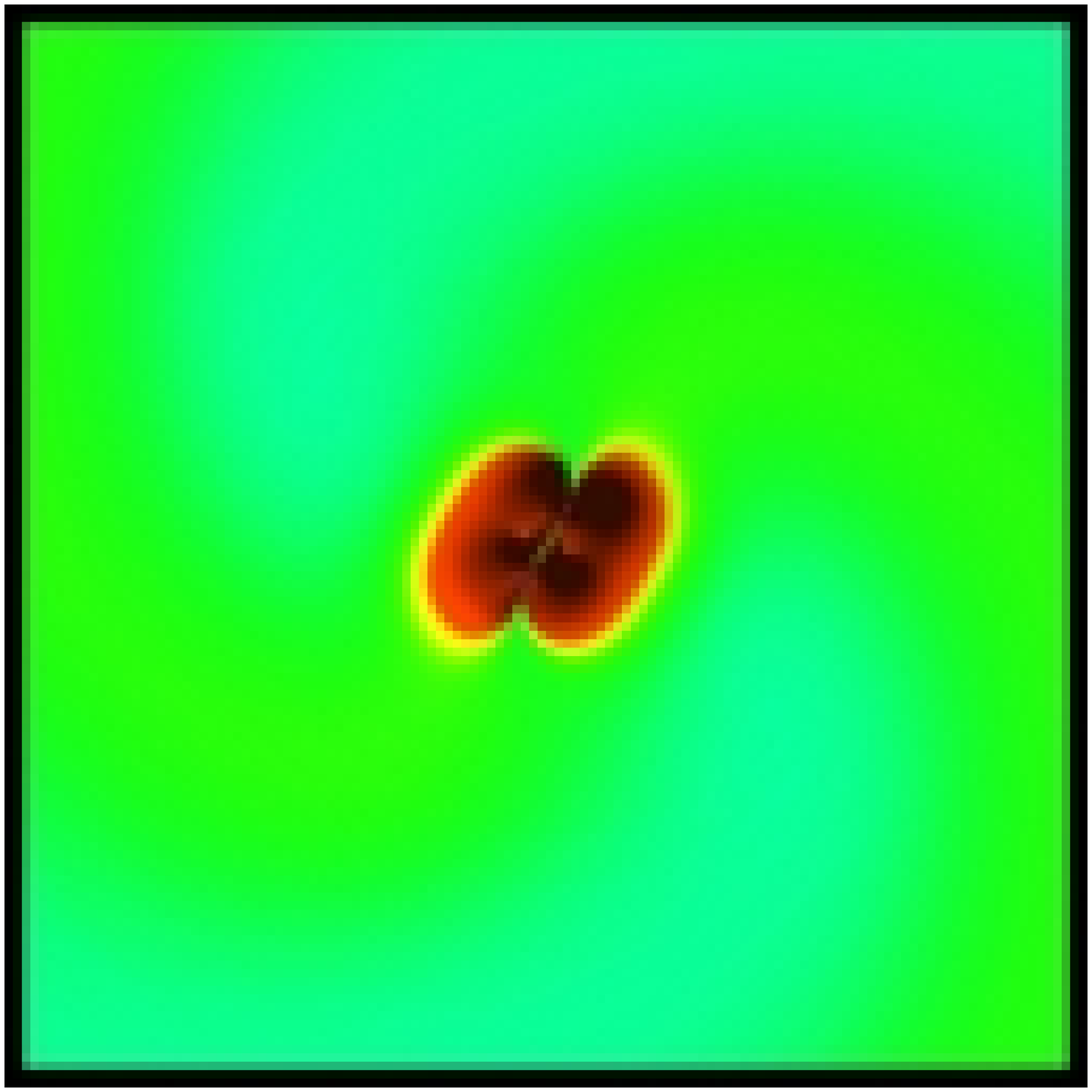} \hspace{-0.4cm}
\includegraphics[width=1.27in,clip]{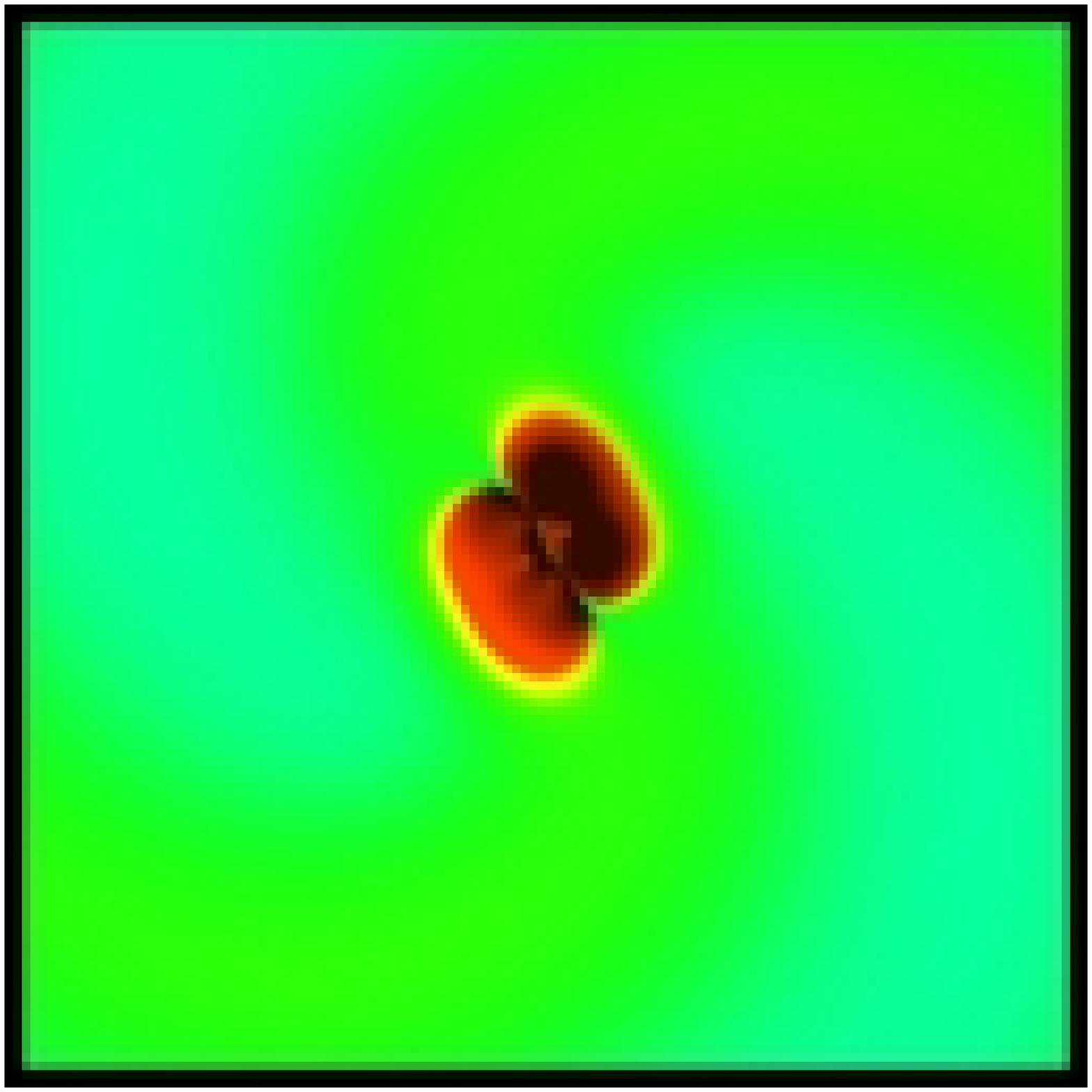} \\\vspace{-0.1cm}
\includegraphics[width=1.27in,clip]{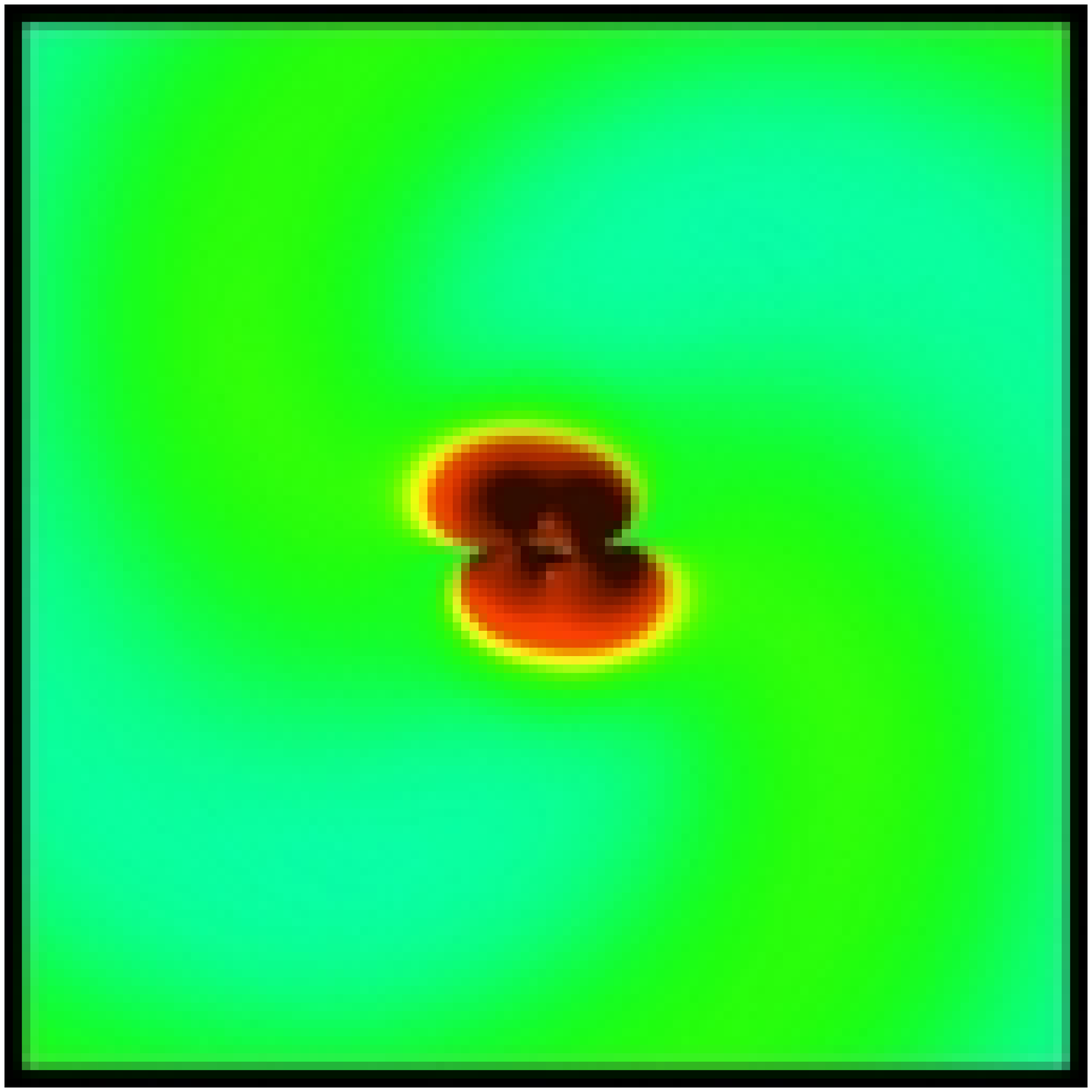} \hspace{-0.4cm}
\includegraphics[width=1.27in,clip]{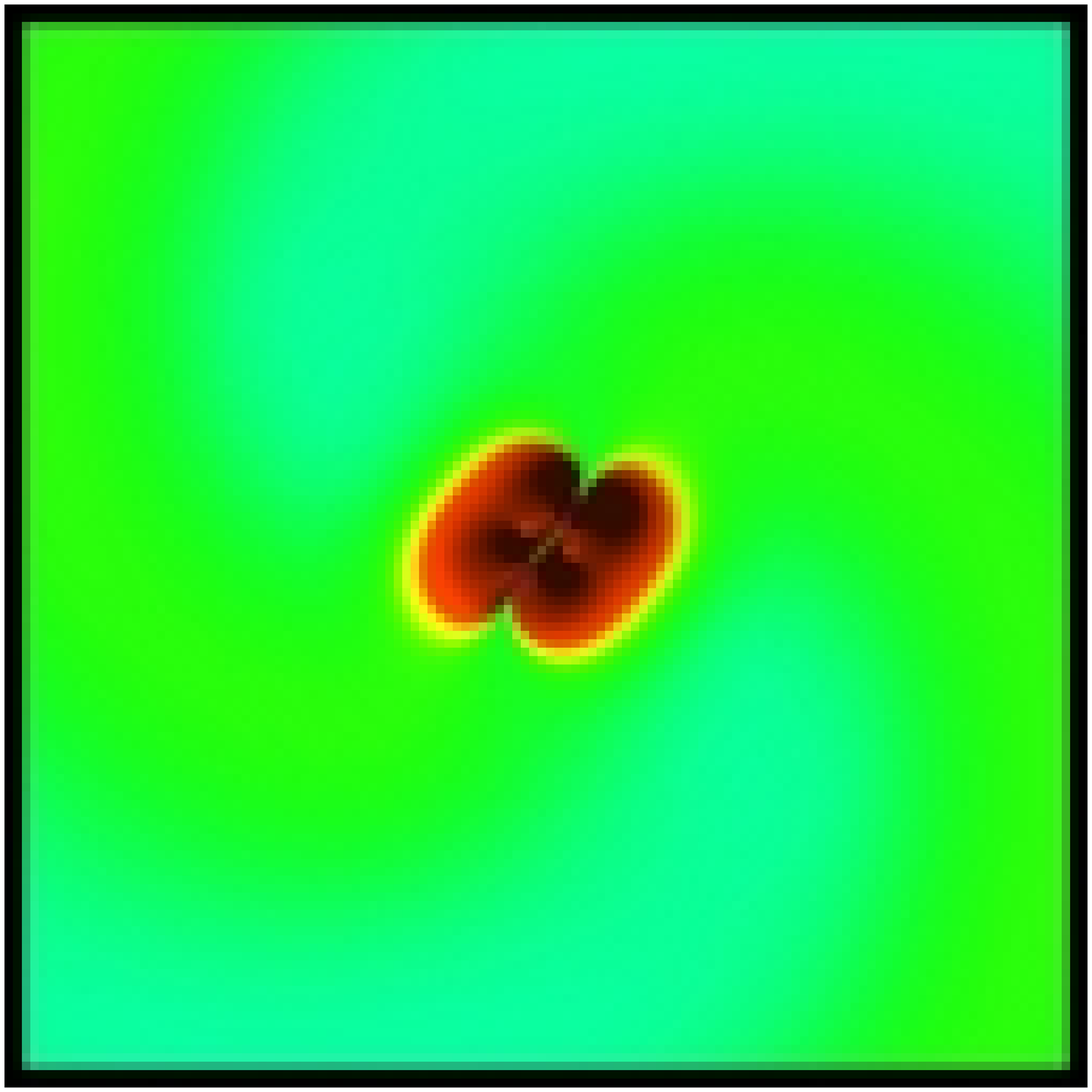} \hspace{-0.4cm}
\includegraphics[width=1.27in,clip]{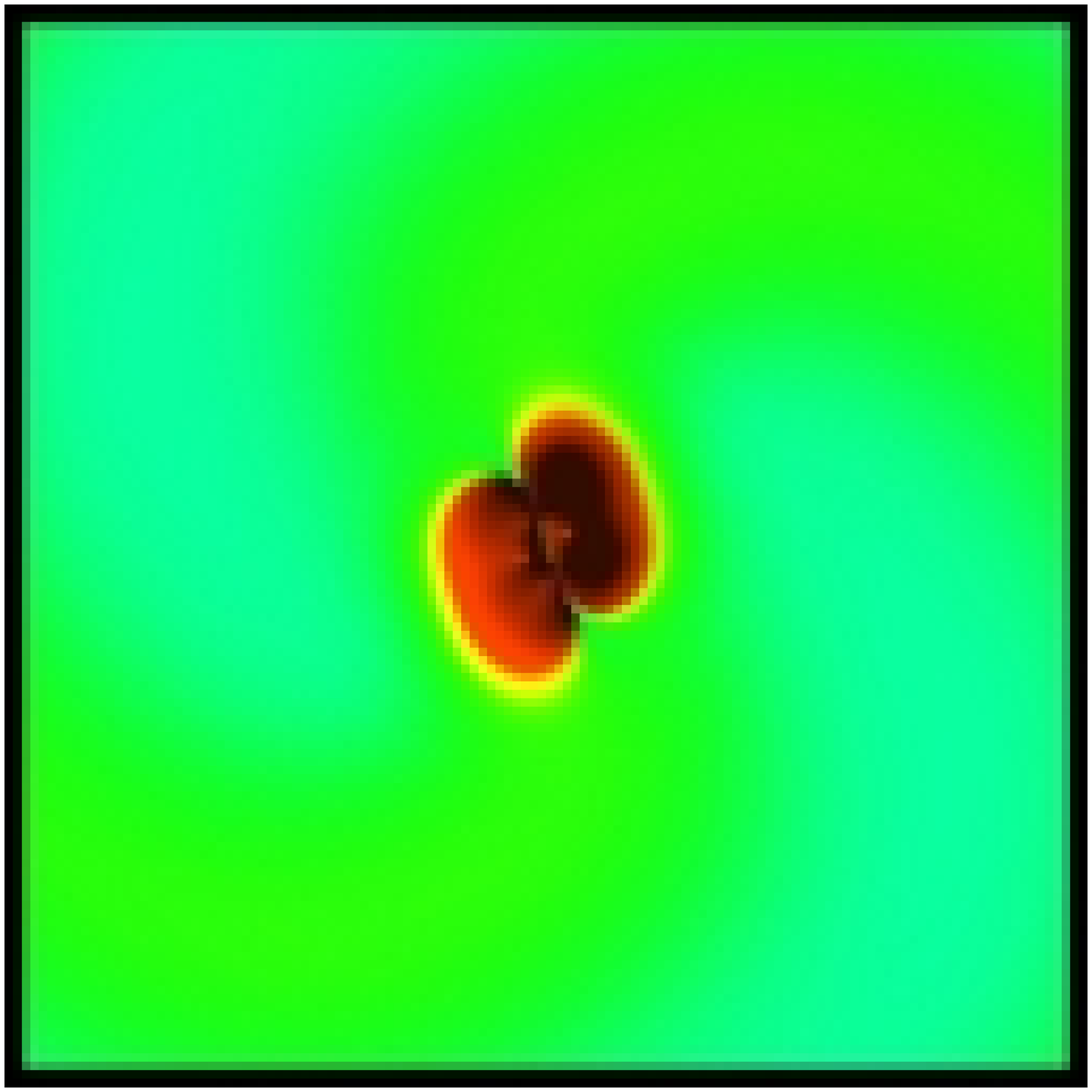} \hspace{-0.4cm}
\includegraphics[width=1.27in,clip]{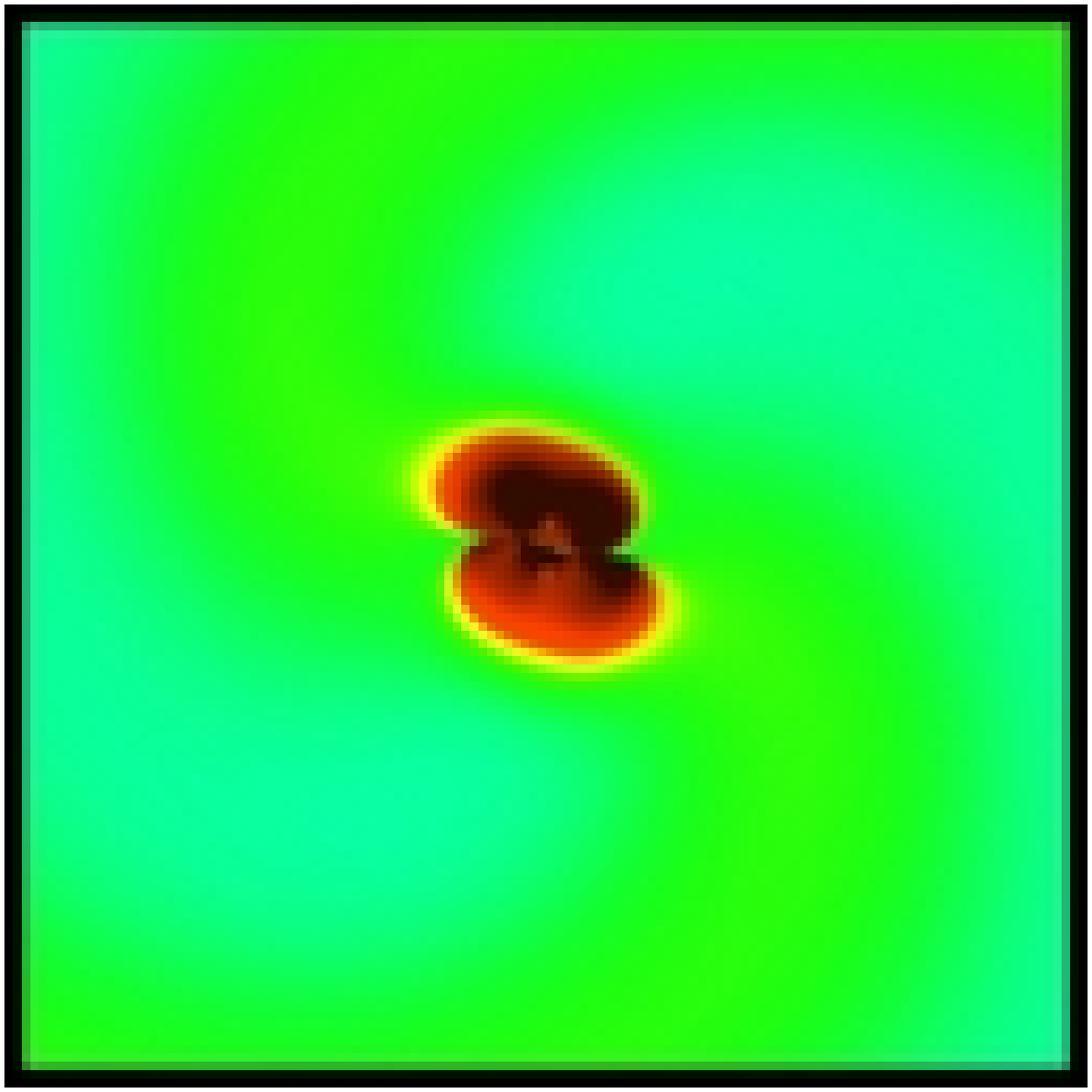} \hspace{-0.4cm}
\includegraphics[width=1.27in,clip]{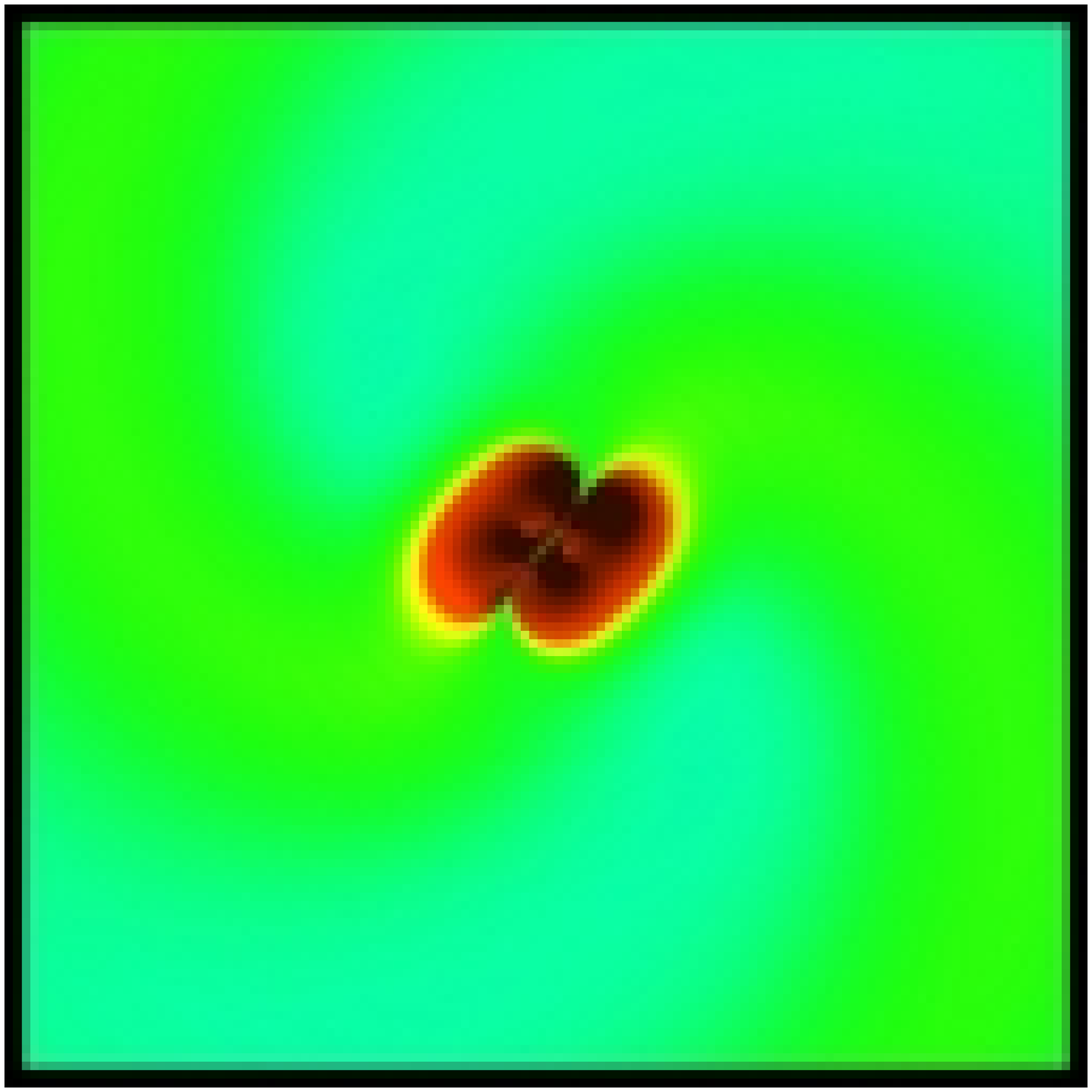} \\\vspace{-0.1cm}
\includegraphics[width=1.27in,clip]{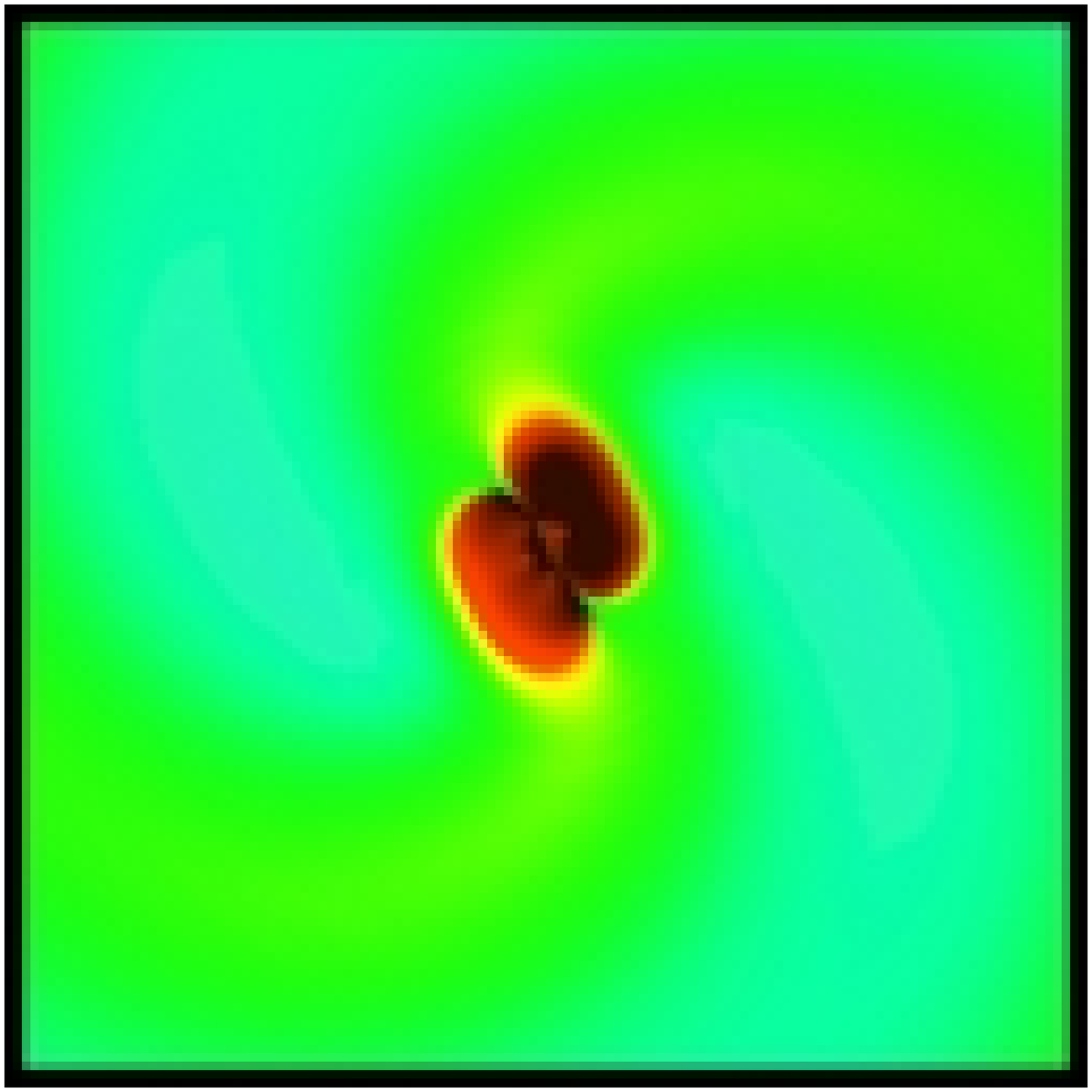} \hspace{-0.4cm}
\includegraphics[width=1.27in,clip]{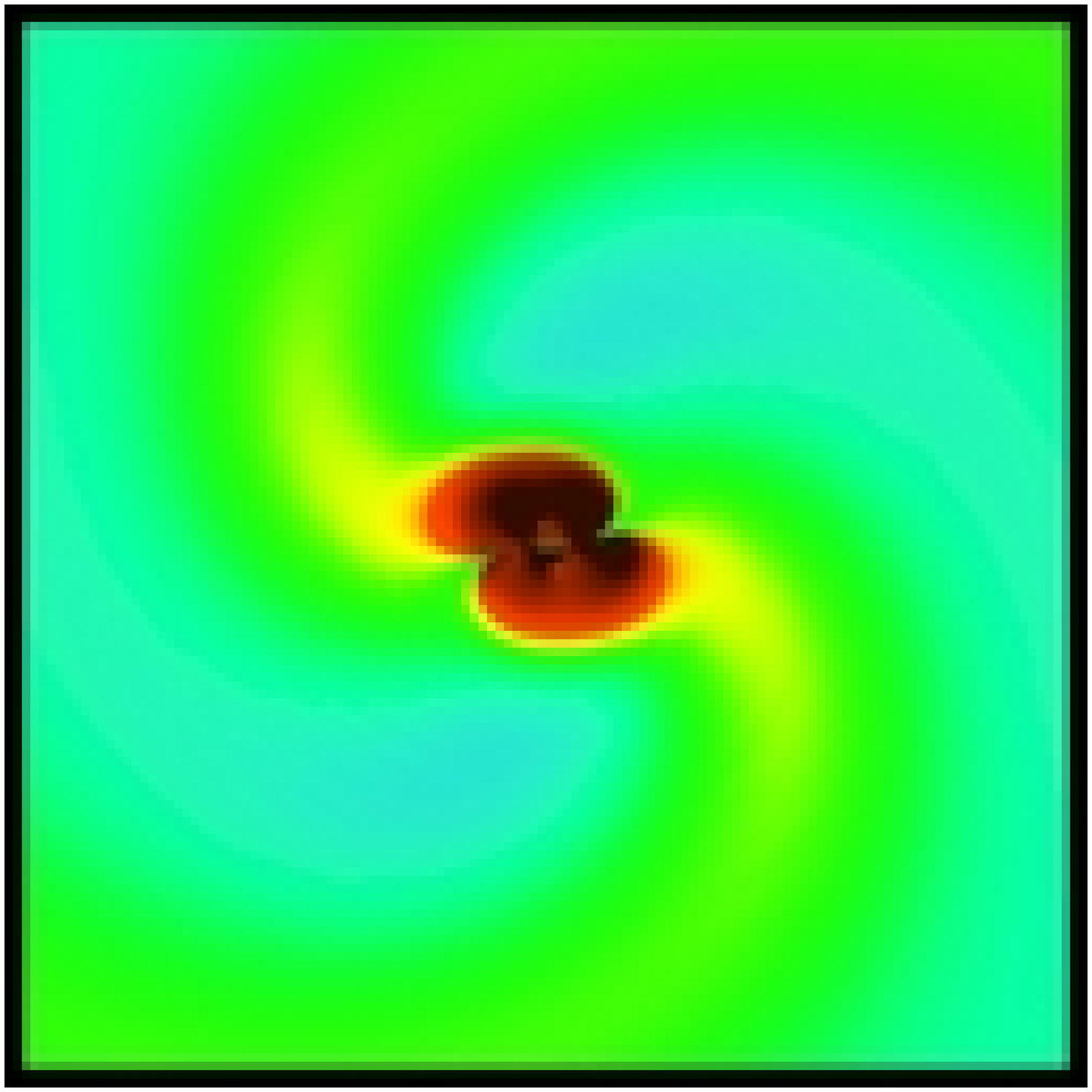} \hspace{-0.4cm}
\includegraphics[width=1.27in,clip]{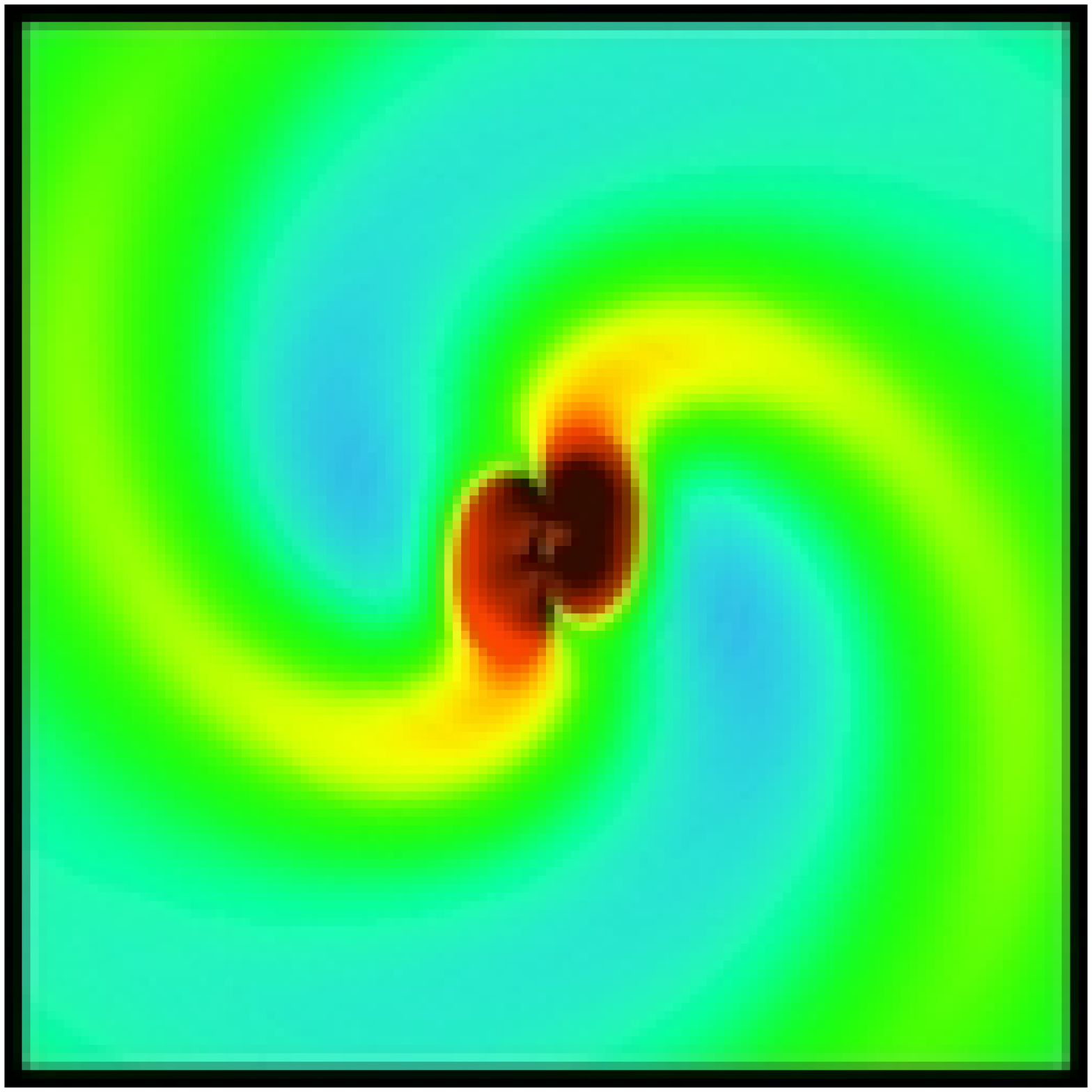} \hspace{-0.4cm}
\includegraphics[width=1.27in,clip]{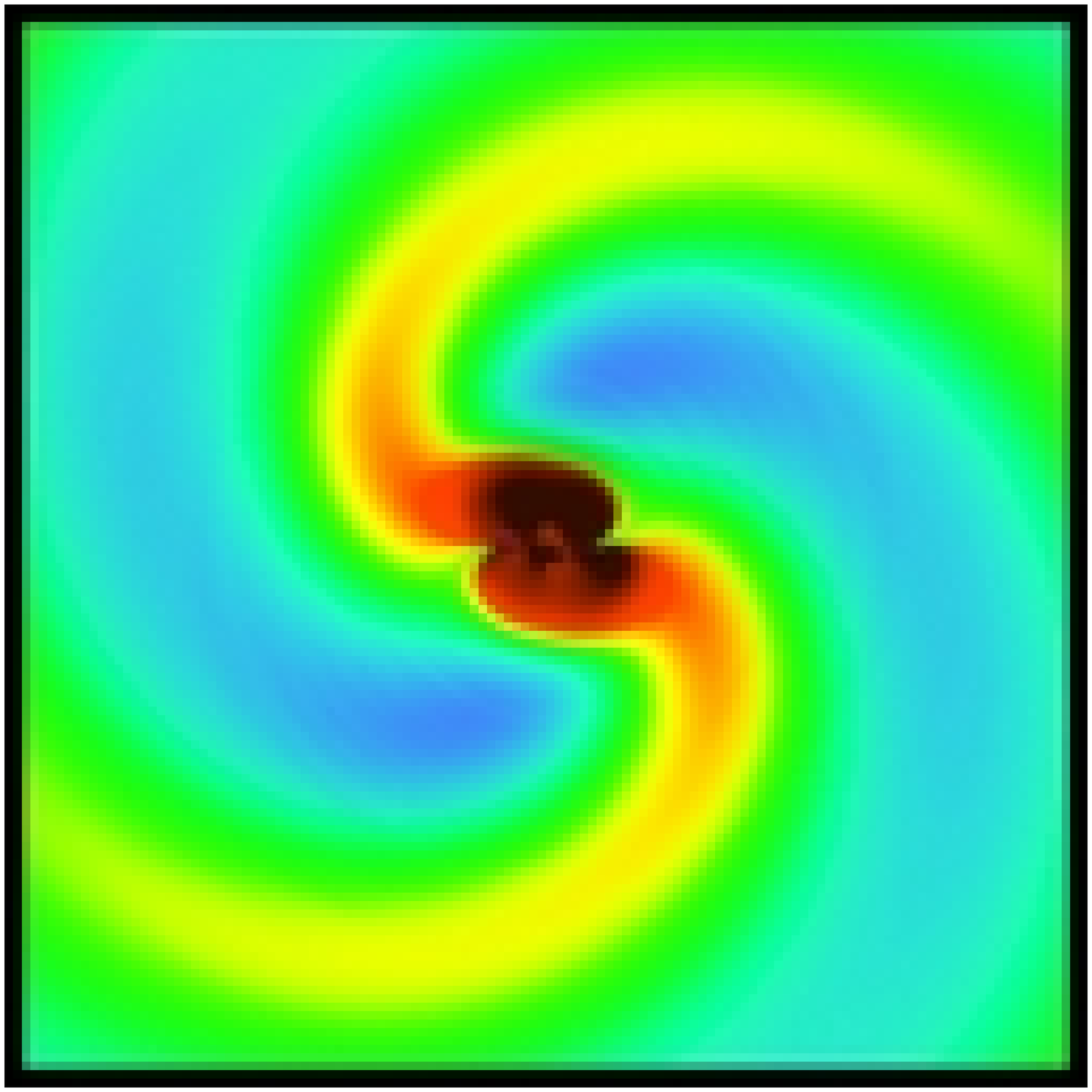} \hspace{-0.4cm}
\includegraphics[width=1.27in,clip]{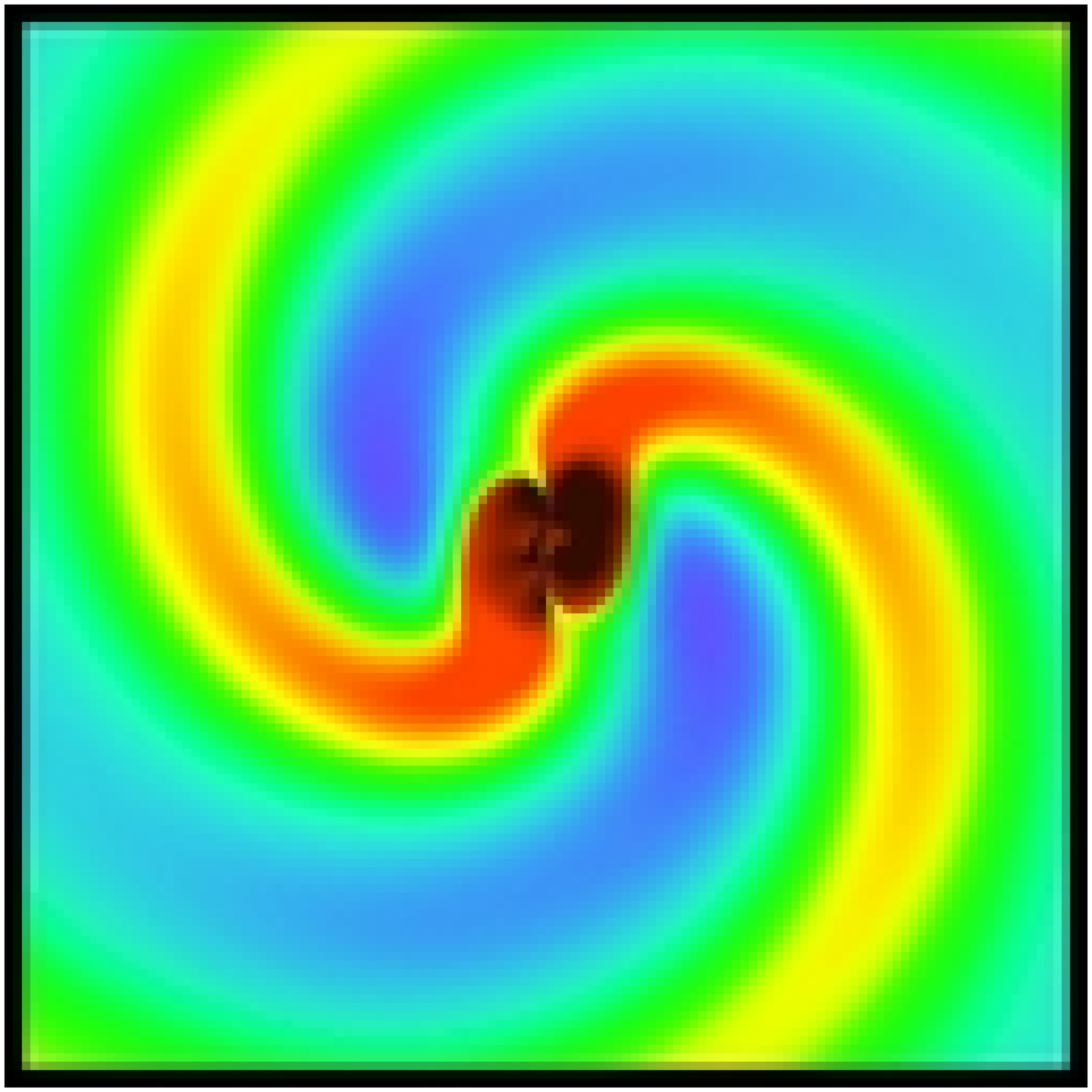} \\\vspace{-0.1cm}
\includegraphics[width=1.27in,clip]{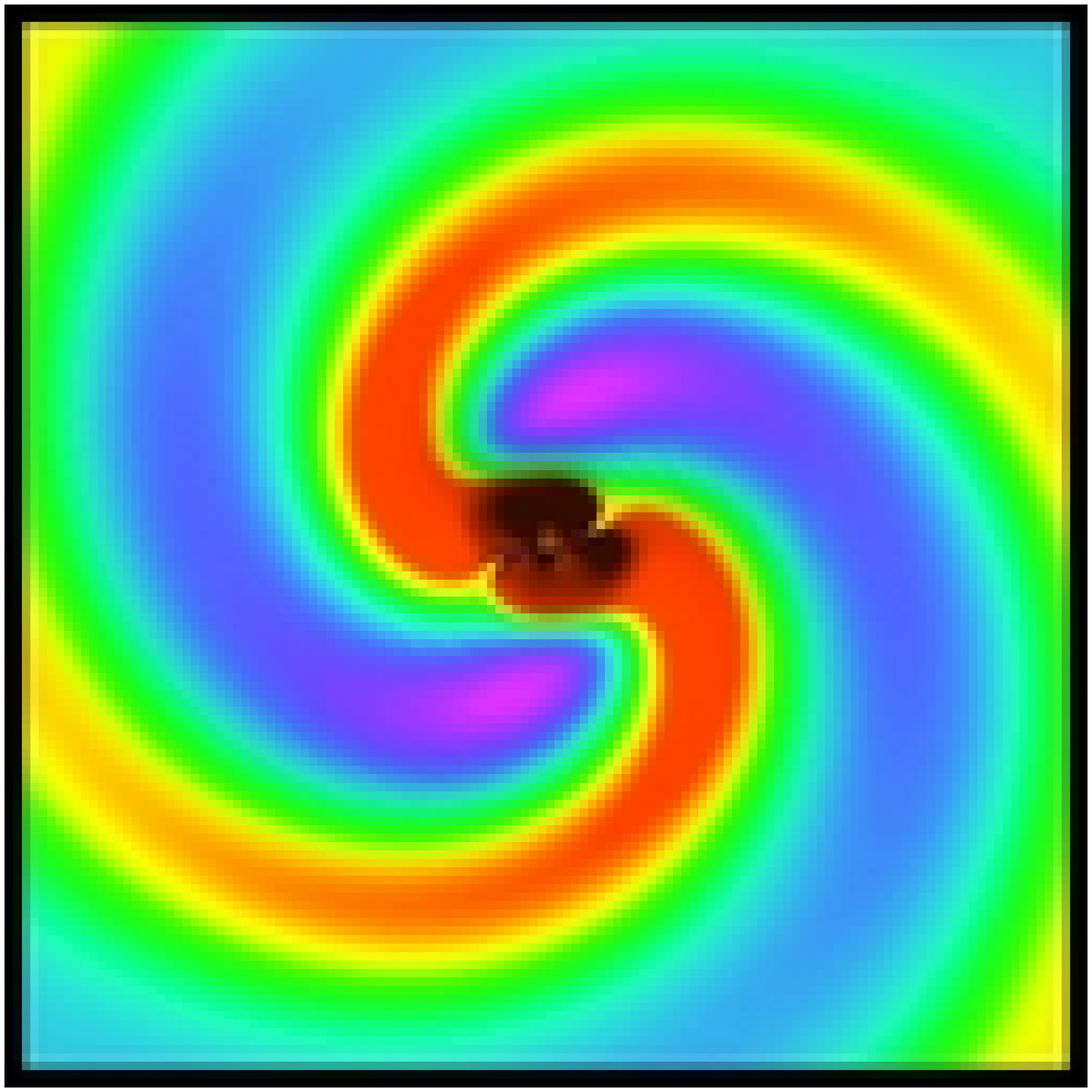} \hspace{-0.4cm}
\includegraphics[width=1.27in,clip]{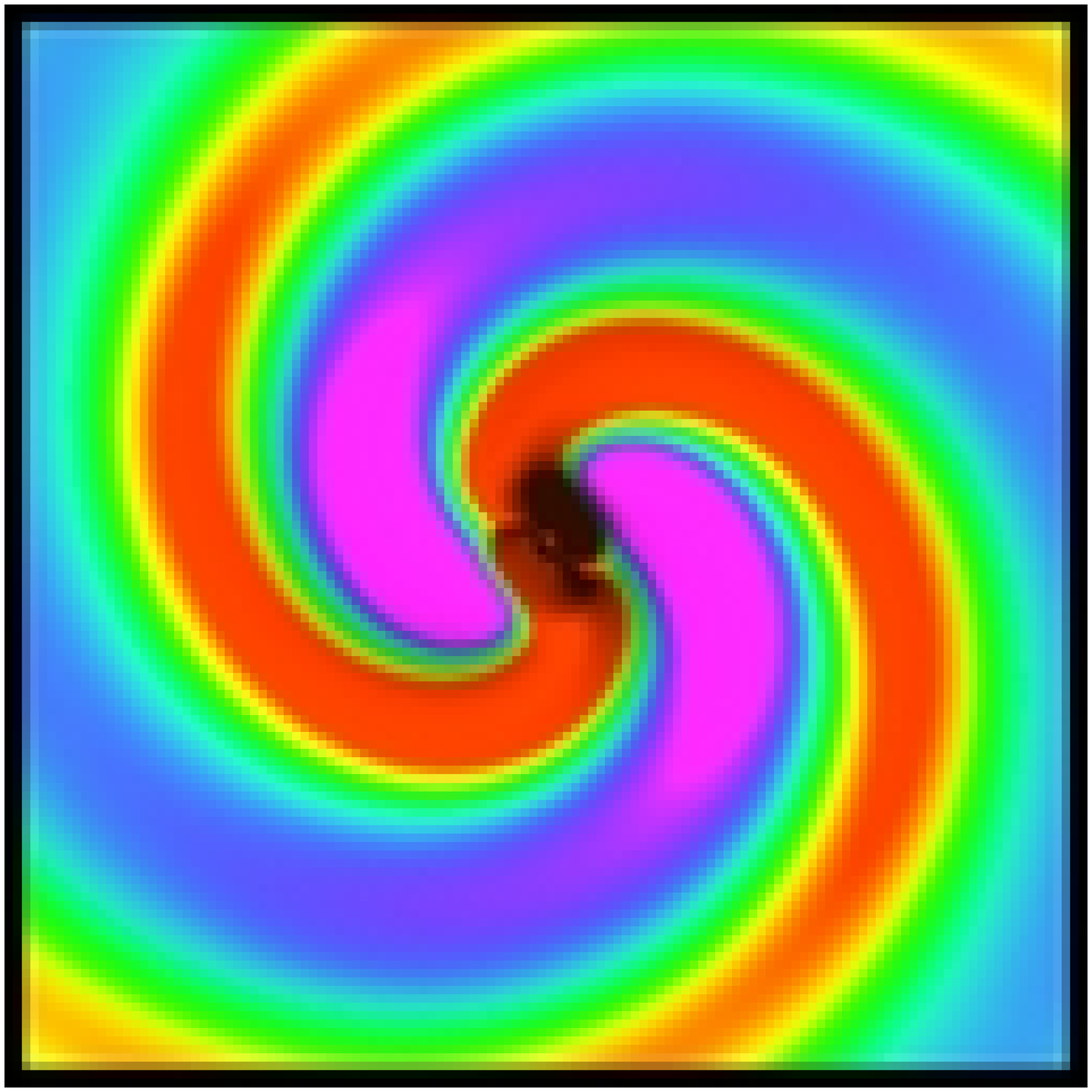} \hspace{-0.4cm}
\includegraphics[width=1.27in,clip]{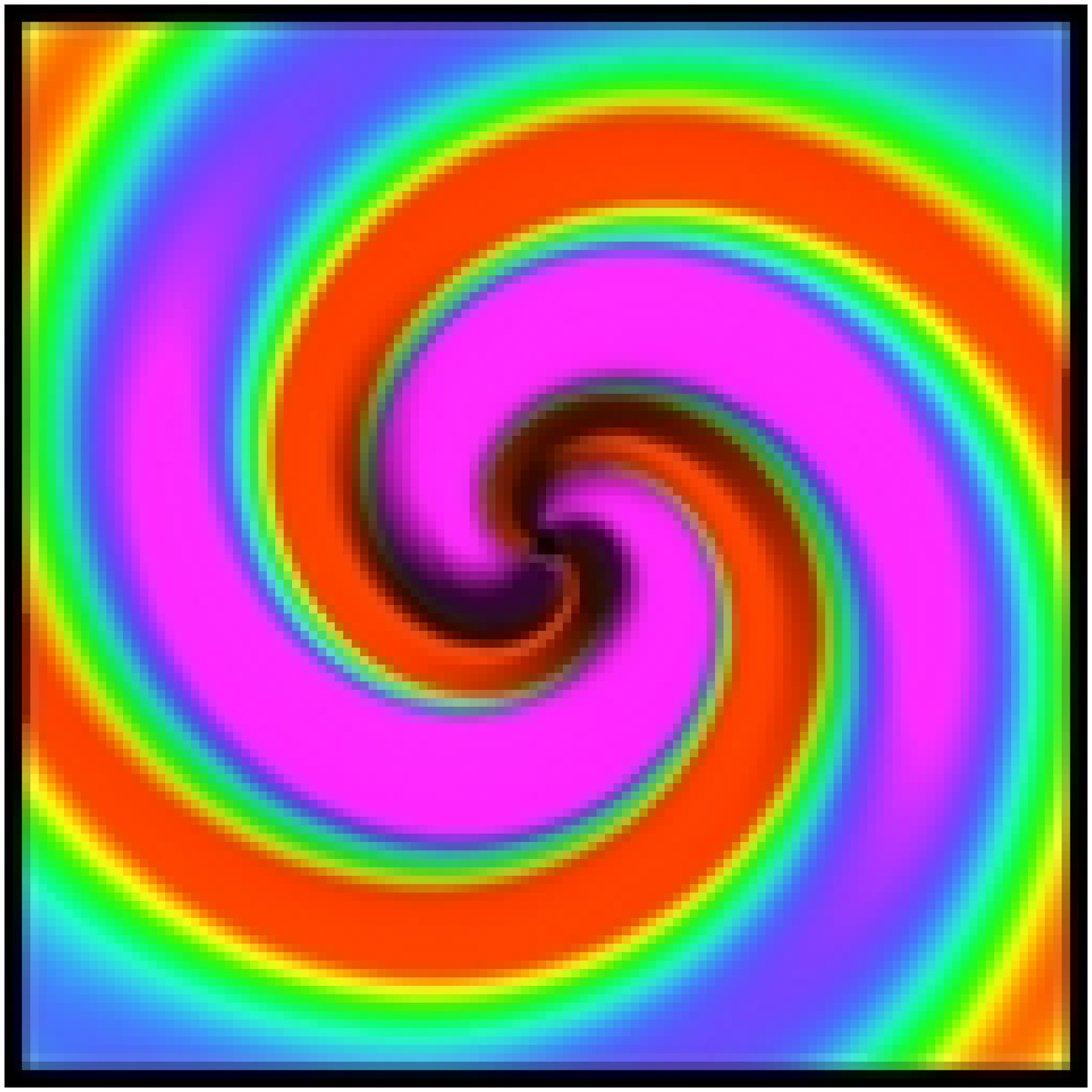} \hspace{-0.4cm}
\includegraphics[width=1.27in,clip]{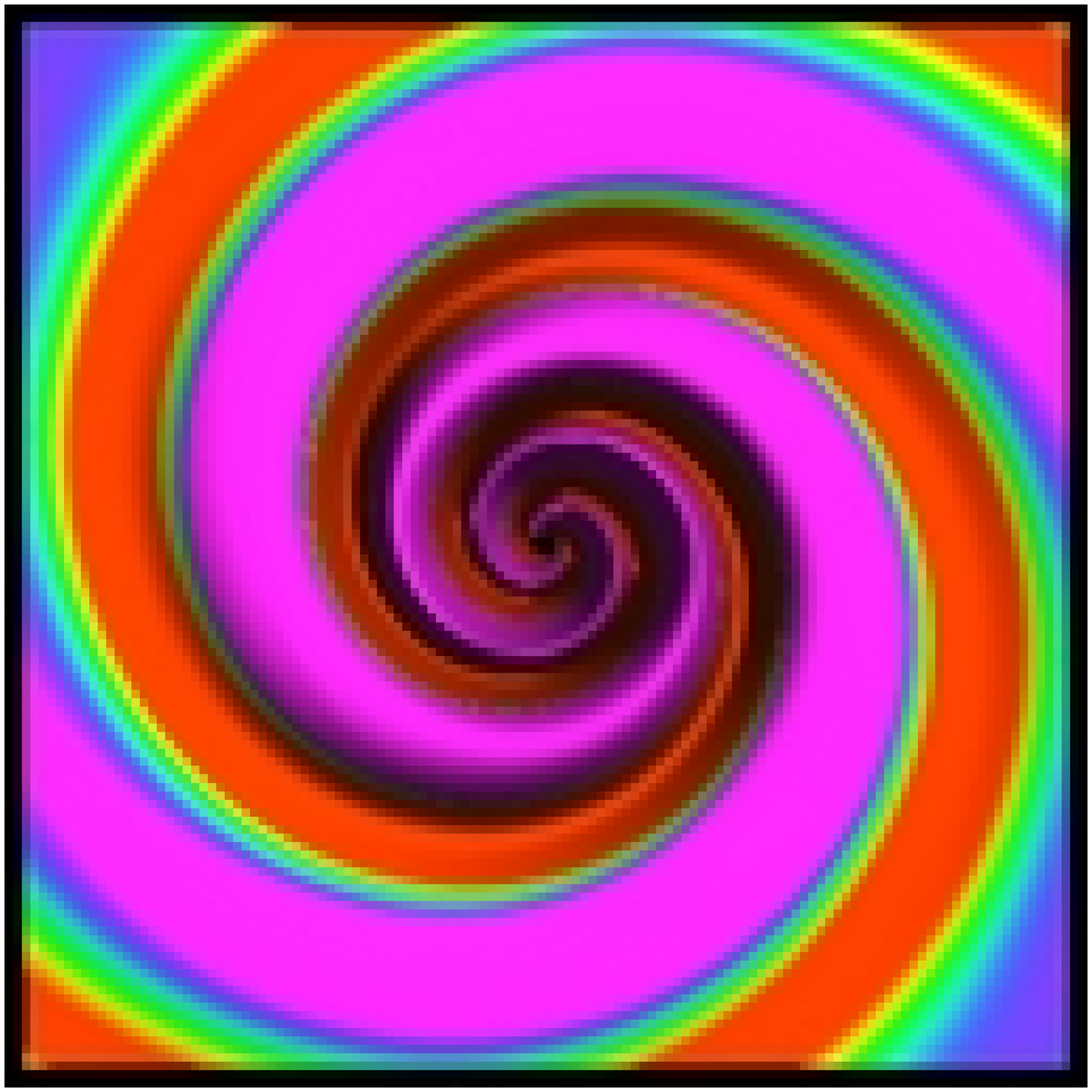} \hspace{-0.4cm}
\includegraphics[width=1.27in,clip]{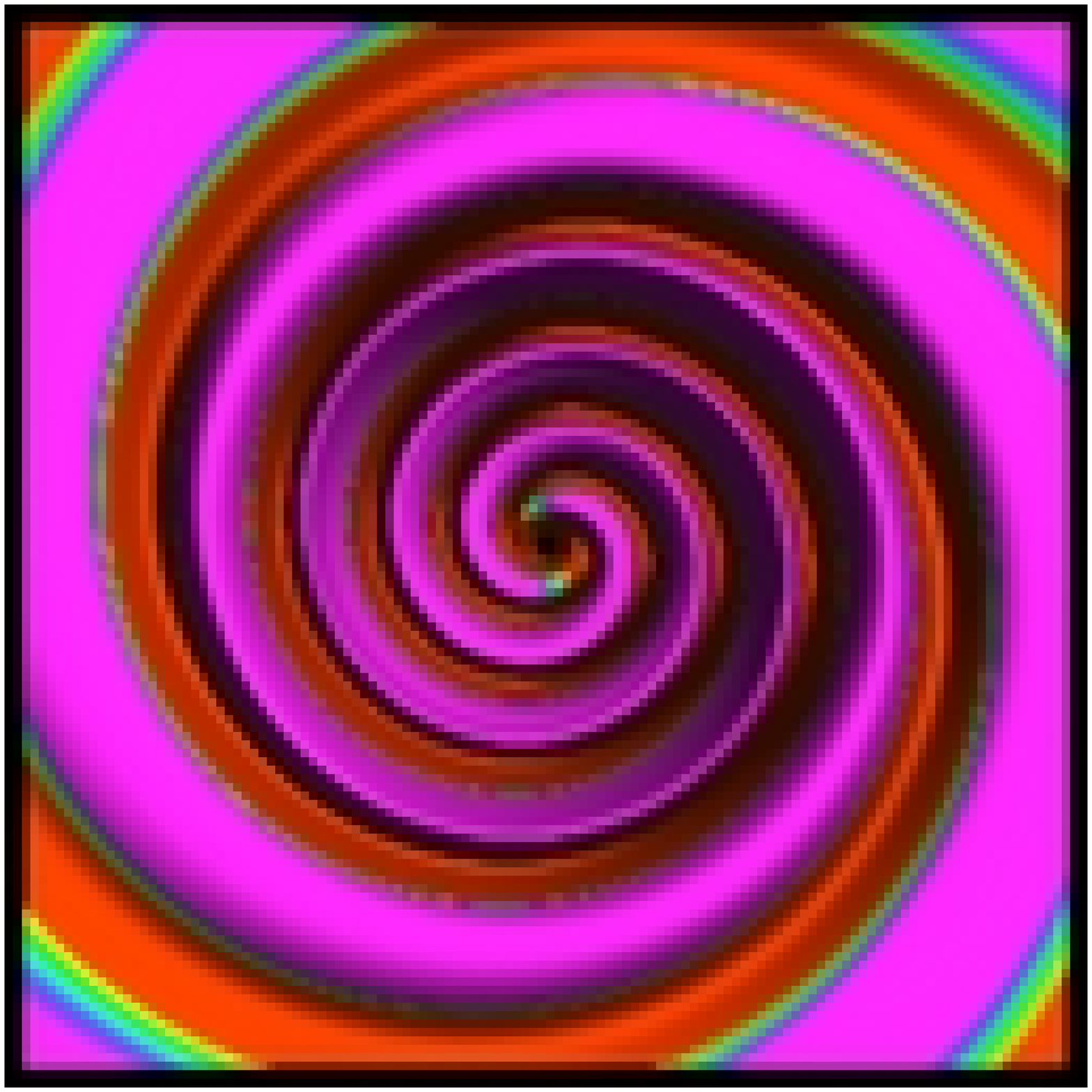} \\\vspace{-0.1cm}
\includegraphics[width=1.27in,clip]{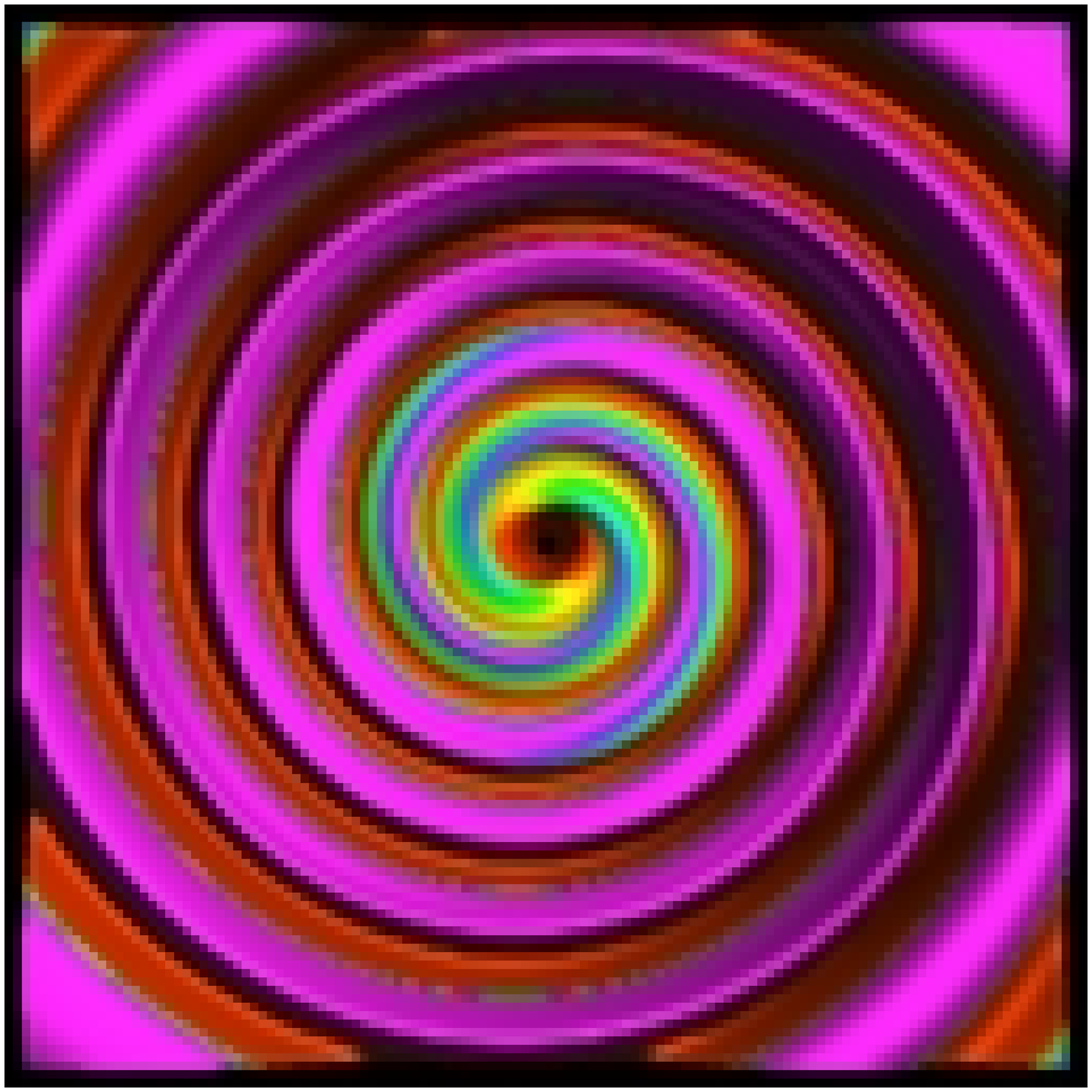} \hspace{-0.4cm}
\includegraphics[width=1.27in,clip]{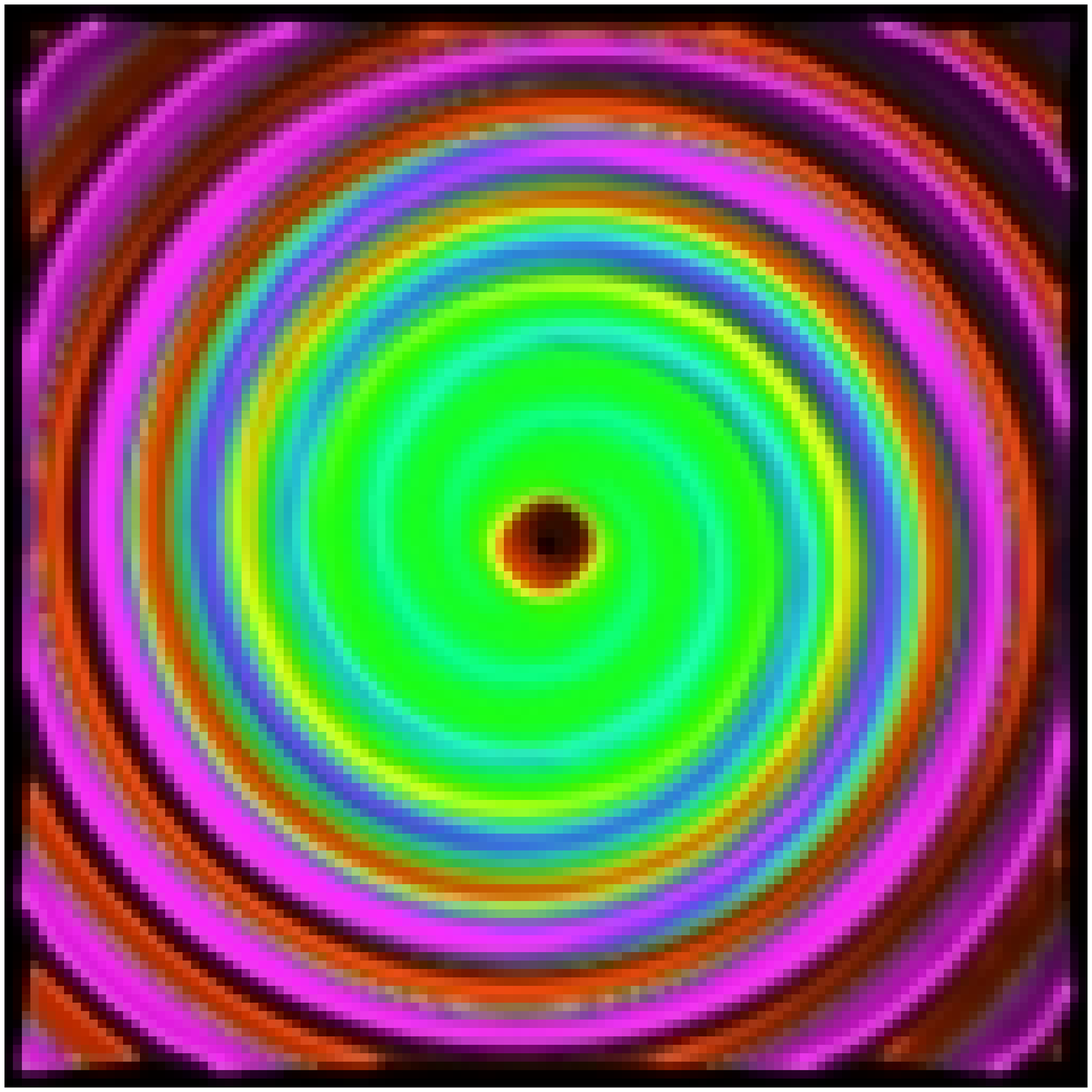} \hspace{-0.4cm}
\includegraphics[width=1.27in,clip]{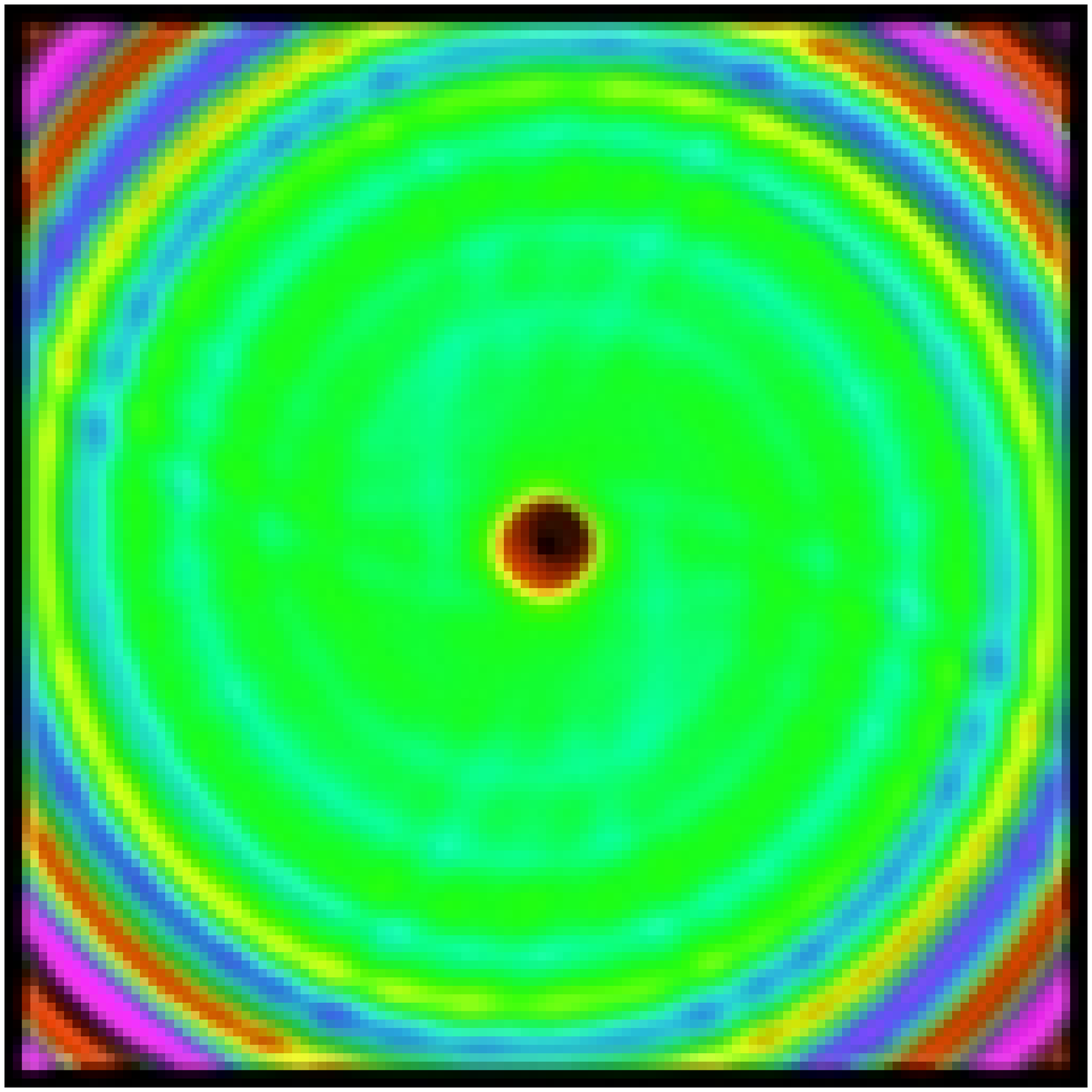} \hspace{-0.4cm}
\includegraphics[width=1.27in,clip]{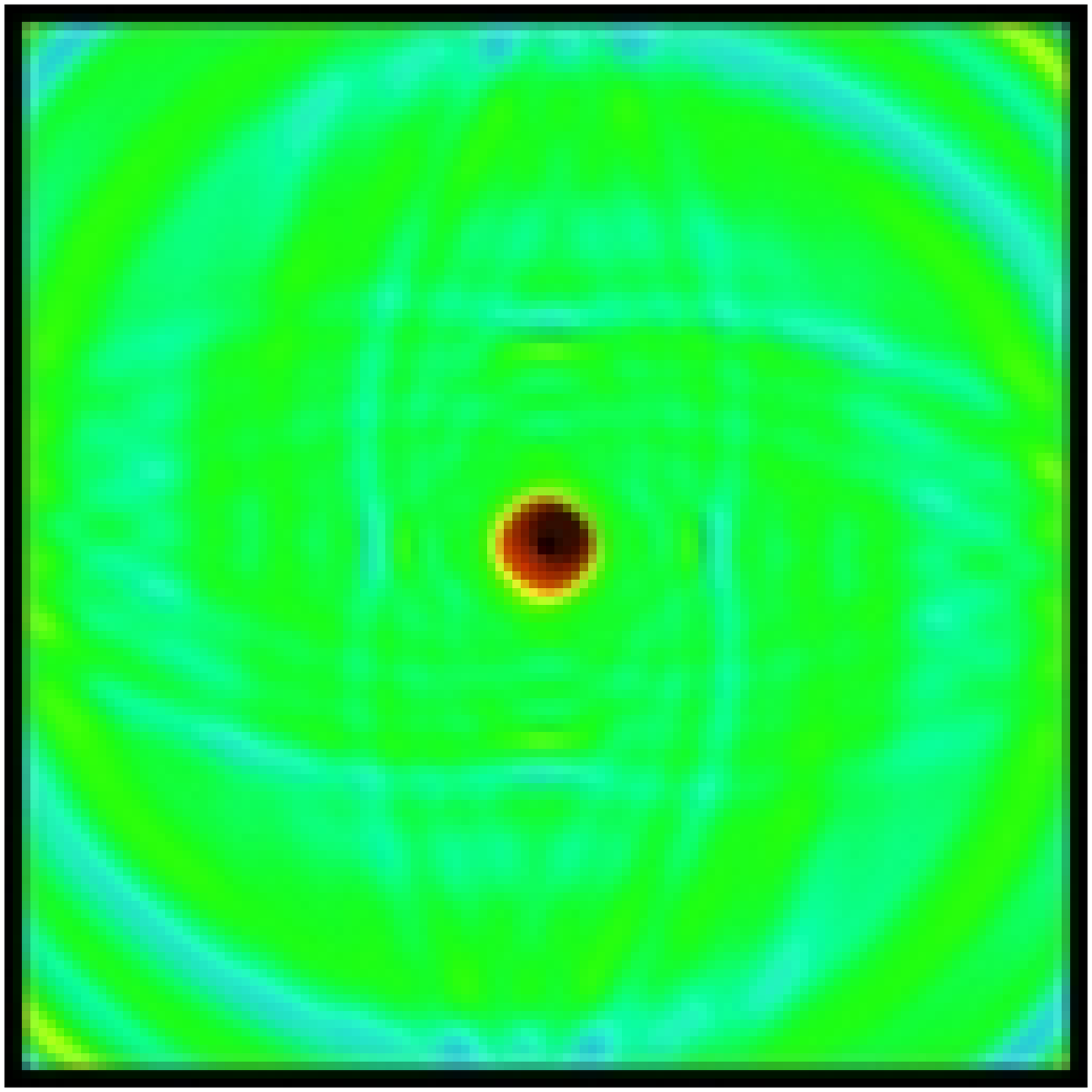} \hspace{-0.4cm}
\includegraphics[width=1.27in,clip]{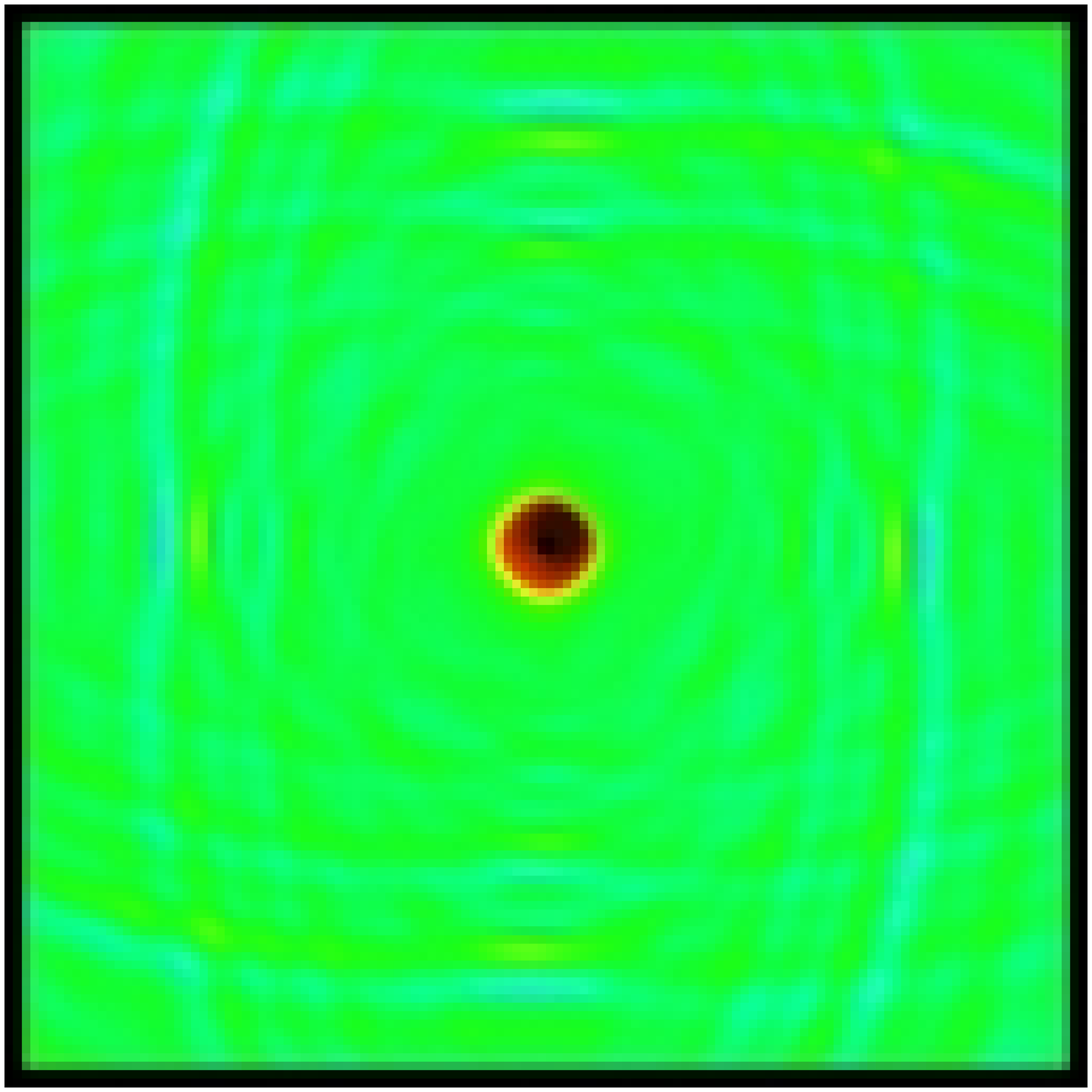}
\caption{A depiction of the gravitational waves emitted during
the merger of two equal mass black holes (specifically ``d=19''
Cook-Pfeiffer initial data~\cite{Buonanno:2006ui}). Shown is a color-map
of the real component of the Newman-Penrose scalar $\Psi_4$ multiplied 
by $r$ along a slice through the orbital plane, which far from
the blackholes is proportional to the second time derivative of the
plus polarization (green is $0$, toward violet (red) positive (negative)). 
The time sequence is from top to bottom, and
left to right within each row. Each imagine is $25M$ apart, and
a common apparent horizon is first detected at $t=529M$ 
(i.e., the ``merger''), which is a little after the frame in row 5, column 3. 
In the first several frames the
spurious radiation associated with the initial data, and how quickly it leaves the domain, is clearly evident.
The width/height of each box is around $100M$.
\label{psi4}}
\end{figure}

\begin{figure}
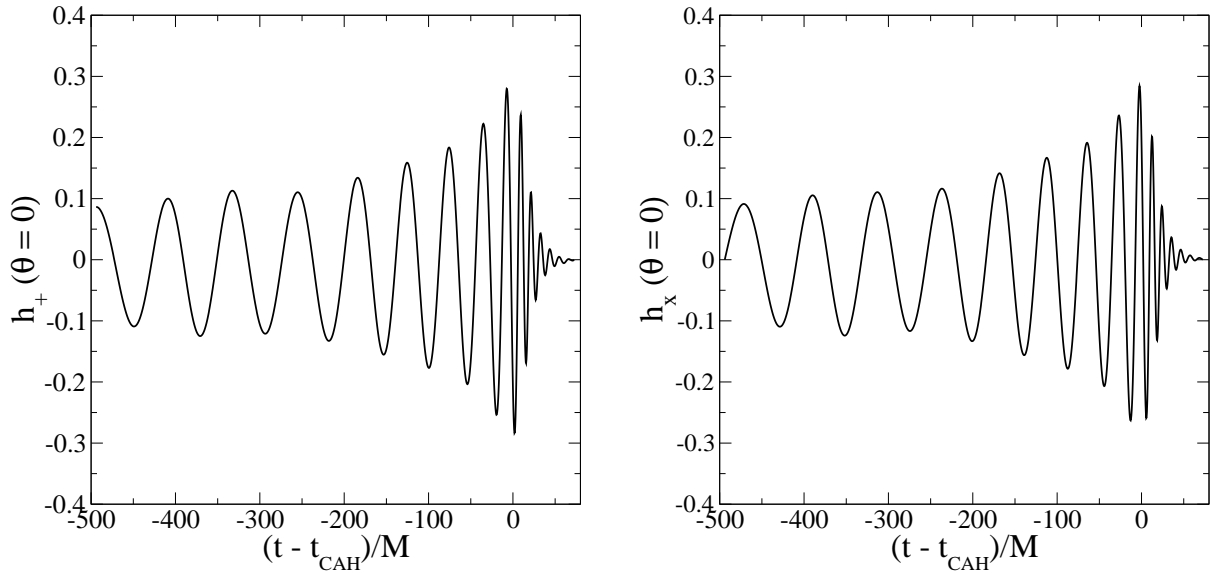

\includegraphics[width=3.0in,clip]{hpd19_no} \hspace{0.5cm}
\includegraphics[width=3.0in,clip]{hcd19_no} 
\caption{The plus (left) and cross (right) polarizations of the
waveform (multiplied by coordinate distance $r$ from the source,
and by the total mass $M$ of the spacetime to non-dimensionalize)
from the simulation shown in Fig.\ref{orbits}, though
here measured along the axis normal to the orbital plane. $t_{CAH}$
is the time when a common apparent horizon is first detected.
\label{hpc}}
\end{figure}

\begin{figure}
\vspace{0.5cm}
\includegraphics[width=5.0in,clip]{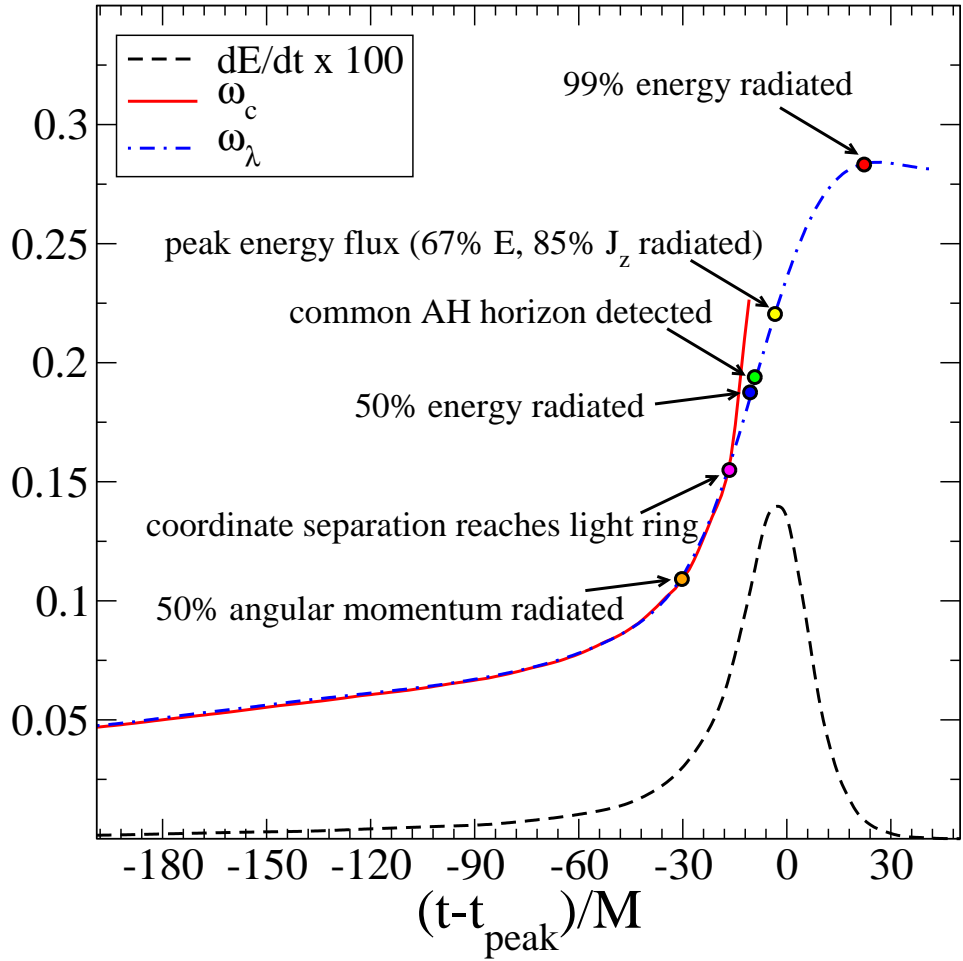} 
\caption{Several phases of the merger as a function of time (horizontal
axis) and orbital/wave angular frequency (vertical axis), from~\cite{Buonanno:2006ui}. $\omega_c$
is the orbital angular frequency of the apparent 
horizons in coordinate space (multiplied by the total mass $M$ to 
non-dimensionalize); this curve ends once
a common apparent horizon forms. $\omega_\lambda$ is the 
instantaneous frequency of the emitted gravitational wave divided
by 2 (and normalized by $M$ again). $dE/dt$ is the luminosity
of the wave integrated over the wave extraction 2-sphere.
$J_z$ is the component of the angular momentum of the gravitational
waves normal to the orbital plane. The ``light ring'' here is defined
as the coordinate location of the unstable equatorial photon
orbit of the final Kerr black hole. One cannot define this precisely
or unambigously in the binary spacetime, though it is interesting
that the orbital and gravitational wave frequencies decouple
roughly at this separation. This is also the time when the EOB approach advocates
stitching together the inspiral waveform from resummed PN calculations
to a ringdown signal. 
\label{phases}}
\end{figure}

For illustrative purposes, Fig.'s \ref{orbits}-\ref{phases} show 
some of the simulation results, taken from evolution of Cook-Pfeiffer initial data,
described in detail in~\cite{Buonanno:2006ui}.
Fig.~\ref{orbits} is a plot of the orbital trajectory,
Fig.~\ref{psi4} shows the real component of the Newman-Penrose scalar
$\Psi_4$ in the orbital plane (which far from the source
represents the second time derivative of the ``plus'' polarization
of the usual gravitational wave strain), Fig.~\ref{hpc}
shows the plus and cross polarizations of the waveform
extracted on the axis normal to the orbital plane,
and Fig.~\ref{phases} shows the instantaneous gravitational wave
frequency (divided by 2) and energy flux
versus time together with labels depicting
some phases of the merger.


\subsubsection{Gravitational wave structure}\label{bbh_equal}

Decomposed into a spin-weight 2 spherical harmonic basis, 
the waveform throughout the evolution is dominated by
the quadrupole ($\ell=2,|m|=2$) component. The next
leading order component ($\ell=4,|m|=4$) has
an amplitude less than $1/10^{th}$ the quadrupole
mode during the inspiral phase, growing briefly to 
$1/5^{th}$ of it near the peak of the emitted
energy flux (and note that the energy content of a
mode is proportional to the {\em square} of its amplitude)~\cite{Buonanno:2006ui,Berti:2007fi}.
Moreover, the {\em quadrupole
formula} seems to describe the physics of 
gravitational wave {\em production} quite accurately throughout
the orbital phase, in that the coordinate 
motion of the apparent horizons taken from the simulation and
plugged verbatim into the
quadrupole formula for two point-masses shows remarkable
agreement with the full numerical waveform~\cite{Buonanno:2006ui,Pretorius:2007jn}.   
Several studies have now also shown that the higher order PN and EOB
methods can reproduce, to within the various
uncertainties of the comparison (including numerical
error, mapping parameters between the two descriptions,
when to begin the comparison, etc.),
up until close to merger~\cite{Buonanno:2006ui,Baker:2006ha,Hannam:2007ik,
Pan:2007nw,Buonanno:2007pf}. The most accurate
study to date~\cite{Hannam:2007ik} of exactly
{\em when} the waveform from a particular PN approach 
begins to deviate from the numerical signal to within 
the errors of the simulation---$0.3\%$ in the phase
and $1\%$ in the amplitude over 18 cycles (9 orbits)
of inspiral---showed that despited a relatively large amplitude
disagreement of $7\%$ between the restricted 3.5PN Taylor waveforms,
the cumulative phase difference was $0.15$ after $13$ cycles,
suggesting for the given numerical accuracy only the
last $4.5$ orbits of the inspiral would require numerical
solution.

The transition from inspiral to ringdown does not last
very long, only on the order of $10-20M$. There is also
no  noticeable ``plunge'' in the orbital motion from the time
there are two distinct black holes to a single one (see Fig.\ref{orbits}). 
However, in the Fourier transform of the waveform there seems to be
a distinct change in the slope of the spectrum
from the leading order PN prediction of $-7/6\approx-1.2$ to somewhere
between $-0.6$ and $-0.8$ before
asymptoting to the dominant ringdown frequency~\cite{Buonanno:2006ui,Baumgarte:2006en}.

The ringdown portion of the waveform is dominated by the 
the fundamental harmonic ($n=0$) of the quadrupole moment $(\ell=2,m=2)$
of the final black hole's quasi-normal
modes (QNMs)~\cite{Buonanno:2006ui,Berti:2007fi}.
The first two overtones of the quadrupole
mode have amplitudes close to the fundamental mode,
though they decay rapidly and are thus only
discernible early on during the inspiral. Higher order multiple
modes are also present, though as with the waveform itself
at a much reduced amplitude compared to the quadrupole mode.
An interesting property of the waveform is that from
the moment of the peak in the flux onwards it can quite
accurately be represented as a sum of QNMs. One reason
why this is interesting is
that here one would expect to be furthest into
the regime where ``non-linear effects'' are most apparent,
yet the wave can be described as coming from a linearly
perturbed black hole. Proponents of the EOB approach
predicted this behavior, and in fact have further
suggested that with a sufficient number of QNM overtones
and harmonics that the entire post-inspiral portion
of the waveform may be described as a ringdown.
This prescription has been carried out quite successfully 
for the extreme mass ratio case~\cite{Damour:2007xr},
and a range of non-spinning near equal mass mergers (with mass
ratios from $1:1$ to $4:1$) ~\cite{Buonanno:2007pf},
though may not be as straight-forward (or possible at all)
for general configurations with spin. 

\subsection{Unequal mass, minimal eccentricity and spin}\label{bbh_unequal}

Relaxing the condition of equal mass from the
configuration discussed in Sec.\ref{bbh_equal},
several qualitative features of the merger and corresponding
waveform change~\cite{Herrmann:2006ks,Baker:2006vn,Gonzalez:2006md,Berti:2007fi}. 
First, the equal mass case maximizes
the total energy emitted
and also maximizes the final spin of the remnant black hole.
To a good approximation the total energy radiated decreases by a factor
$(\eta/\eta_1)^2$, and the final spin decreases
linearly in $\eta$ via $a/M_f\approx 0.089 + 2.4\eta$, where the symmetric mass ratio 
$\eta=q/(1+q)^2$, $\eta_1=1/4$, $q=M/m$ with $q\geq 1$,
and $M_f=m+M$~\cite{Gonzalez:2006md,Berti:2007fi}.
The second difference is that although the quadrupole mode
still dominates in the waveform, certain higher multipole
modes, in particular the $\ell=3,|m|=3$ component, become non-negligible~\cite{Berti:2007fi}.
The simple explanation for this is quadrupole-formula physics again:
the reduced quadrupole {\em moment} of the effective source energy distribution
now has higher multipole {\em modes} when
expressed in terms of a spherical harmonic decomposition,
and this will be reflected in the structure of the gravitational 
waves emitted. The final significant difference is that there is
an asymmetric beaming of the gravitational radiation in the
orbital plane due to the mass difference. If not for the inspiral
this would average to zero over a complete orbit, however the inspiral,
combined with fact that the radiation eventually ceases due to merger,
results in the asymmetry. This imparts a ``kick'', or recoil
of the final black hole within the orbital plane to compensate
for the net linear momentum carried away by the radiation.
The dependence of recoil speed $v$ on mass ratio can be
approximated by the Fitchett formula~\cite{fitchett} $v=A\eta^2\sqrt{1-4\eta}(1+B\eta)$,
with $A\approx1.20\times10^4$ and $B\approx-0.93$~\cite{Gonzalez:2006md}. The
maximum is upwards of $175 km/s$ achieved around a mass ratio 
$\approx 3:1$. Note that the
direction of recoil, which in this case occurs within the orbital plane, depends 
on the ``initial'' phase of the orbit, and thus for astrophysical sources
can be regarded as a uniform random variable.

\subsection{Equal mass, non-negligible spin, minimal eccentricity}

Simulations of binary black hole spacetimes where the initial black holes
have spin angular momentum have to date largely focused on equal 
mass black holes, and with the spin vectors 
having ``non-generic'' 
alignments~\cite{Campanelli:2006uy,Campanelli:2006fg,Campanelli:2006fy,
Herrmann:2007ac,Koppitz:2007ev,Pollney:2007ss,Campanelli:2007ew,
Choi:2007eu,Baker:2007gi,
Campanelli:2007cg,Vaishnav:2007nm,Schnittman:2007ij,Brugmann:2007zj}: either both 
black holes were given spins aligned
and/or anti-aligned with the orbital angular 
momentum or the spin vectors were set equal in magnitude but opposite in direction
and lying within the orbital plane. In all these configurations
the net angular momentum vector is aligned with the orbital angular
momentum, and thus there will be no precession of the orbital
plane during evolution (this, ignoring radiation-reaction,
is a consequence of conservation of angular momentum, which incidentally
is also why precession {\em does} occur in cases where the orbital
and net angular momentum are misaligned). A couple of exceptional
studies examining more generic spin configurations have been
presented in~\cite{Campanelli:2006fy,Campanelli:2007ew}.

There have not yet been the kinds of detailed or systematic studies
of inspiral with spin as described for the non-spinning
case in Sec.\ref{bbh_equal} in terms of understanding
the mulptipole structure of the waves, comparison with
PN inspiral waveforms, extraction of the QNMs, etc.
Though at least one can describe certain qualitative features of the merger.
Also, one of the more sought-after answers 
has been to the question of what the range of magnitudes of the recoil
velocity are when spin is included, and many of the above cited 
papers have recently addressed this.
In the next subsection we will outline the basics of what
changes during merger with spin, and the subsection
following that will describe the recoil results.

\subsubsection{Qualitative features of a merger of spinning black holes}
When the black holes are given spin, several aspects
of the merger can changed compared
to the non-spinning, equal mass case.
First, the net amount of energy/angular momentum radiated can
change significantly, and consequently the final spin and
mass of the remnant black hole. If the component of the net spin in 
the direction of the orbital angular momentum has the same (opposite)
sign as the orbital angular momentum, then typically more (less)
energy and angular momentum will be radiated compared to the
non-spinning case. One explanation for this comes from the PN
description of the spin-orbit interaction(see for example~\cite{Kidder:1995zr}), where in
the aligned (anti-aligned) case this interaction
term results in a repulsive (attractive) force between the black holes,
thus causing them to orbit for a longer (shorter) amount
of time emitting more (less) net radiation before merger.
As an example,~\cite{Campanelli:2006uy} (see also~\cite{Pollney:2007ss}) 
considered the merger of two equal mass black holes
with spin parameters $a=0.76$; for the case where the
two spin vectors were aligned with the orbital angular
momentum $\approx6.7\%$ of the rest-mass energy was radiated, leaving
a black hole with a spin of $\approx0.89$, whereas in the anti-aligned
case $\approx2.2\%$ energy was emitted, and the final black
hole had a spin of only $\approx0.44$ (in the direction of the
orbital angular momentum). Components of spin in the
orbital plane have a much smaller effect on the dynamics of the
orbit, and consequently the amount of energy emitted; for example,
the configurations that result in the largest recoil velocities
described in the next section are equal mass, have zero-net spin angular momentum with
the spin vectors lying in the orbtial plane, and in this
case the total energy and angular momentum radiated is very
close to the amount for the equivalent non-spinning 
case~\cite{Herrmann:2007ac,Campanelli:2007cg}.

A second significant effect of spin in a merger is that the
spin vectors and orbital angular momentum vector will in 
general precess, and near the time of merger by potentially
large enough amounts to cause spin and orbital plane ``flips''.
In PN-terms this can be thought of as due to spin-spin and spin-orbit
interactions~\cite{Kidder:1995zr}. A more Newtonian way of thinking
about these interactions is that a spinning black hole 
effectively has a quadrupole moment, and thus the exterior gravitational
field of the second black hole will in general exert a torque
on the first black hole (and vice-versa), causing precession
of the spins, and hence the orbital plane to conserve angular 
momentum (ignoring radiation). 
The only full numerical study to date of these effects 
were presented in~\cite{Campanelli:2007ew,Campanelli:2006fy},
though it will certainly not be long before more systematic
studies are available from several groups.

\subsubsection{Recoil velocities}\label{sec_max_k}

Any property of an orbit resulting in an asymmetric beaming pattern
for the gravitational waves could, via conservation of linear momentum,
impart a kick to the remnant black hole. As discussed in
Sec.~\ref{bbh_unequal} unequal masses produce such an asymmetry,
and so can individual black hole spins. An obvious example
is the asymmetry that would be produced by precession of the orbital plane, and
if the precession time scale is shorter than the orbital and
inspiral time scales (which it is near merger) then there will
not be enough time to average the momentum beamed in any one
direction to zero before merger, thus resulting in a net momentum
flux in some direction. For near-equal mass mergers this can produce larger
kicks that an unequal mass ratio alone---typically around several
hundred $km/s$. A less obvious source of asymmetry, though
one resulting in the largest kicks of up to $4000km/s$~\cite{Herrmann:2007ac,Campanelli:2007cg,Campanelli:2007ew,Baker:2007gi,Brugmann:2007zj},
are equal mass black holes with equal but opposite spin vectors
lying within the orbital plane. At a first glance this is a rather surprising
configuration for producing a large recoil, as
it is not obvious where the asymmetry in the energy emission
is. Thus it will be instructive to spend a bit more time
describing this configuration in the next couple of
paragraphs, and see how the kick can be understood as a frame 
dragging (or gravito-magnetic) effect. More technical explanations
of the source of the kick can be found in~\cite{Brugmann:2007zj,Schnittman:2007ij}
A discussion of the astrophysical
implications of such large recoil velocities is deferred 
to Sec.\ref{sec_imp}.

\begin{figure}
\begin{center}
\includegraphics[width=6.5in,clip=true]{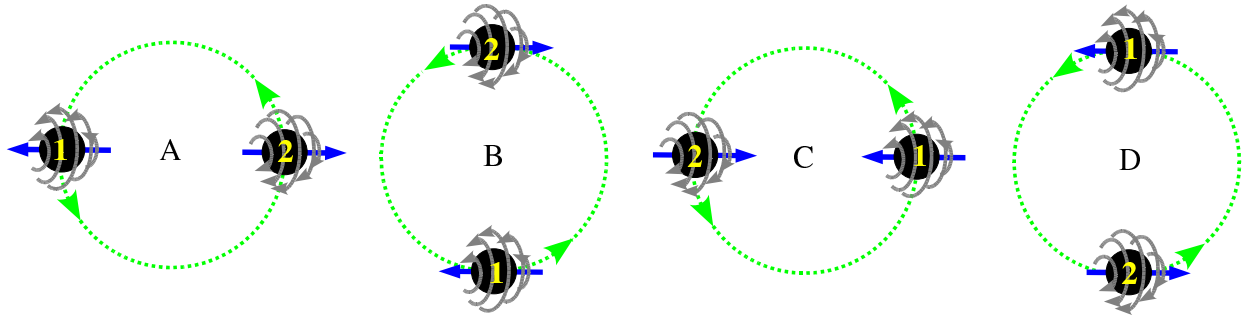}
\end{center}
\caption{A depiction of the orbital configuration resulting
in the largest kick velocities. The orbital angular momentum points
out of the paper in this case, and the spin vectors for each black hole
is in the orbital plane within the paper as shown by the solid blue vectors. 
The grey curved lines illustrate the dragging of spacetime about the
black hole caused by its spin.
\label{kick_pic}
}
\end{figure}

Consider the orbit depicted in Fig.\ref{kick_pic}, and imagine 
what the effect of rotation of black hole 2 on the motion
of black hole 1 would be. The rotation of black hole 2 causes
spacetime to be ``dragged'' about it following the right hand rule:
grasp the spin vector with your right hand so that the thumb
points in the direction of spin; then the direction in which the
rest of your fingers curl about the axis indicates how spacetime
is whirled about due to the spin of the black hole. Using
this image, notice that at phases (A) and (C) within
the orbit black hole 2 can not impart any effective velocity to black hole 1.
However, at phase (B) the dragging of the spacetime caused
by black hole 2 will cause black hole 1 to move in the negative
$z$ direction, where $z$ is in the direction of the orbital angular
momentum (i.e. it will move into the paper in the illustration), and the opposite
at phase (D). The same analysis of the effect of the rotation of
black hole 1 on black hole 2 shows that with this particular 
configuration of spins {\em both} black holes will at each
instant have the {\em same} velocity induced in the direction
normal to the orbital plane by the other black hole. In otherwords,
one can think of this as causing the entire orbital plane to oscillate
normal to the plane with the orbital frequency, or equivalently,
the trajectory of each orbit will be tilted by equal but opposite
angles relative to the original orbital plane.
This normal-motion by itself does not produce much radiation, however,
it does cause the more copious amounts of radiation caused by
the circular orbital motion to get blue-shifted in one direction
normal to the orbital plane while at the same time being red-shifted 
in the other. Averaged over one orbit, and ignoring radiation reaction,
the net doppler shift in any one direction is zero. However, as
the orbital radius begins to shrink due to radiation reaction,
the flux and the magnitude of the doppler shift increases until
about the time of merger. Up until this time the net momentum radiated
in the $z$ direction will be a function depending sinusoidally on the phase
of the orbit, and slowly increasing in amplitude. Depending on
where in the orbit the merger occurs ultimately determines the magnitude
and direction of the kick normal to the plane. 

Of course, the structure of spacetime in the vicinity of
two spinning black hole just about to merger will be quite non-trivial,
preventing one from unambiguously localizing the positions of the black holes
or, when and where the radiation is being produced, so the preceding
description of the production of a kick is somewhat cartoonish.
However, one can apply it at face value to come up with an order
of magnitude estimate for the kick which is in the correct
ball park, as well as account for the linear dependence
of the kick on the spin $a$ of the black holes and
sinusoidal dependence on initial orbital phase, as follows.
When far apart, at any instant in time black hole 1~(2)
will not have any component of orbital angular momentum in the direction
of the spin axis of black (2)~1. Thus, one can approximate the
instantaneous velocity imparted by frame dragging as that which
a particle, dropped from rest at infinity, would have falling toward
the black hole. This is a particle on a zero angular momentum orbit, 
which will have an instantaneous
$z$ velocity of 
\be
v_z(r,\theta) \approx \frac{2 m a \sin(\theta)}{r^2}
\ee
where $\theta$ is the angle relative to the spin axis of the black hole
with spin parameter $a$ and (total) mass $m$, and $r$ is the distance to 
the black hole. We have used the Boyer-Lindquist form of the Kerr
metric in the above, and only kept the term to leading order in $r$.
From this expression one immediately sees where the linear dependence
in $a$ and sinusoidal dependence on the phase arises. To estimate
the maximum possible kick velocity, note that this would be produced
if the doppler-shifting of the radiation ceases at maximum 
velocity, i.e. when $sin(\theta)=1$. Assume that this occurs
when the black holes merge, and this happens when the two black holes
``touch'', so when $r\approx4m$. Thus,
\be
v_{zmax} \approx \frac{j}{8},
\ee
where $j\equiv a/m$ is the dimensionless spin parameter.
The energy density $e$ of a gravitational wave is proportional to 
the square of the wave frequency; thus the doppler shifted energy
density will be proportional to $(1\pm2v_{zmax})e$,
and so the net momentum density radiated at this moment
will be $\delta p = 4 v_{zmax} e$. This accumulates over
the last part of an orbit, during which a total of $E=\epsilon M$ ($M=2m$)
of energy is emitted in gravitational waves. Therefore the net
momentum radiated in $z$ would be $\delta P = 4 v_{zmax} E$, giving
an estimate for the maximum recoil speed $\delta P/M$ as
\be
v_{recoil,max} \approx \frac{j\epsilon}{2}.
\ee
For a concrete number, take $\epsilon\approx0.01$ (which is not
too unreasonable given that the net energy emitted in the
last orbit/merger/ringdown is around $0.035$), then
for an extremal ($j=1$) black hole this gives 
$v_{recoil,max}\approx 1500km/s$. 

Note that a similar line of argument can give an intuitive
understanding of the effective repulsive (attractive)
force between binaries with spin axis aligned (anti-aligned)
to the orbital angular momentum, again due to
frame dragging. For empirical formulas giving the
net recoil for various spin and mass configurations
see~\cite{Baker:2007gi,Campanelli:2007cg}.

\subsection{Equal mass, large eccentricity, minimal spin}\label{sec_ecc}
An equal mass, zero spin but sizable eccentricity case
was in fact the first complete merger event simulated in~\cite{paper2}.
Adding eccentricity to the orbit was not intentional, rather this
was an artifact of the initial data method, which is as follows.
Boosted, highly compact concentrations of scalar field energy
are chosen for the initial conditions, which then quickly undergo
gravitational collapse and form black holes. Any remnant
scalar field energy quickly accretes into the black holes or radiates away from the
vicinity of the orbit, leaving behind, for all intents and purposes,
a vacuum black hole binary spacetime.
For a given initial separation of the
scalar field pulses, a single boost parameter $k$
controls the initial data---one scalar field pulse is placed at $(x,y,z)=(d,0,0)$
and given a boost $k^i=(0,k,0)$, while the second
is placed at $(-d,0,0)$ and given a boost $(0,-k,0)$.
It turns out for sufficiently
close separation (as used in the simulations) the resultant
black hole binary has non-negligible eccentricity regardless
of $k$. Furthermore, probably due to the scalar field dynamics
and accretion, the effective vacuum binary black hole orbit
that could be ascribed to the black holes has an apoapsis much
further out that the initial scalar field pulses. The consequences
are that for the smaller values of $k$ which result
in strong interaction of the black holes early on, the black holes
have much more kinetic energy than what black holes
on a slow, adiabatic inspiral at the same separation would have.
This offers an explanation for why the interesting 
threshold ``zoom-whirl'' behavior~\cite{zoom_whirl,zoom_whirl_b}
explained in the next paragraph could be observed
using this class of initial data, though at the
time this was puzzling as it was (incorrectly) assumed
the binary was in the adiabatic inspiral regime where any
``radiation-reaction'' effects would always force the binaries
to be closer on average from one orbit to the next.

\begin{figure}
\begin{center}
\includegraphics[width=8.0cm,clip=true]{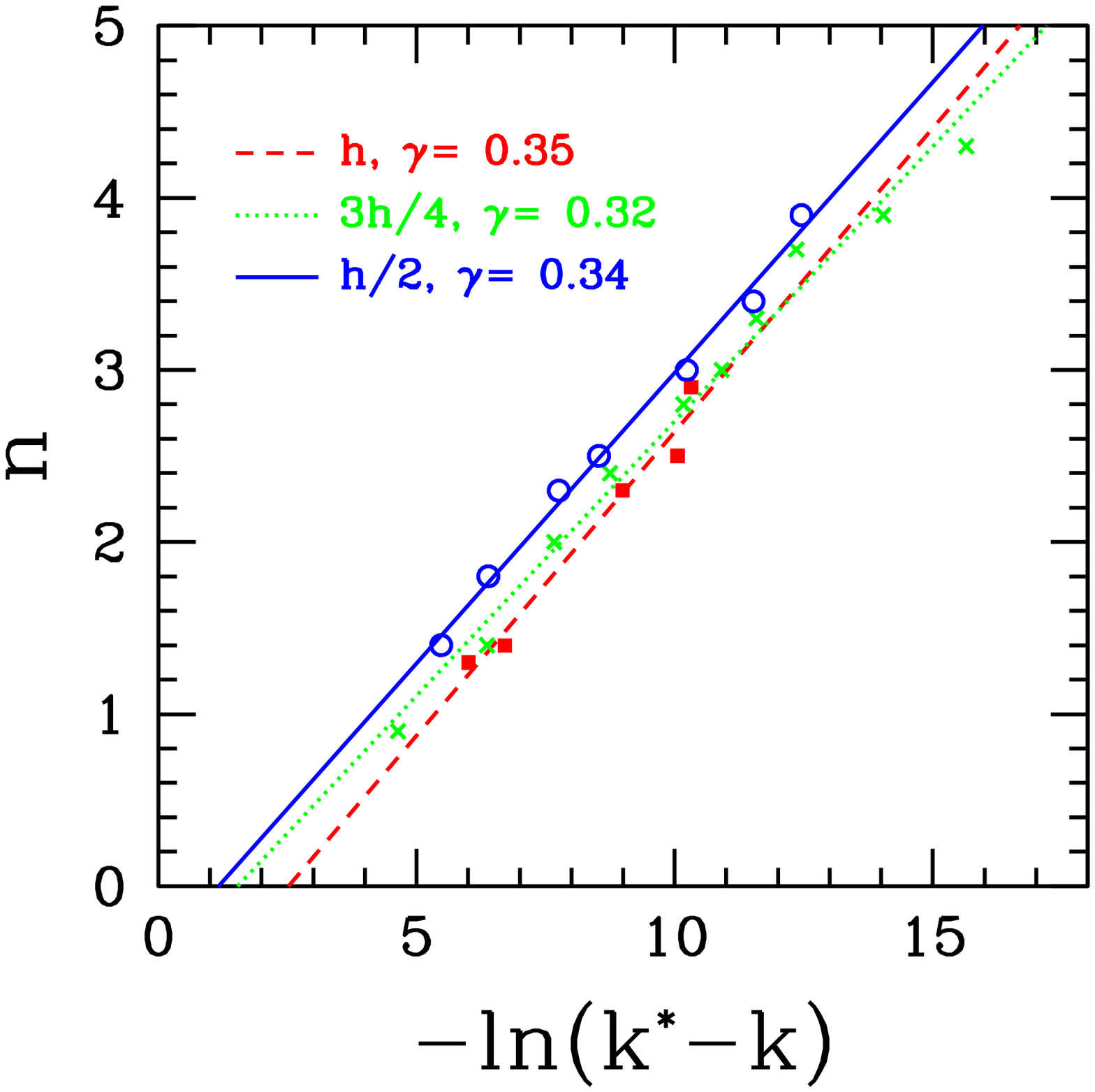}
\hspace{0.5cm}
\includegraphics[width=8.0cm,clip=true]{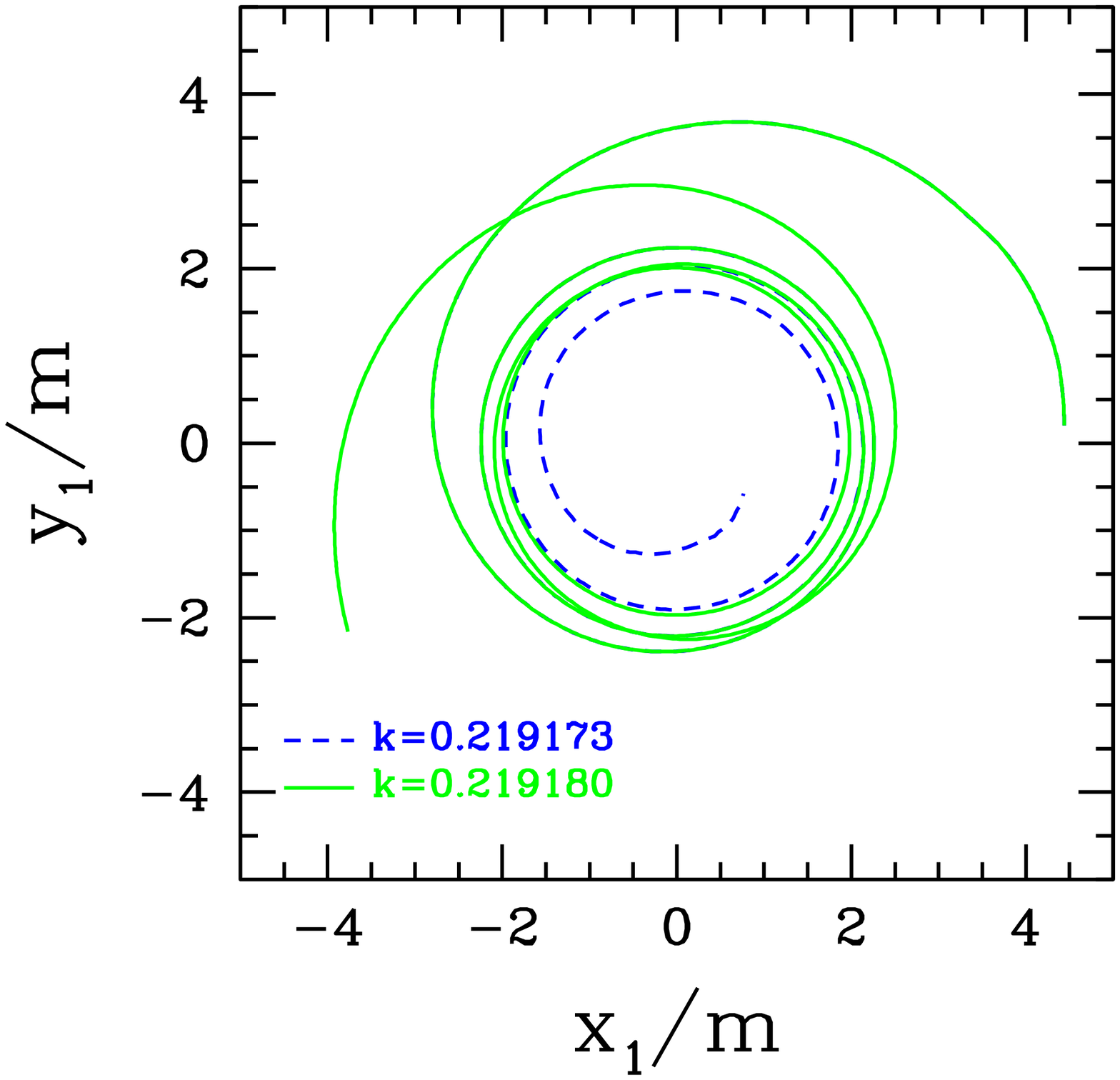}
\end{center}
\caption{Left: The number of orbits $n$ versus logarithmic
distance of the initial boost parameter $k$ from the
immediate merger threshold $\kstar$, for evolutions
that did result in a merger. Results from three
resolutions are plotted with characteristic mesh
spacings $h$ (lowest resolution), $3h/4$ and $h/2$ (highest)
to schematically illustrate convergence. For each
resolution, a least-squares
fit to the data is shown assuming the relation (\ref{n_gamma}).
Right: plots of the orbital motion from the two
higher resolution simulations ($h/2$) tuned closest
to threshold (only the coordinate motion of a single
black hole---initially at positive $x$---is shown for clarity).
The dashed curve is
the case resulting in a merger, and the curve ends
once a common apparent horizon is first detected, while
for the solid curve the black holes separate again and
here the curve ends when the simulation was stopped.
\label{fig_gamma_merge_orbit}
}
\end{figure}

\begin{figure}
\begin{center}
\includegraphics[width=8cm,clip=true]{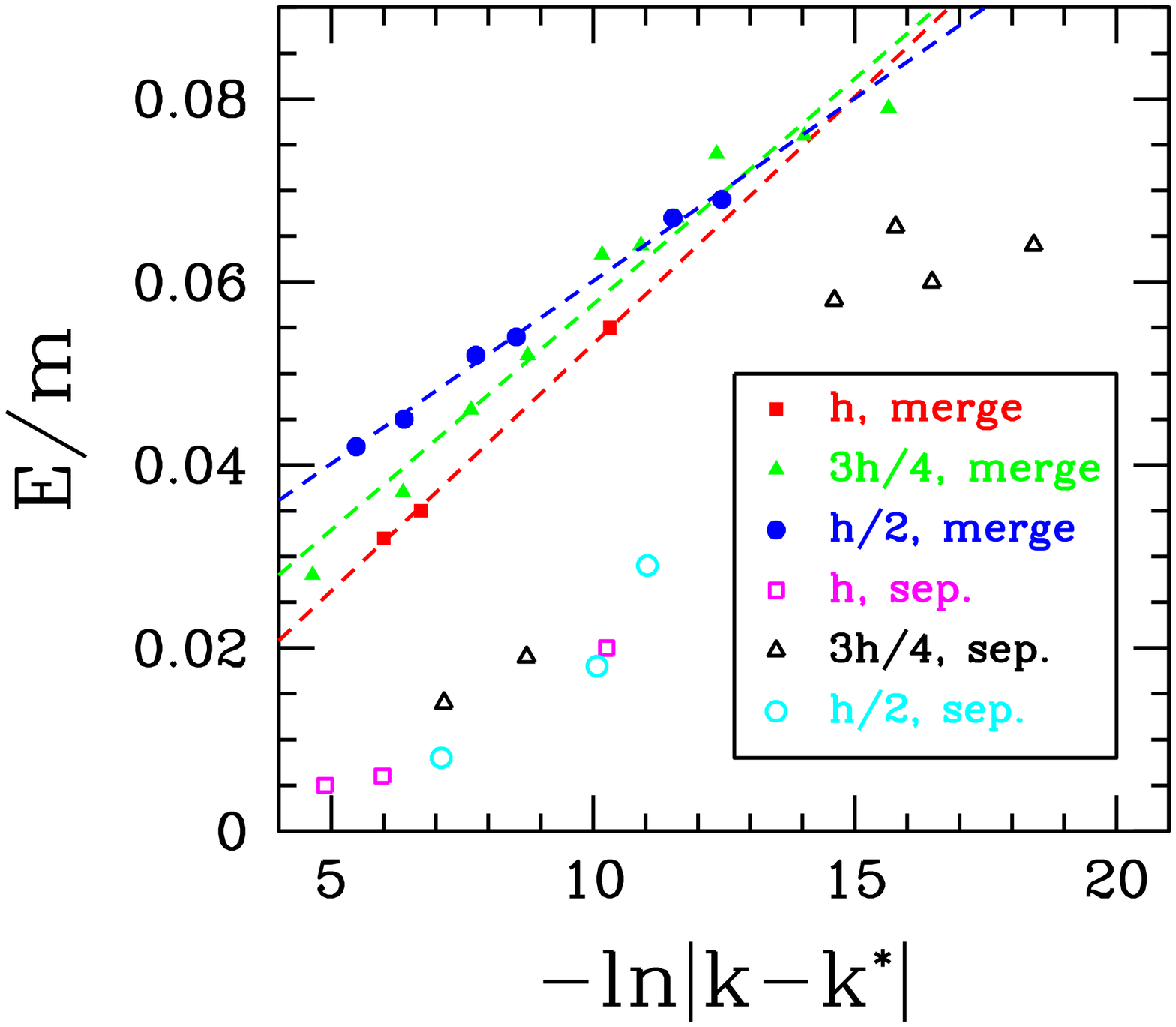}
\hspace{0.5cm}
\includegraphics[width=7.5cm,clip=true]{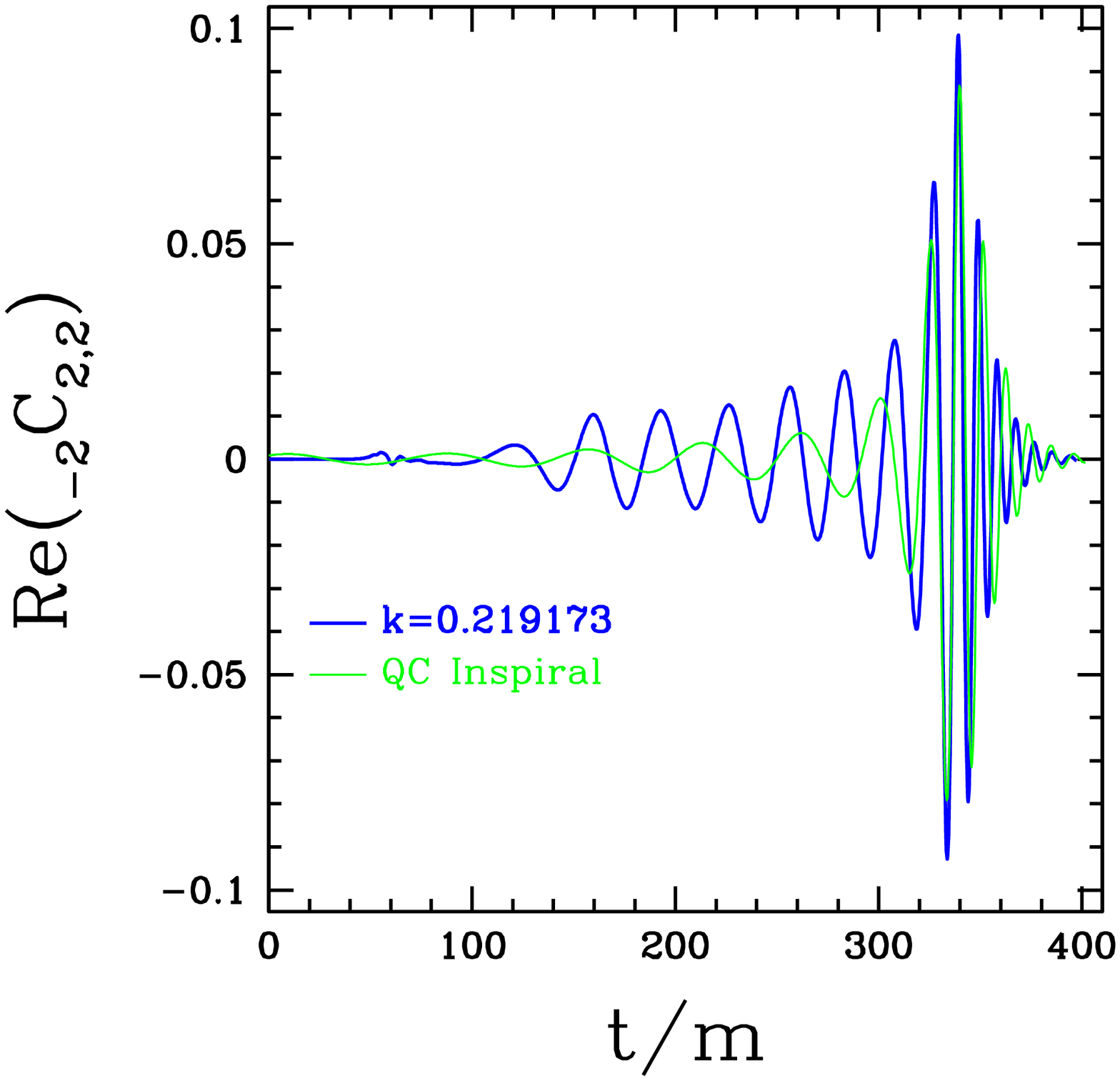}
\end{center}
\caption{
Left: the total energy radiated in gravitational waves
plotted as a function of logarithmic distance
from the immediate merger threshold. Data
from both super and sub critical cases are 
shown (and from each of the three characteristic resolutions run),
though for clarity only the former have added
regression lines.
Right: The gravitational waves emitted during
a merger event. The real part
of the dominant spin weight -2, $\ell=2$, $m=2$
spherical harmonic component of $\Psi_4$ is shown, and for
interest the corresponding representation of the signal from the quasi-circular
inspiral simulation depicted in Fig.~\ref{hpc} is also shown,
time-and-phase-shifted so that the waveforms match at peak
amplitude.
\label{fig_Lm2_c22r_super_dE_dt}
}
\end{figure}

Imagine what should happen as the boost parameter $k$ is
varied. At one extreme, $k=0$, there will be a
head on collision; at the other, $k\approx 1$, the black hole
trajectories will be deflected by some amount,
though ultimately they will fly apart and separate.
At intermediate values of $k$ there should be a
significant amount of close-interaction of the black holes,
and then they will either merger or separate (to possibly
merge at some time in the future).
What appears to happen near the threshold
value of $k$ between these two distinct end-states
is the black holes evolve toward an {\em unstable near-circular orbit}, 
remain in that configuration for an amount of time sensitively
related to the initial conditions, then either plunge
toward coalescence or separate. 
Specifically, for the {\em single} class of initial
conditions examined in~\cite{cqg_review,Pretorius:2007jn}, the number of
orbits $n$ observed near threshold is found to scale as
\begin{equation}\label{n_gamma}
e^{n} \propto |k-\kstar|^{-\gamma}
\end{equation}
with $\gamma \approx 0.34\pm0.02$---see Fig.'s ~\ref{fig_gamma_merge_orbit}
and ~\ref{fig_Lm2_c22r_super_dE_dt} that depict this
scaling relation for cases that merge,
near-threshold orbits,
a sample of the gravitational waves and 
the energy emitted energy as a function
of $k$.
Note that due to energy loss via gravitational radiation 
the threshold cannot be ``sharp'', i.e. if the time $t_m(k)$ to
merger is plotted as a function of $k$, this will {\em not} have a
discontinuous step at $k=\kstar$. There will be a maximum
number of orbits $N$ for a given class of initial conditions,
and what from a distance might appear like a step function will
be resolved into a smooth transition over a region of
size $\delta k \approx e^{-N/\gamma}$. Also note that the initial
conditions need to be highly fine tuned to obtain even a few
whirl-orbits. Thus, when close encounters of near equal mass black holes
on hyperbolic or highly eccentric orbits occur in 
nature (which might occasionally happen in a dense environment
such as a globular cluster), it is highly unlikely that it will
be a near-immediate-threshold encounter. 

The behavior just described is very similar to that of equatorial {\em geodesic}
motion on a Kerr background, where geodesics near
the threshold of capture approach the unstable circular geodesic
orbits of the background spacetime,
and exhibit similar scaling behavior of the number
of orbits versus distance from threshold as in (\ref{n_gamma}). In
that case, the scaling exponent $\gamma$ is inversely proportional
to the Lyapunov (or instability) exponent of the corresponding
unstable circular geodesic~\cite{CL}, and in fact numerically
has a value quite similar to that in the analagous equal mass 
scenario; for more information see the discussion in~\cite{Pretorius:2007jn}.
In contrast to near-equal mass binary black hole
encounters in the universe, extreme mass ratio inspirals of a compact
object into a supermassive black hole are expected to be numerous
enough that a significant number of zoom-whirl type orbits
will be seen with LISA~\cite{Gair:2004iv,AmaroSeoane:2007aw}. 

\subsection{Generic}

To date, there has not been any published systematic numerical studies
of fully generic initial binary conditions, namely with
varying mass ratios, spin magnitudes and orientations, and orbital
eccentricity. The reason is simply that this field is still young, 
though given the rapid rate at which new results have been released
over the past couple of years it should not be long before a rather
detailed knowledge of a large class of astrophysically relevant
merger spacetimes is available.

\section{Implications, prospects and questions}\label{sec_imp}

This article concludes with a discussion of some of
the implications of current results from the newly uncovered merger phase
of the two body problem and what questions still need to be addressed.
As before, the discussion is broken up into the rest-mass dominated
regime of relevance to astrophysical black holes, and the kinetic
energy dominated regime of the black hole scattering problem.

\subsection{Black holes in our universe}

Even though the merger phase has not yet presented 
any unexpected or bizarre phenomenology, that there are
finally concrete numbers and waveforms associated with
an ever growing set of initial conditions allows many consequences
of the merger to be seriously explored. The key numbers
are the amount of energy and net momentum
lost to gravitational waves, and knowledge of the structure
of the waves gives data analysts the information to build
trust-worthy template banks. Note that the topics discussed in the
next several sections hardly exhaust all the possible
consequences and applications of black hole mergers, and certainly
much of the future work on the ``two body'' problem in general
relativity will include a thorough examination, numerical
and otherwise, of mergers in astrophysical environments.

\subsubsection{Consequences of radiated energy}
An equal mass, non-spinning merger releases close to $\epsilon=4\%$ of the
rest-mass energy of the system into gravitational waves in
the last plunge/orbit, merger and ringdown. With spins,
depending on the relative alignment, this can increase or
decrease by roughly a factor of two. For an unequal mass
system with mass ratio $q=M/m$ ($q\geq 1$) the energy will
be reduced by slightly less than a factor of $q$ for small
$q$, though approaching a factor of $q^2$ for large $q$ (see Sec.~\ref{bbh_unequal}).
Thus for a ``major merger'', with $q$ a few or less, a significant
amount of the total gravitating energy of the system is effectively
lost on a very short time-scale.
To get an idea of just how short, 
define the {\em light crossing frequency} $f_{lc}$ of
a system to be the frequency at which light could cross back and
forth between the black holes separated by a distance
$R$, i.e. $f_{lc}=c/2R$. Converting to the units
given for the orbital frequency for an equal mass merger
in ({$\ref{omega_orbit}$), this is
\be
f_{lc} \approx 51{\rm kHz}\left(\frac{R_s}{R}\right)
                          \left(\frac{\msun}{M}\right)
\ee
Note that this is the {\em fastest} frequency at which any
causal process in the close vicinity of the binary could
operate, and by comparison with ({$\ref{omega_orbit}$)
one can see that near coalescence ($R\rightarrow R_s$) the
orbital frequency becomes a sizable fraction of this
maximum possible frequency. The time-scale over which
the final burst of energy is released will therefore be much shorter
that almost any other astrophysical process that could
be happening close to the binary. A likely non-vacuum
environment for a binary is a circumbinary gas disk.
Thus, a near-term effect of the passing gravitational waves
on the particles in the disk is that essentially 
instantaneously the central mass they are orbiting
will drop by a fraction $\epsilon$\cite{Milosavljevic:2004cg}; said another way,
if they were initially following circular orbits in a
potential with mass $M$, they will suddenly be
on eccentric orbits about
a potential of mass $M(1-\epsilon)$. Such a rapid
perturbation of the  disk could set up asymmetric waves
and warping in the disk which could conceivably produce
a weak but prompt electromagnetic counterpart to the merger event,
though no detailed calculations of this have yet
been performed. A related phenomena 
accompanies the secular evolution of the disk as the inspiral
and merger occurs~\cite{Milosavljevic:2004cg}.
Early on in the inspiral phase the viscous timescale
in the disk is shorter than the inspiral rate, allowing the
inner edge of the disk with radius roughly twice
the binary semi-major axis~\cite{Macfadyen:2006jx,Hayasaki:2006fq} 
to ``follow'' the inspiral.
However, eventually the inspiral time becomes
much shorter than the viscous time, leading to an essentially
non-accreting
disk about the final black hole that is much further out
than what a steady-state accretion disk would be (the innermost
stable circular orbit). The subsequent inward migration
of the disk and turn-on of accretion will produce
a strong X-ray afterglow on a timescale 
of $\approx 7(1+z)(M/10^6\msun)^{1.32}yr$ (with $z$ the cosmological
redshift)
that could be seen by future X-ray 
observatories~\cite{Milosavljevic:2004cg,Dotti:2006zn}. 

\subsubsection{Consequences of radiated momentum} 
One of the more significant potential astrophysical consequences
of a merger is when asymmetric radiation of linear momentum
occurs, resulting
in the recoil of the remnant black hole as discussed
in section~\ref{sec_res}. The largest recoil speeds of several thousand
kilometers per second for near-equal mass mergers are high enough to be able to 
eject the remnant from even the most massive galactic halo. 
If such large kicks are common, it would seem to be
in contradiction with the observation of supermassive
blackholes in most galaxies, and hierarchical
structure formation scenarios~\cite{Rees:1984si,Ferrarese:2004qr,White:1977jf,Springel:2005nw}. Note that the large recoils require each black hole
to be spinning by a fair amount ($a>\approx0.3$)---X-ray observations of
relativistic line broadening in a few AGN suggest spins
{\em are} high, with 
$a>0.9$\cite{Iwasawa:1996uh,Fabian:2002gj,Reynolds:2002np,Brenneman:2006hw}. 
As with the effects
of energy loss discussed in the previous section, large
recoils would also require major mergers, as the kick
scales roughly as $1/q^2$ for large mass ratio $q$~\cite{Baker:2007gi,Campanelli:2007cg}. 
Recent estimates using the effective-one-body model,
calibrated with some full numerical results,
suggest that for mergers with spin magnitudes of $a=0.9$, {\em uniformally} distributed 
spin configurations and mass ratios $1\leq q\leq 10$,
only about $3\%$ ($10\%$) of mergers result in kicks
greater that $1000km/s$ ($500km/s$)~\cite{Schnittman:2007sn}. 
Small kicks could still eject black holes
from galaxies in the early universe when halos were much less massive,
though the presence of supermassive black holes at the
centers of most galaxies might be a more robust consequence
of structure formation than naively thought in light of kicks.
One study examined the effect of natal kicks in a scenario
where supermassive black holes are formed in a primary halo
via capture of intermediate mass black holes from surrounding
secondary halos, and found that kick velocities imparted
to mergers of the intermediate mass black holes had
little affect on the growth of the supermassive black 
hole~\cite{Micic:2005gj}. Another study following black hole
formation through a simplified merger tree model found
that even when a high probability of ejection was assigned
to merger events, still more than $50\%$ of galaxies today
retain their supermassive black holes~\cite{Schnittman:2007nb}.
An examination of the effect of large recoil velocities
on the predicted event rates for LISA to detect supermassive
mergers suggested the event rate would drop by 
{\em at most} $60\%$ if
the seed black holes were light ($\approx 10^2\msun$, from Population III stars)
or by at most $15\%$ if the seeds were heavier ($\approx 10^4\msun$,
from direct gas collapse of primordial 
disks)\footnote{in both cases configurations were assumed to be most
favorable for large kicks, which is probably a significant
overestimate}~\cite{sesana:2007zk}.

The study~\cite{Schnittman:2007sn} mentioned in the preceding paragraph
on the probability of kicks assumed a uniform probability distribution for 
the spin orientation
in the progenitor binary system.
However, the distribution of initial spins is likely highly 
non-uniform---\cite{Bogdanovic:2007hp}
have shown that in gas-rich mergers, torques from the surrounding
gas will tend to align the black hole's spin vectors with the net
angular momentum vector of the gas, a configuration which results in
much more modest recoils of $<200km/s$. Thus, only a 
small fraction of supermassive mergers would result in the black hole
leaving the host galaxy. For those that do, there could
be significant electromagnetic counterparts due to the recoil, as
the black hole will drag the inner part (tens of thousands
of Schwarzschild radii) of any accretion
disk with it~\cite{Loeb:2007wz}. However, given that such recoils
would preferentially occur in gas-{\em poor} environments, accretion-related
counterparts might also be correspondingly dim. A search
for doppler shifted emission lines from quasars in the
Sloan Digital Sky Survey, which would be a signature of
an ejected accretion disk, placed upper
limits of incidence of recoiling black holes
in quasars at $4\%$ ($0.35\%$) for kicks greater
that $500km/s$ ($1000km/s$) in the line-of-sight ~\cite{Bonning:2007vt}.
Note that similar
arguments~\cite{Bogdanovic:2007hp} also place doubt on a common 
explanation that X-shaped jets from radio-loud AGN are the
result of spin re-alignment from a recent merger 
event~\cite{ekers,DennettThorpe:2001vy,Merritt:2002hc,Komossa:2003wz,Vir Lal:2006ce,Cheung:2007bv}.
Kicks of smaller velocities that temporarily displace the black
hole from the galactic center could also have interesting
consequences, since this will transfer energy to stars in the
nucleus, softening a steep density cusp~\cite{Merritt:2004xa}.
Also, modest recoil velocities will have a pronounced effect
on the black hole population of globular 
clusters, effectively depleting the clusters of a large
fraction of their black holes and leading to a ``rogue'' population
of wandering black holes in the galactic 
halo~\cite{Miller:2002pg,O'Leary:2005tb,HolleyBockelmann:2007eh}.

\subsubsection{Implication of waveform structure for detection efforts}

The relative simplicity of the merger waveform, assuming the trend
in new results continue and no complicated and lengthy structures
in the merger phase occur generically, is on one hand a boon
for gravitational wave astronomy, and on the other hand a curse.
On the positive side is that the simple transition from inspiral
to ringdown should allow the construction of high fidelity
hybrid or fully analytic template banks, such as recently
presented in~\cite{Buonanno:2007pf}. This will ensure,
if the circumstances of the sources are consistent
with the assumptions of the models, that
the waves from the highest possible fraction of events passing
earth during the operation of the various instruments will be
detected. The downside is that, the
less structure and shorter the waves, the more difficult it will
be to discriminate between different events, and the less
confidence with which one could claim observation of a {\em particular}
event, statistics of source populations, etc. 
Indeed, in~\cite{Baumgarte:2006en,Pan:2007nw} it was shown that high 
fitting factors can easily be achieved between 
a numerical source model and some member from a ``wrong'' template family.
Note that this problem is only a significant issue for
sources where the part of the waveform dominating the SNR
comes from the last few cycles of inspiral, merger and ringdown, i.e.
essentially the final burst (for LIGO this will be the merger
of tens to hundreds of solar mass black holes, for LISA
around $10^7-10^8 \msun$ supermassive black holes). 
When the inspiral portion of the event
is visible to detectors it could be in band for hundreds to thousands
of cycles, and in that case even small differences in the phase
evolution relative to a given template could drastically affect the 
SNR.
One way to deal with this problem for burst-like sources is to 
use a small, core template
family to search for a given source, then have an expanded
control group of template families with systematic deviations
from the core family that will be used to place confidence levels
and/or error-bars on any conclusions reached with the core family.
For example, say one wants to test the hypothesis that all stellar
mass mergers occur in environments where the orbits have circularized
well before the time of merger. The core template family for
this search would thus be a set of zero-eccentricity inspiral events,
and the control group would be a set of similar inspirals 
with eccentricity. The point here is not
to go into details of data analysis, rather it is to emphasize how
important it will be to investigate non-standard, unexpected or unusual
scenarios, not just for the hope of a serendipitous detection
of a surprising event, but to strengthen the science that
could be done with the usual and more mundane detections.
This is perhaps most true for what will be the first triumph
of detection of a binary black hole merger---confirmation of the
{\em existence} of black holes. It is easy to lapse into a mental
image of a black hole as this dark, concrete object with a surface,
rather than {\em the boundary demarking a region where the 
geometric structure of spacetime is undergoing gravitational
collapse---an instrinsically dynamical regime where space and
time itself are funneled to a singular state outside of the
grasp of contemporary physical theories}. To claim that such
a remarkable scenario exists in the universe demands
strong evidence, and part of this evidence would be
being able to quantify exactly how distinctive the signatures
of mergers of binary black holes within Einstein's theory of gravity are. 
For example, could compact boson
star ~\cite{Ruffini:1969qy,Colpi:1986ye,Palenzuela:2007dm},
``gravistar''~\cite{Chirenti:2007mk}, or other exotic object mergers
produce inspiral signals that would be
detected using a black hole inspiral template family? How different
would a metric theory of gravity have to be to produce observable
differences from Einstein's theory (yet be consistent in the
weak field)? Could merger events contain the signatures of
certain extra-dimension scenarios? The list of such questions
is endless, though a reasonable subset will need to be addressed 
if only to bolster our confidence about what
possible future detections could tell us about general relativity
and how acurately it describes spacetime.

\subsection{The black hole scattering problem}\label{sec_scatt}

There is no-known natural mechanism in the universe that
can accelerate black holes to ultra-relativistic velocities,
and hence the black hole scattering problem is largely a thought
experiment that can probe a very interesting regime
of Einsteins' theory. However, over the past few years several ideas
have emerged suggesting that an understanding of this problem could have relevance
to high-energy particle physics experiments---this will
briefly be discussed in the following section.
As with rest-mass dominated collisions until recently, full solutions
to the metrics describing ultra-relativistic collisions
have eluded analytic attempts to obtain them.
Most of what is known about this regime comes
from studies of the collision of infinite-boosted Schwarzschild
black holes, each described by the Aichelberg-Sexl metric~\cite{Aichelburg:1970dh}. 
In an impact with zero or small impact parameter, 
an apparent horizon has been found
at the moment of impact~\cite{penrose,D'Eath:1976ri,D'Eath:1992hb,D'Eath:1992hd,D'Eath:1992qu,Eardley:2002re,yoshino_nambu,berti_et_al,yoshino_rychkov,cardoso_et_al}; this is not a trivial statement,
as the Aichelberg-Sexl metric does not contain an event horizon,
and in this extreme case one might imagine that a naked singularity
would form, in particular by drawing parallels to collisions
of infinite plane gravitational waves~\cite{Khan:1971vh}.
For zero impact parameter, perturbative studies~\cite{D'Eath:1976ri,D'Eath:1992hb,D'Eath:1992hd,D'Eath:1992qu} have given a description
of the gravitational waves released in the process. Now, as
is turning out with merger simulations in the rest-mass dominated regime, it may be
that full (presumably numerical) solutions of the scattering
problem will only add some details to our understanding of
the process, though of course this will not be known until
the solutions are discovered. Furthermore, as described in 
Sec.\ref{sec_bhs}, the threshold of immediate merger could have
very interesting behavior associated with it, where essentially
all of the energy in the spacetime is converted to gravitational
waves. Note that in the infinite boost limit the threshold
of immediate merger also corresponds to 
a threshold of black hole {\em formation}. If, as conjectured
by Choptuik~\cite{Choptuik:1992jv}, the threshold of black hole
formation has a universal solution, then at critical impact parameter
the structure of spacetime should be described by the Abrahams-Evans
axisymmetric critical collapse vacuum solution~\cite{Abrahams:1993wa}.

The black hole scattering problem will be difficult
to simulate numerically. First of all, it is unclear
whether generalized harmonic or BSSN evolutions could 
be used without modification in this regime. Attempted
evolution of boosted exact Schwarzschild solutions with Lorentz
$\gamma$-factors above around $1.7$ with the generalized
harmonic code used in~\cite{paper1} suggested that at the
very least new gauge conditions will be needed for long-term
stable evolution. The BSSN 
code in~\cite{Sperhake:2006cy} has been used to evolve boosted
Bowen-York black holes with somewhat larger $\gamma$ factors,
though such initial data contains gravitational waves, and 
apparently a considerable amount with larger boosts~\cite{uli}.
A further issue with evolving highly boosted black holes 
is the tremendous computational resources that would
be required. Consider a single blackhole with total
energy $E$, which will be the characteristic length scale
of the geometrically non-trivial portion of the spacetime.
The black hole would have a rest mass of $E/\gamma$, which would
be the smallest length scale in the direction transverse to
the boost direction. Length contraction in the direction of the
boost will compress the horizon by an addition factor of $\gamma$
in that direction.
Furthermore, using a ``naive'' boost of the Schwarzschild solution,
certain components of the metric get scaled by factors of $\gamma^2$.
Thus, in all, to obtain a numerical solution with a grid-based
method will require a mesh spacing a factor of $\gamma^4$ smaller
than a similar rest-mass dominated problem and obtain similar
accuracy. To be well within
the kinetic energy dominated regime would require $\gamma\approx10-100$,
implying around $10^4-10^8$ times the computational resources.
Of course, this is rather simplistic accounting, and certainly
with some ingenuity a couple of orders of magnitude could
be shaved off the estimated cost.

\subsection{High energy particle experiments and black hole collisions}\label{sec_he}

Recently proposed
extra-dimension scenarios~\cite{arkani_hamed_et_al,randall_sundrum},
offer the intriguing possibility that
the Planck scale could be within reach of energies
attainable by the Large Hadron Collider
(LHC)~\cite{banks_fischler,giddings_thomas,dimopoulos_landsberg,kanti,gingrich,gingrich2}.
This implies that the LHC may be able to probe the quantum gravity regime,
and that black holes could be produced in substantial quantities
by the particle collisions. Similarly, cosmic ray collisions
with the earth would produce black holes~\cite{feng_shapere}, and this
may be detected with current or near-future cosmic ray
experiments~\cite{landsberg}. In the collision of two particles
with super-Planck kinetic energies, gravity dominates the interaction,
and thus to a good approximation the collision can be modeled
as the ultra-relativistic collision of two black holes~\footnote{Though without
a full theory of the physical laws that would operate in this
regime, such statements are a bit hand-waving.}.
Another intriguing application of ultra-relativistic black hole
collisions is in 5 dimensional AdS spacetime, and how that
might relate to the collision of gold ions at the Relativistic
Heavy Ion Collider (RHIC).
At RHIC, gold ions are accelerated to Lorentz gamma factors of around 100
before colliding. It is believed that during the collision
a quark-gluon plasma (QGP) is formed. The present data supports
this idea~\cite{Arsene:2004fa,Adcox:2004mh,Back:2004je,Adams:2005dq} though
there are some puzzles, in particular
why the QGP is strongly interacting, behaving almost
like an ideal fluid (the energies of RHIC collisions are in the
regime where the asymptotic freedom of QCD should be manifest, implying
one should have a weakly interacting QGP). One suggested
method for deriving properties of the QGP is via the AdS/CFT
correspondence of string
theory~\cite{Maldacena:1997re,Witten:1998qj,Gubser:1998bc}.
Specifically, the
supposition is that $N=4$ super-Yang-Mills (SYM) theory
at strong coupling, though different
in many respects from QCD, can describe some of the features of
a ``real-world'' QGP, and that a practical way of calculating the
relevant SYM state
is using the AdS/CFT map applied to the corresponding
process in 5D AdS spacetime. It has been suggested that the
AdS equivalent process is the collision of two black holes~\cite{Nastase:2005rp},
and in~\cite{Friess:2006kw} the quasinormal ringing of a
perturbed 5D AdS black hole,
which represents the final stage of a black hole collision and
may describe aspects of
thermalization and collective flow of the QGP,
was used to provide in-the-ballpark estimates of the thermalization
time and elliptic flow coefficients of an anisotropic heavy ion
collision.

Thus there is considerable motivation to study black hole
collisions in higher dimensional spacetimes, and in spacetimes
with different asymptotic structures in regimes where the
asymptotics are expected to affect the physics of the
collision (in particular for collisions in AdS the black holes
need to be ``large'' in terms of the length scale imparted by the
cosmological constant). Full five (and higher) dimensional
numerical simulations of collisions, in particular ultra-relativistic
ones, will require computers several generations more powerful
than current ones. However, if the analogy between geodesic
behavior and the full problem at the threshold of immediate
merger described in Sec.~\ref{sec_ecc} holds, then
for an application to the LHC all that would be needed is a
head-on collision simulation, which could be reduced
to an 2D simulation~\footnote{Of course, the ``catch-22'' here
is that without doing the full n-dimensional problem it
will not be known whether the analogy holds}. Furthermore,
if the threshold geodesic behavior of Myers-Perry 
solutions~\cite{Myers:1986un} are 
an adequate description of the analogous problem in
higher dimensions, it turns out that to a good approximation
for dimensions greater than $5$
the energy emitted in an ultra-relativistic collision
will be given by~\cite{cm}
\be
E(b)=E_0 \left(\Theta(b)-\Theta(b-b^*)\right),
\ee
where $\Theta(b)$ is the unit step function, $b$ the impact parameter, 
$b^*$ the critical
impact parameter for the geodesic (which is close to the
Schwarzschild radius of the equivalent black hole),
and $E_0$ the energy emitted during the head-on 
collision (estimates of which can be found in~\cite{yoshino_rychkov},
for example). In otherwords, when black holes form, 
as much energy is lost to gravitational waves as in
the head-on collision case, regardless of the impact parameter.
This ``missing energy'', in addition to Hawking radiation
emitted when the black holes evaporate, could be used
to detect this hypothesized scenario at the LHC.

\section{Conclusion}

The two body problem in general relativity is a fascinating,
rich problem that is just beginning to be fully revealed
by recent breakthroughs in numerical relativity. At the 
same time, a new generation of gravitational wave detectors
promise to offer us a view of the universe via the gravitational
wave spectrum for the first time. Black hole mergers
are a promising source for gravitational waves, and detecting
them would provide direct evidence for these remarkable objects,
while providing much information about their environments.
Suggestions that there might be more than four spacetime
dimensions offers the astonishing possibility that
black holes could be produced by proton collisions
at $TeV$ energies, which will be reached by the Large Hadron
Collider, planned to begin operation within a year. Given
all this, it is difficult not to be excited about what
might be learnt about the universe from the smallest to largest
scales during the next decade, and that black hole collisions
could have something important to say at both extremes.

\bigskip

\noindent{\bf{\em Acknowledgments:}}
I would like to thank Alessandra Buonanno, Matthew Choptuik, Gregory Cook, Charles Gammie, 
David Garfinkle, Steven Gubser, Carsten Gundlach, 
Luis Lehner, Jeremiah Ostriker, Don Page, David Spergel and Ulrich Sperhake for many
stimulating conversations related to some of the discussion presented
here.

\bibliography{references}

\end{document}